\newif\ifhal
\halfalse
\haltrue

\ifhal
\documentclass[reprint,%
]{sigplanconf}
\else
\documentclass[]{sigplanconf}
\fi

\usepackage[utf8]{inputenc}
\usepackage[T5,T1]{fontenc}
\usepackage{amsmath}
\usepackage{mathtools}
\usepackage{amssymb}
\usepackage[retainorgcmds]{IEEEtrantools}
\usepackage{amsthm}
\usepackage{nicefrac}
\usepackage{stmaryrd}
\usepackage[bb=boondox]{mathalfa}
\usepackage{mathpartir}
\usepackage[pass]{geometry}
\usepackage{enumitem}
\usepackage{balance}

\newif\ifdirtytricks
\dirtytricksfalse
\newif\ifanonymous
\anonymousfalse

\usepackage{setup}

\begin{document}

\ifhal\else\toappear{}\fi
\copyrightyear{2016}
\conferenceinfo{ICFP\,'16}{September 18--22, 2016, Nara, Japan}
\copyrightdata{978-1-4503-4219-3/16/09}
\copyrightdoi{http://dx.doi.org/10.1145/2951913.2951928}

\setlength{\pdfpageheight}{\paperheight}
\setlength{\pdfpagewidth}{\paperwidth}



\title{
  Set-Theoretic Types for Polymorphic Variants

\authorinfo{Giuseppe Castagna$^1$ \and Tommaso Petrucciani$^{1,2}$ \and Kim \Nguyen$^3$}
           {$^1$CNRS,  Univ Paris Diderot, Sorbonne Paris Cité,  Paris,
France\\\qquad
$^2$DIBRIS, Università degli Studi di Genova, Genova, Italy
\\\qquad
$^3$LRI, Universit\'e Paris-Sud, Orsay, France\qquad}
}{
\vspace{-1.5cm}
}

\maketitle

\begin{abstract}
Polymorphic variants are a useful feature of the OCaml language whose
current definition and implementation rely on kinding constraints to
simulate a subtyping relation via unification. This yields an awkward
formalization and results in a type system whose behaviour is in some
cases unintuitive and/or unduly restrictive.

In this work, we present an alternative formalization of polymorphic
variants, based on set-theoretic types and subtyping, that yields
a cleaner and more streamlined system.  Our formalization is more
expressive than the current one (it types more programs while preserving
type safety), it can internalize some meta-theoretic properties, and it
removes some pathological cases of the current implementation resulting
in a more intuitive and, thus, predictable type system.
More generally, this work shows how to add full-fledged union types to
functional languages of the ML family that usually rely on the
Hindley-Milner type system.  As an aside, our system also improves
the theory of semantic subtyping, notably by proving completeness for
the type reconstruction algorithm.

\end{abstract}




\section{Introduction}\label{intro}

Polymorphic variants are a useful feature of OCaml, as they balance
static safety and code reuse capabilities with a remarkable
conciseness. They were
originally proposed as a solution to add union types to Hindley-Milner
(HM) type systems \cite{Garrigue2002}. Union types have several
applications and make it possible to deduce types that are finer grained than
algebraic data types, especially in languages with pattern
matching. Polymorphic variants cover several of the applications of union
types, which explains their success; however they
provide just a limited form of union types: although they offer some sort
of subtyping and value sharing that ordinary variants do not, it is still not possible
to form unions of values of generic types, but just finite
enumerations of tagged values. This is obtained by superimposing on the HM type system
a system of kinding constraints, which is used to simulate
subtyping without actually introducing it. In general, the current
system reuses the ML type system---including unification for type
reconstruction---as much as possible. This is the source of several
trade-offs which yield significant complexity, make polymorphic variants hard to
understand (especially for beginners), and jeopardize
expressiveness insofar as they forbid several useful applications that
general union types make possible.

We argue that using a different system, one that departs drastically
from HM, is advantageous. In this work we advocate the use of
full-fledged union types (i.e., the original motivation of polymorphic
variants) with standard set-theoretic subtyping. In particular we
use \emph{semantic subtyping}~\cite{Frisch2008}, a type system where
$(i)$~types are interpreted as set of values, $(ii)$~they are enriched
with unrestrained unions, intersections, and negations interpreted as
the corresponding set-theoretic operations on sets of values, and
$(iii)$~subtyping corresponds to set containment.  Using set-theoretic
types and subtyping yields a much more natural and easy-to-understand
system in which several key notions---e.g., bounded quantification and
exhaustiveness and redundancy analyses of pattern matching---can be
expressed directly by types; conversely, with the current
formalization these notions need meta-theoretic constructions (in the
case of kinding) or they are meta-theoretic properties not directly
connected to the type theory (as for exhaustiveness and redundancy).

All in all, our proposal is not very original: in order to have the
advantages of union types in an implicitly-typed language, we simply
add them, instead of simulating them roughly and partially
by polymorphic variants. This implies to generalize notions such as   instantiation and generalization to
cope with subtyping (and, thus, with unions). We chose not to start from scratch, but
instead to build on the existing: therefore we show how to add
unions as a modification of the type checker, that is, without
disrupting the current syntax of OCaml. Nevertheless, our results can
be used to add unions to other implicitly-typed languages of the
ML family.

We said that the use of kinding constraints instead of full-fledged
unions has several practical drawbacks and that the
system may therefore result in unintuitive or overly restrictive behaviour. We
illustrate this by the following motivating examples in OCaml.

\paragraph{\textsc{Example 1:} loss of polymorphism.} Let us consider the identity function and its application to a polymorphic variant in OCaml (``\Caml!#!'' denotes the interactive toplevel prompt of OCaml, whose input is ended by a double semicolon and followed by the system response):
\begin{CamlBlock}
# let id x = x;;
val id : 'a -> 'a = <fun>
# id `A;;
- : [> `A ] = `A
\end{CamlBlock}
The identity function \texttt{id} has type%
\ifdirtytricks{} \else\footnote{Strictly speaking, it is a \emph{type scheme}: cf. Definition~\ref{def:variants-k-type-schemes}.}\fi
$\Scheme{\alpha.\alpha\to\alpha}$ (Greek letters denote type
variables). Thus, when it is applied to the polymorphic variant
value \Caml:`A: (polymorphic variants values are literals prefixed by a backquote),
OCaml statically deduces that the result will be of type ``at
least \Caml!`A!'' (noted \Caml![> `A]!), that is, of a type greater than or equal to the type whose only value is \Caml!`A!. Since the only value
of type \Caml![> `A]!
is \Caml!`A!, then the value \Caml!`A! and the
expression \Caml:id `A: are completely interchangeable.\footnote{Strictly speaking,  \Caml!`A! is the only value \emph{in all instances} of \Caml![> `A]!: as shown in Section~\ref{mlvariants} the type  \Caml![> `A]! is actually a constrained type variable.}
For instance, we can
use \Caml!id `A! where an expression of type ``\emph{at most} \Caml!`A!''
(noted \Caml![< `A]!) is expected:
\begin{CamlBlock}
# let f x = match x with `A -> true;;
val f : [< `A ] -> bool = <fun>
# f (id `A);;
- : bool = true
\end{CamlBlock}
Likewise \Caml!`A! and \Caml!id `A! are equivalent in any context:
\begin{CamlBlock}
# [`A; `C];;
- : [> `A | `C ] list = [`A; `C]
# [(id `A); `C];;
- : [> `A | `C ] list = [`A; `C]
\end{CamlBlock}
We now slightly modify the definition  of the identity function:
\begin{CamlBlock}
# let id2 x = match x with `A | `B -> x;;
val id2: ([< `A | `B ] as 'a) -> 'a = <fun>
\end{CamlBlock}

Since \Caml!id2! maps \texttt{x} to \texttt{x}, it still is the identity function---so it has type
$ \alpha\,{\to}\,\alpha $---but since its argument is matched
against \Caml!`A | `B!, this function can only be applied to arguments of
type ``at most \Caml!`A | `B!'', where ``\texttt{|}'' denotes a union.
Therefore, the type variable $\alpha$ must
be constrained to be a ``subtype'' of \Caml!`A | `B!, that is,
$\Scheme{(\alpha \leq \TagLit{A} \KeywOr \TagLit{B}). \alpha \to \alpha} $, expressed by the
OCaml toplevel as \Caml!([< `A | `B ] as 'a) ->$$ 'a!.

A priori, this should not change the typing of the application of the (newly-defined) identity to  \Caml!`A!, that is, \Caml!id2 `A!. It should still be statically known to have type ``at least \Caml!`A!'',
and hence to be the value \Caml!`A!. However, this is not the case:
\begin{CamlBlock}
# id2 `A;;
- : [< `A | `B > `A ] = `A
\end{CamlBlock}
\Caml!id2 `A! is given the type  \Caml![< `A | `B > `A ]! which is parsed as \break \texttt{[\,$($< $($\Backq A|\Backq B$))$ $($> \Backq A$)$\,]} and means ``at least  \Caml!`A!'' (i.e.,  \texttt{[\,$($> \Backq A$)$\,]}) and (without any practical justification) ``at most  \texttt{\Backq A\,|\,\Backq B}'' (i.e.,\break\texttt{[\,$($< $($\Backq A\,|\,\Backq B$))$\,]}). As a consequence \Caml!`A! and \Caml!id2 `A! are no longer considered statically equivalent:
\begin{CamlBlock}
# [(id2 `A); ^`C^];;
Error: This expression has type [> `C ] but an
expression was expected of type [< `A | `B > `A ].
The second variant type does not allow tag(s) `C
\end{CamlBlock}
Dealing with this problem requires the use of awkward explicit coercions that hinder any further use of subtype polymorphism.



\paragraph{\textsc{Example 2:} roughly-typed pattern matching.}
The typing of pattern-matching expressions on polymorphic variants can prove to be imprecise. Consider:
\begin{CamlBlock}
# let f x = match id2 x with `A -> `B | y -> y;;
val f : [ `A | `B ] -> [ `A | `B ] = <fun>
\end{CamlBlock}
the typing of the function above is tainted by two
approximations: $(i)$~the domain of the function should
be \Caml:[< `A | `B ]:, but---since the argument \Keyw{x} is passed to
the function \Caml!id2!---OCaml deduces the type \Caml:[ `A | `B ]: (a shorthand for
 \Caml:[< `A | `B > `A | `B ]:), which is less precise: it loses subtype
polymorphism; $(ii)$ the return type states that \Caml|f| yields
either \Caml|`A| or \Caml|`B|, while it is easy to see that only the
latter is possible (when the argument is \Caml|`A| the function
returns \Caml|`B|, and when the argument is \Caml|`B| the function
returns the argument, that is, \Caml|`B|). So the type system deduces
for \Caml|f| the type \Caml:[ `A | `B ] ->$$ [ `A | `B ]: instead of the
more natural and precise \Caml:[< `A | `B ] ->$$ [> `B ]:.

To recover the correct type, we need to state explicitly
that the second pattern will only be used when \Keyw{y} is \Caml:`B:,
by using the alias pattern \Caml:`B as y:.
This is a minor inconvenience here, but writing the type for \Keyw{y}
is not always possible and is often more cumbersome.

Likewise, OCaml unduly restricts the type of the function
\begin{CamlBlock}
# let g x = match x with `A -> id2 x | _ -> x;;
val g : ([< `A | `B > `A ] as 'a) -> 'a = <fun>
\end{CamlBlock}
as it states \Keyw{g} can only be applied to \Caml|`A| or \Caml|`B|;
actually, it can be applied safely to, say,  \Caml|`C|
or any variant value with any other tag.
The system adds the upper bound \Caml:`A | `B: because \Keyw{id2} is applied to \Keyw{x}.
However, the application is evaluated only when $ \Keyw{x} = \TagLit{A} $:
hence, this bound is unnecessary.
The lower bound \Caml|`A| 
is unnecessary too.

The problem with these two functions is not specific to variant
types. It is more general, and it stems from the lack of full-fledged
connectives (union, intersection, and negation) in the types,
a lack which neither allows the system to type
a given pattern-matching branch by taking into account the cases
the previous branches have already handled (e.g., the typing of the
second branch in \Caml|f|), nor allows it to use the information provided by a
pattern to refine the typing of the branch code (e.g., the typing of the
first branch in \Caml|g|).

As a matter of fact, we can reproduce
the same problem as for \Keyw{g}, for instance, on lists:
\begin{CamlBlock}
# let rec map f l = match l with
    | [] -> l
    | h::t -> f h :: map f t;;
val map : ('a -> 'a) -> 'a list -> 'a list = <fun>
\end{CamlBlock}
This is the usual \texttt{map} function, but it is given an overly
restrictive type, accepting only functions with equal domain and
codomain. The problem, again, is that the type system does not use the
information provided by the pattern of the first branch to deduce that
that branch always returns an empty list (rather than a generic
$\alpha$ \texttt{list}). Also in this case alias patterns could be used
to patch this specific example, but do not work in general.

\paragraph{\textsc{Example 3:} rough approximations.}

During type reconstruction for pattern matching, OCaml uses
the patterns themselves to determine the type of the matched expression.
However, it might have to resort to approximations:
there might be no type which corresponds precisely to the set of values
matched by the patterns.
Consider, for instance, the following function \cite[from][]{Garrigue2004a}.
\begin{CamlBlock}
# let f x = match x with
    | (`A, _) -> 1 | (`B, _) -> 2
    | (_, `A) -> 3 | (_, `B) -> 4;;
val f : [> `A | `B ] * [> `A | `B ] -> int
\end{CamlBlock}
The type chosen by OCaml states that the function can be applied to any pair
whose both components have a type greater than \Caml:`A | `B:.
As a result, it can be applied to \Caml:(`C, `C):,
whose components have type \Caml:`A | `B | `C:.
This type therefore makes matching non-exhaustive:
the domain also contains values that are not captured by any pattern
(this is reported with a warning).
Other  choices could be made to ensure exhaustiveness, but they all pose different problems:
choosing \Caml![< `A | `B ] * [< `A | `B ] ->$$ int! makes the last two branches
become redundant; choosing instead a type such as \Caml![> `A | `B ] * [< `A | `B ] ->$$ int!
(or vice versa) is unintuitive as it breaks symmetry.

These rough approximations arise from the lack of full-fledged union types.
Currently, OCaml only allows unions of variant types. If
we could build a union of product types, then we could pick the type
\Caml!([< `A | `B ] * [> ]) | ([> ] * [< `A | `B ])!
(where \Caml|[> ]| is ``any variant''): exactly the set we need.
More generally, true union types (and singleton types for constants) remove the need of any approximation for the set of values matched by the patterns of a \Keyw{match} expression, meaning we are never forced
to choose---possibly inconsistently in different cases---between
exhaustiveness and non-redundancy.

Although artificial, the three examples above provide a good over\-view of the kind of problems of the current formalization of polymorphic variants. Similar, but more ``real life'',  examples of problems that our system solves can be found on the Web \citep[e.g.,][]{caml-difference,caml-filtering,caml-typing,caml-difference2,nicollet11,taw06}.

\paragraph{Contributions.}

The main technical contribution of this work is the definition of a type system
for a fragment of ML with polymorphic variants and pattern
matching. Our system aims to replace the parts of the current type checker of OCaml that deal with these features.
This replacement would result in a conservative extension of
the current type checker (at least, for the parts that concern variants and pattern matching), since our system types (with the same or more
specific types) all programs OCaml currently does; it would also be more expressive since it accepts  more programs,
while preserving type safety.
The key of our solution is the addition of semantic subtyping---i.e., of unconstrained set-theoretic unions, intersections, and negations---to the
type system. By adding it only in the type checker---thus,
without touching the current syntax of types the
OCaml programmer already knows---it is possible to solve all problems
we illustrated in Examples 1 and 2.
By a slight extension of the
syntax of types---i.e., by permitting unions ``\texttt{|}'' not only of variants but of any two types---and no further modification we can solve
the problem described in Example~3.
We also show that adding intersection
and negation combinators, as well as singletons, to the syntax of types
can be advantageous (cf.\ Sections~\ref{extensions-overloading} and \ref{conclusion}). Therefore, the general contribution of our work is to show a way to add full-fledged union, intersection, and difference types to implicitly-typed languages that use the HM type system.

Apart from the technical advantages and the gain in expressiveness, we
think that the most important advantage of our system is that it is
simpler, more natural, and arguably more intuitive than the current
one (which uses a system of kinding constraints).
Properties such as
``\emph{a given branch of a \Keyw{match} expression will be executed
for all values that can be produced by the matched expression,
that can be captured by the pattern of the branch,
and that cannot be captured by the patterns of the preceding branches}''
can be expressed precisely and
straightforwardly in terms of union, intersection, and
negation types (i.e., the type of the matched expression, intersected
by the type of the values matched by the pattern, minus the union of all
the types of the values matched by any preceding pattern: see rule
\Rule{Ts-Match} in Figure~\ref{fig:variants-s-typing}).
The reason for this is that in our system we can express much more
information at the level of types, which also means we can do without the
system of kinding constraints. This is made possible by the presence
of set-theoretic type connectives.  Such a capability allows the type
system to model pattern matching precisely and quite intuitively: we
can describe exhaustiveness and non-redundancy checking in terms of
subtype checking, whereas in OCaml they cannot be defined at the level
of types. Likewise, unions and intersections allow us to encode bounded
quantification---which is introduced in OCaml by structural
polymorphism---without having to add it to the system. As a
consequence, it is in general easy in our system to understand the origin
of each constraint generated by the type checker.

Our work  also presents several side contributions. First, it extends the
type reconstruction of~\citet{Castagna2015} to pattern matching
and let-polymorphism and, above all, proves it to be sound and complete
with respect to our system
(reconstruction in \citet{Castagna2015} is only proven sound).
Second, it provides a technique for a finer typing of
pattern matching that applies to types other than polymorphic variants
(e.g., the typing of \Caml!map! in Example~2) and
languages other than OCaml (it is implemented in the development
branch of \CDuce~\cite{Benzaken2003,CDUCE}). Third, the \VariantsK{} system we define in
Section~\ref{mlvariants} is a formalization of polymorphic variants
and full-fledged pattern matching as they are currently implemented in
OCaml: to our knowledge, no published formalization is as complete as \VariantsK.




\paragraph{Outline.}

Section~\ref{language} defines the syntax and semantics
of the language
we will study throughout this work. Sections~\ref{mlvariants} and~\ref{setvariants} present two different type systems for this language.

In particular, Section~\ref{mlvariants} briefly describes the \VariantsK{} type system we have developed
as a formalization of how polymorphic variants are typed in OCaml. Section~\ref{setvariants} describes the \VariantsS{} type system,
which employs set-theoretic types with semantic subtyping: we first give a deductive presentation of the system,
and then we compare it to \VariantsK{}
to show that \VariantsS{} can type every program that the \VariantsK{} system can type.
Section~\ref{recon} defines a type reconstruction algorithm that is sound and complete
with respect to the \VariantsS{} type system.

Section~\ref{extensions} presents three extensions or modifications of the system:
the first is the addition of overloaded functions;
the second is a refinement of the typing of pattern matching,
which we need to type precisely
the functions \Keyw{g} and \Keyw{map} of Example 2;
the third is a restriction which solves a discrepancy
between our model and OCaml
(the lack of type tagging at runtime in the OCaml implementation).

Finally, Section~\ref{related} compares our work
with other formalizations of polymorphic variants
and with previous work on systems with set-theoretic type connectives,
and Section~\ref{conclusion} concludes the presentation and points out some directions for future research.

For space reasons we omitted all the proofs as well as some definitions. They can be found in the Appendix%
\ifhal.\else\ contained in the extended version available online~\cite{CPN16full}.\fi

\section{The language of polymorphic variants}\label{language}
In this section, we define the syntax and semantics of the language
with polymorphic variants and pattern matching that we study in this work.
In the sections following this one  we will define two
different type systems for it (one with kinds in Section~\ref{mlvariants}, the other with set-theoretic
types in Section~\ref{setvariants}), as well as a type reconstruction algorithm (Section~\ref{recon}).

\subsection{Syntax}
\label{sec:variants-k-syntax}

We assume that there exist
a countable set $ \ExprVars $ of \emph{expression variables},
ranged over by $ x $, $ y $, $ z $, \dots,
a set $ \Constants $ of constants, ranged over by $ c $,
and a set $ \Tags $ of tags, ranged over by $ \Tagg $.
Tags are used to label variant expressions.

\begin{Definition}[Expressions]
\label{def:variants-expressions}
    An \emph{expression} $ e $ is a term inductively generated by the following grammar:
    \begin{alignat*}{2}
        \Bnf{ e }
            & & \Bnf{{} \= {}}
            & \Bnf{
                x \| c \| \Abstr{x.e} \| \Appl{e}{e}
                \| (e,\! e) \| \Tagg(e) \|
                \Match{e with (p_i {\to} e_i)_{i \in\! I}}}
    \end{alignat*}
    where $ p $ ranges over the set $ \Patterns $ of patterns, defined below.
    We write $ \Expressions $ to denote the set of all expressions.
\end{Definition}

We define $ \Fvar(e) $ to be the set of expression variables
occurring free in the expression $ e $,
and we say that $ e $ is \emph{closed} if and only if $ \Fvar(e) $ is empty.
As customary, we consider expressions up to \textalpha-renaming
of the variables bound by abstractions and by patterns.

The language is a \LambdaCalculus{}
with constants, pairs, variants, and pattern matching.
Constants include a dummy constant \Unit{} (`unit')
to encode variants without arguments;
multiple-argument variants are encoded with pairs.
Matching expressions specify one or more branches (indexed by a set $ I $)
and can be used to encode \Keyw{let}-expressions:
$
    \Let{x = e_0 in e_1} \eqdef \Match{e_0 with x \to e_1}
    \: .
$

\begin{Definition}[Patterns]
\label{def:variants-patterns}
    A \emph{pattern} $ p $ is a term
    inductively generated by the following grammar:
    \[
      \Bnf{ p
        \= \Wildcard \| x \| c \| (p, p) \| \Tagg(p)
        \| \PAnd{p \& p} \| \POr{p | p} }
    \]
    such that
    $ (i) $ in a pair pattern $ (p_1, p_2) $
    or an intersection pattern $ \PAnd{p_1 \& p_2} $,
    $ \Capt(p_1) \cap \Capt(p_2) = \varnothing $;
    $ (ii) $ in a union pattern $ \POr{p_1 | p_2} $,
    $ \Capt(p_1) = \Capt(p_2) $,
    where $ \Capt(p) $ denotes the set of expression variables
    occurring as sub-terms in a pattern $ p $
    (called the \emph{capture variables} of $ p $).
    We write $ \Patterns $ to denote the set of all patterns.
\end{Definition}

Patterns have the usual semantics.
A wildcard ``\Wildcard'' accepts any value and generates no bindings;
a variable pattern accepts any value and binds the value to the variable.
Constants only accept themselves and do not bind.
Pair patterns accept pairs
if each sub-pattern accepts the corresponding component,
and variant patterns accept variants with the same tag
if the argument matches the inner pattern
(in both cases, the bindings are those of the sub-patterns).
Intersection patterns require the value to match both sub-patterns
(they are a generalization of the alias patterns $ p \mathrel{\Keyw{as}} x $
of OCaml),
while union patterns require it to match either of the two
(the left pattern is tested first).

\subsection{Semantics}
\label{sec:variants-k-semantics}

We now define a small-step operational semantics for this calculus.
First, we define the values of the language.

\begin{Definition}[Values]
\label{def:variants-values}
    A \emph{value} $ v $ is a closed expression
    inductively generated by the following grammar.
    \[
      \Bnf{ v
        \= c \| \Abstr{x. e} \| (v, v) \| \Tagg(v) }
    \]
\end{Definition}

We now formalize the intuitive semantics of patterns
that we have presented above.

Bindings are expressed in terms of \emph{expression substitutions},
ranged over by $ \varsigma $:
we write $ \SubstDots[v_1/x_1 ... v_n/x_n] $ for the substitution
that replaces free occurrences of $ x_i $ with $ v_i $, for each $ i $.
We write $ e \varsigma $
for the application of the substitution $ \varsigma $ to an expression $ e $;
we write $ \varsigma_1 \cup \varsigma_2 $
for the union of disjoint substitutions.

The semantics of pattern matching we have described is formalized
by the definition of $ \Matching{v / p} $
given in Figure~\ref{fig:variants-semantics-matching}.
In a nutshell, $ \Matching{v / p} $ is the result of matching
a value $ v $ against a pattern $ p $.
We have either $ \Matching{v / p} = \varsigma $,
where $ \varsigma $ is a substitution defined on the variables in $ \Capt(p) $,
or $ \Matching{v / p} = \MatchFail $.
In the former case, we say that $ v $ matches $ p $
(or that $ p $ accepts $ v $); in the latter, we say that matching fails.

\begin{figure}

    \begin{align*}
        \Matching{v / \Wildcard}
            & = \EmptySubst
            \\
        \Matching{v / x}
            & = \SubstSingle[v / x]
            \\
        \Matching{v / c}
            & =
            \begin{cases}
                \EmptySubst & \text{if } v = c \\
                \MatchFail  & \text{otherwise}
            \end{cases}
            \\
        \Matching{v / (p_1, p_2)}
            & =
            \begin{cases}
                \varsigma_1 \cup \varsigma_2 &
                    \text{if } v = (v_1, v_2)
                    \text{ and } \forall i . \:
                    \Matching{v_i / p_i} = \varsigma_i \\
                \MatchFail & \text{otherwise}    
            \end{cases}
            \\
        \Matching{v / \Tagg(p_1)}
            & =
            \begin{cases}
                \varsigma_1 &
                    \text{if } v = \Tagg(v_1)
                    \text{ and } \Matching{v_1 / p_1} = \varsigma_1 \\
                \MatchFail & \text{otherwise}
            \end{cases}
            \\
        \Matching{v / \PAnd{p_1 \& p_2}}
            & =
            \begin{cases}
                \varsigma_1 \cup \varsigma_2 &
                    \text{if } \forall i . \: \Matching{v / p_i} = \varsigma_i
                    \\
                \MatchFail & \text{otherwise}
            \end{cases}
            \\
        \Matching{v / \POr{p_1 | p_2}}
            & =
            \begin{cases}
                \Matching{v / p_1} &
                    \text{if } \Matching{v / p_1} \neq \MatchFail \\
                \Matching{v / p_2} & \text{otherwise}
            \end{cases}
    \end{align*}
    
    \caption{Semantics of pattern matching.}
    \label{fig:variants-semantics-matching}
\end{figure}

Note that the unions of substitutions in the definition are always disjoint
because of our linearity condition on pair and intersection patterns.
The condition that sub-patterns of a union pattern $ \POr{p_1 | p_2} $
must have the same capture variables
ensures that $ \Matching{v/p_1} $ and $ \Matching{v/p_2} $
will be defined on the same variables.

Finally, we describe the reduction relation.
It is defined by the following two notions of reduction
\label{pag:l-reduction}
\[
  \begin{array}{rcll}
    \Appl{(\Abstr{x. e})}{v} & \leadsto & e \SubstSingle[v/x]
    \\[5pt]
    \Match{v with (p_i \to e_i)_{i \in I}}
      & \leadsto & e_j \varsigma &
      \begin{array}{l}
        \text{if } \Matching{v / p_j} = \varsigma \text{ and} \\
        \forall i < j . \: \Matching{v / p_i} = \MatchFail
      \end{array}
  \end{array}
\]
applied with a leftmost-outermost strategy
which does not reduce inside \textlambda-abstractions
nor in the branches of \Keyw{match} expressions.

The first reduction rule is the ordinary rule for call-by-value
\textbeta-reduction. It states that the application of an abstraction
$ \Abstr{x. e} $ to a value $ v $ reduces to the body $ e $ of the
abstraction, where $ x $ is replaced by $ v $. The second rule states
that a \Keyw{match} expression on a value $ v $ reduces to the branch
$ e_j $ corresponding to the first pattern $ p_j $ for which matching
is successful. The obtained substitution is applied to $ e_j $,
replacing the capture variables of $ p_j $ with sub-terms of $ v $. If
no pattern accepts $ v $, the expression is stuck.

\section{Typing variants with kinding constraints}\label{mlvariants}
In this section, we formalize \VariantsK,
the type system with \Kin{}inding constraints
for polymorphic variants as featured in OCaml;
we will use it to gauge the merits of \VariantsS{},
our type system with \Sem{}et-theoretic types.
This formalization is derived from, and extends, the published systems
based on structural polymorphism \citep{Garrigue2002,Garrigue2015}.
In our ken,
no formalization in the literature
includes polymorphic variants, let-polymorphism, and full-fledged
pattern matching (see Section~\ref{related}), which is why we give
here a new one. While based on
existing work, the formalization is far from being trivial (which with hindsight
explains its absence), and thus we needed to prove all its
properties from scratch.
For space reasons we outline just the features that distinguish our
formalization, namely variants, pattern matching, and type generalization
for pattern capture variables.
The Appendix presents the full definitions and proofs of all properties.

The system consists essentially of the core ML type system with the
addition of a kinding system to distinguish normal type variables
from \emph{constrained} ones. Unlike normal variables,
constrained ones cannot be instantiated into any type, but only
into other constrained variables with compatible constraints. They are
used to type variant expressions: there are no `variant
types' \emph{per se}. Constraints are recorded in \emph{kinds} and
kinds in a \emph{kinding environment} (i.e., a mapping from type variables to kinds) which is included in typing
judgments. An important consequence of using kinding constraints is
that they implicitly introduce (a limited form of) recursive types, since a
constrained type variable may occur in its constraints.

We assume that there exists a countable set $ \TypeVars $
of \emph{type variables},
ranged over by $ \alpha $, $ \beta $, $ \gamma $, \dots.
We also consider a finite set $ \BasicTypes $ of \emph{basic types},
ranged over by $ b $,
and a function $ b_{(\Placeholder)} $ from constants to basic types.
For instance, we might take
$ \BasicTypes = \Set{\Keyw{bool}, \Keyw{int}, \Keyw{unit}} $,
with $ b_{\Keyw{true}} = \Keyw{bool} $, $ b_{\Unit} = \Keyw{unit} $, and so on.

\begin{Definition}[Types]
\label{def:variants-k-types}
    A \emph{type} $ \tau $ is a term
    inductively generated by the following grammar.
    \[
        \Bnf{ \tau \= \alpha \| b \| \tau \to \tau \| \tau \times \tau }
    \]
\end{Definition}

The system only uses the types of core ML:
all additional information is encoded in the kinds of type variables.

Kinds have two forms: the \emph{unconstrained kind} ``$ \KUnkind $''
classifies ``normal'' variables, while variables used to type variants
are given a \emph{constrained kind}.
Constrained kinds are triples
describing which tags may or may not appear
(a \emph{presence} information) and which argument types are
associated to each tag (a \emph{typing} information). The presence
information is split in two parts, a lower and an upper bound. This is
necessary to provide an equivalent to both covariant and contravariant
subtyping---without actually having subtyping in the system---that
is, to allow both variant values and functions defined on variant
values to be polymorphic.

\begin{Definition}[Kinds]
\label{def:variants-k-kinds}
    A \emph{kind} $ \kappa $ is
    either the \emph{unconstrained kind} ``$ \KUnkind $''
    or a \emph{constrained kind}, that is, a triple $ (L, U, T) $ where:
    \begin{compactitemize}
        \item $ L $ is a finite set of tags $ \Set{\Tagg_1, \dots, \Tagg_n} $;
        \item $ U $ is either a finite set of tags
            or the set $ \Tags $ of all tags;
        \item $ T $ is a finite set of pairs of a tag and a type, written
            $ \{ \Tagg_1\colon \tau_1, $ $ \dots, \Tagg_n\colon \tau_n \} $
            (its domain $ \Dom(T) $ is the set of tags occurring in it);
    \end{compactitemize}
    and where the following conditions hold:
    \begin{compactitemize}
        \item $ L \subseteq U $, $ L \subseteq \Dom(T) $, and,
            if $ U \neq \Tags $, $ U \subseteq \Dom(T) $;
        \item tags in $ L $ have a single type in $ T $, that is,
            if $ \Tagg \in L $, whenever both $ \Tagg\colon \tau_1 \in T $
            and $ \Tagg\colon \tau_2 \in T $, we have $ \tau_1 = \tau_2 $.
    \end{compactitemize}
\end{Definition}

In OCaml, kinds are written with the typing information
inlined in the lower and upper bounds.
These are introduced by \Caml|>| and \Caml|<| respectively and, if missing,
$ \varnothing $ is assumed for the lower bound and $ \Tags $ for the upper.
For instance, \Caml![> `A of int | `B of bool ] as 'a! of OCaml
is represented here by assigning to the variable $ \alpha $ the kind $
    (\Set{\TagLit{A}, \TagLit{B}}, \Tags,
        \Set{\TagLit{A}\colon\!\Keyw{int}, \TagLit{B}\colon\!\Keyw{bool}})
$;
\Caml![< `A!\;\Caml!of!\;\Caml!int | `B!\;\Caml!of!\;\Caml!bool!\:\Caml!] as 'b! corresponds to $ \beta $ of kind $
    (\varnothing, \Set{\TagLit{A}, \TagLit{B}},
        \Set{\TagLit{A}\colon \Keyw{int}, \TagLit{B}\colon \Keyw{bool}})
$; finally \Caml![< `A of int | `B of bool & unit > `A ] as 'c! corresponds to
$ \gamma $ of kind $
    (\Set{\TagLit{A}}, \Set{\TagLit{A}, \TagLit{B}},
        \Set{\TagLit{A}\colon \Keyw{int}, \TagLit{B}\colon \Keyw{bool},
        \TagLit{B}\colon \Keyw{unit}})
$.

\begin{Definition}[Type schemes]
\label{def:variants-k-type-schemes}
    A \emph{type scheme} $ \sigma $ is of the form
    $ \KScheme{A. K > \tau} $, where:
    \begin{compactitemize}
        \item $ A $ is a finite set $ \Set{\alpha_1, \dots, \alpha_n} $
            of type variables;
        \item $ K $ is a kinding environment, that is,
            a map from type variables to kinds;
        \item $ \Dom(K) = A $.
    \end{compactitemize}
\end{Definition}

We identify a type scheme $ \KScheme{\varnothing. \varnothing > \tau} $, which
quantifies no variable, with the type $ \tau $ itself. We consider
type schemes up to renaming of the variables they bind and disregard
useless quantification (i.e., quantification of variables that do not
occur in the type).

Type schemes single out, by quantifying them, the variables of a type
which can be instantiated. In ML without kinds, the quantified variables
of a scheme can be instantiated with any type. The addition of kinds changes this:
variables with constrained kinds may only be instantiated into other
variables with \emph{equally strong} or \emph{stronger} constraints. This relation on constraints is
formalized by the following entailment relation:
\[
    \KKindEntail{ (L, U, T) |= (L', U', T') }
    \iff
    L \supseteq L' \land U \subseteq U' \land T \supseteq T'
    \, ,
\]
where $ \KKindEntail{ \kappa_1 |= \kappa_2 } $ means that $\kappa_1$ is a constraint stronger than $\kappa_2$. This relation is used to select the type substitutions (ranged over by $\theta$) that are \emph{admissible}, that is, that are sound with respect to kinding.

\begin{Definition}[Admissibility of a type substitution]
\label{def:variants-k-type-substitutions2}
    A type substitution $ \theta $ is
    \emph{admissible} between two kinding environments $ K $ and $ K' $,
    written $ \KAdmissible{K |- \theta: K'} $,
    if and only if,
    for every type variable $ \alpha $ such that $ K(\alpha) = (L, U, T) $,
    $ \alpha \theta $ is a type variable
    such that $ K'(\alpha \theta) = (L', U', T') $
    and $ \KKindEntail{(L', U', T') |= (L, U, T \theta)} $.
\end{Definition}

In words, whenever $ \alpha $ is constrained in $ K $, then $ \alpha \theta $ must be a type variable constrained in $ K' $ by a kind that entails the substitution instance of the kind of $ \alpha $ in $ K $.

The set of the instances of a type scheme are now obtained by applying only  admissible substitutions.
\begin{Definition}[Instances of a type scheme]
\label{def:variants-k-instances}
    The set of \emph{instances} of a type scheme $ \KScheme{A. K' > \tau} $
    in a kinding environment $ K $ is
    \[
        \Inst_K(\KScheme{A. K' > \tau})
        =
        \SetC{\tau \theta | \Domain(\theta) \subseteq A \: \land \:
                            \KAdmissible{ K, K' |- \theta: K } }
        \: .
    \]
\end{Definition}

As customary, this set is used in the type system rule to type expression variables:
\begin{mathpar}
    \MyInfer[Tk-Var]
        { \tau \in \Inst_K(\Gamma(x)) }
        { \KTyping{ K; \Gamma |- x: \tau } }
        {}
\end{mathpar}

Notice that typing judgments are of the form
$ \KTyping{ K; \Gamma |- e: \tau } $:
the premises include a type environment $ \Gamma $ but also, which is new,
a kinding environment $ K $
(the $ \mathbb{K} $ subscript in the turnstile symbol is to distinguish this
relation from $\SSubscr{\vdash}\,$, the relation for the set-theoretic type system of the next section).

The typing rules for constants, abstractions, applications, and pairs are straightforward. There remain the rules for variants and for pattern matching, which are the only interesting ones.

\begin{mathpar}
    \MyInfer[Tk-Tag]
        { \KTyping{ K; \Gamma |- e: \tau } \\
          \KKindEntail{ K(\alpha) |=
            (\Set{\Tagg}, \Tags, \Set{\Tagg\colon \tau}) } }
        { \KTyping{ K; \Gamma |- \Tagg(e): \alpha } }
        {}
\end{mathpar}
The typing of variant expressions uses the kinding environment.
Rule \RuleRef{Tk-Tag} states that $ \Tagg(e) $ can be typed
by any variable $ \alpha $ such that $ \alpha $ has a constrained kind in $ K $
which entails the ``minimal'' kind for this expression.
Specifically, if $ K(\alpha) = (L, U, T) $,
then we require $ \Tagg \in L $ and $ \Tagg\colon \tau \in T $,
where $ \tau $ is a type for $ e $.
Note that $ T $ may not assign more than one type to $ \Tagg $,
since $ \Tagg \in L $.

The typing of pattern matching is by far the most complex part of the type system and it is original to our system.
\begin{mathpar}
    \MyInfer[Tk-Match]
        { \KTyping{ K; \Gamma |- e_0: \tau_0 } \\
          \KExhaustive{ K |- \tau_0 < \SetC{p_i | i \in I} } \\\\
          \forall i \in I \quad
          \KPatternTyping{ K |- p_i: \tau_0 => \Gamma_i } \quad
          \KTyping{ K; \Gamma, \Gen_{K;\Gamma}(\Gamma_i) |- e_i: \tau } }
        { \KTyping{ K; \Gamma |-
            \Match{e_0 with (p_i \to e_i)_{i \in I}}: \tau } }
        {}
\end{mathpar}
Let us describe each step that the rule above implies. First the rule deduces the type $ \tau_0 $ of the matched expression
($ \KTyping{ K; \Gamma |- e_0: \tau_0 } $).
Second, for each pattern $ p_i $, it generates the type environment $ \Gamma_i $
which assigns types to the capture variables of $ p_i $, assuming $ p_i $ is matched
against a value known to be of type $ \tau_0 $. This is done by deducing the
judgment $ \KPatternTyping{ K |- p_i: \tau_0 => \Gamma_i } $, whose inference
system is mostly straightforward (see Figure~\ref{fig:variants-k-pattern-environment} in the Appendix); for instance, for variable patterns we have:
\begin{mathpar}
    \MyInfer[TPk-Var]
        { }
        { \KPatternTyping{ K |- x: \tau => \Set{x\colon \tau} } }
        {}
\end{mathpar}

The only subtle point of this inference system is the rule for patterns of the form $\Tagg(p)$
\begin{mathpar}
    \MyInfer[TPk-Tag]
        { \KPatternTyping{ K |- p: \tau => \Gamma } \\
          K(\alpha) = (L, U, T) \\\\
          (\Tagg \in U \text{ implies } \Tagg \colon \tau \in T) }
        { \KPatternTyping{ K |- \Tagg(p): \alpha => \Gamma } }
        {}
\end{mathpar}
which---after generating the environment for the capture variables of
$p$---checks whether the type of the matched expression is a variant
type (i.e., a variable) with the right constraints for $ \Tagg $.

Third, the rule \RuleRef{Tk-Match}
types each branch $ e_i $ with type $ \tau $,
in a type environment updated with $ \Gen_{K;\Gamma}(\Gamma_i) $,
that is, with the generalization of the $\Gamma_i$
generated by $\KPatternTyping{K |- p_i: \tau_0 => \Gamma_i} $.
The definition of generalization is standard:
it corresponds to quantifying all the variables
that do not occur free in the environment $ \Gamma $.
The subtle point is the definition of the free variables of a type
(and hence of an environment), which we omit for space reasons.
It must navigate the kinding environment $ K $ to collect all variables
which can be reached by following the constraints;
hence, the $ \Gen $ function takes as argument $ K $ as well as $ \Gamma $.

Finally, the premises of the rule also include the exhaustiveness condition $
    \KExhaustive{ K |- \tau_0 < \SetC{p_i | i \in I} }
$, which checks whether every possible value that $ e_0 $ can produce
matches at least one pattern $ p_i $.
The definition of exhaustiveness is quite convoluted.

\begin{Definition}[Exhaustiveness]
\label{def:variants-k-exhaustiveness}
    We say that a set of patterns $ P $ is \emph{exhaustive}
    with respect to a type $ \tau $ in a kinding environment $ K $,
    and we write $ \KExhaustive{K |- \tau < P} $, when,
    for every $ K' $, $ \theta $, and $ v $,
    \[
            (\KAdmissible{ K |- \theta: K' } \land
            \KTyping{ K'; \varnothing |- v: \tau \theta }) \implies
            \exists p \in P, \varsigma . \: \Matching{ v / p } = \varsigma
        \: .
    \]
\end{Definition}


In words, $ P $ is exhaustive when every value that can be typed with any
admissible substitution of $ \tau $ is accepted by at least one
pattern in $ P $. OCaml does not impose exhaustiveness---it just signals
non-exhaustiveness with a warning---but our system does. We do so in
order to have a simpler statement for soundness and to facilitate the
comparison with the system of the next section. We do not discuss how
exhaustiveness can be effectively computed; for more information on
how OCaml checks it, see \citet{Garrigue2004a}
and \citet{Maranget2007}.

We conclude this section by stating the type soundness property of the \VariantsK{} type system.

\begin{Theorem}[Progress]
\label{thm:variants-k-progress}
    Let $ e $ be a well-typed, closed expression. Then, either $ e $ is a value or there exists an expression $ e' $ such that $ \Smallstep{ e ~> e' } $.
\end{Theorem}

\begin{Theorem}[Subject reduction]
\label{thm:variants-k-subject-reduction}
    Let $ e $ be an expression and $ \tau $ a type such that $ \KTyping{ K; \Gamma |- e: \tau } $. If $ \Smallstep{ e ~> e' } $, then $ \KTyping{ K; \Gamma |- e': \tau } $.
\end{Theorem}

\begin{Corollary}[Type soundness]
\label{cor:variants-k-soundness}
    Let $ e $ be a well-typed, closed expression, that is, such that $ \KTyping{ K; \varnothing |- e: \tau } $ holds for some $ \tau $. Then, either $ e $ diverges or it reduces to a value $ v $ such that $ \KTyping{ K; \EmptyEnv |- v: \tau } $.
\end{Corollary}

\section{Typing variants with set-theoretic types}\label{setvariants}
We now describe \VariantsS{}, a type system
for the language of Section~\ref{language} based on set-theoretic types.
The approach we take in its design
is drastically different from that followed for \VariantsK{}.
Rather than adding a kinding system to record information
that types cannot express, we directly enrich the syntax of types
so they can express all the notions we need.
Moreover, we add subtyping---using a semantic definition---rather than
encoding it via instantiation.
We exploit type connectives and subtyping to represent variant types
as unions and to encode bounded quantification by union and intersection.

We argue that \VariantsS{} has several advantages
with respect to the previous system.
It is more expressive:
it is able to type some programs that \VariantsK{}
rejects though they are actually type safe, and
it can derive more precise types than \VariantsK.
It is arguably a simpler formalization:
typing works much like in ML except for the addition of subtyping,
we have explicit types for variants,
and we can type pattern matching precisely and straightforwardly.
Indeed, as regards pattern matching,
an advantage of the \VariantsS{} system is that it can express
exhaustiveness and non-redundancy checking as subtyping checks,
while they cannot be expressed at the level of types in \VariantsK.

Naturally, subtyping brings its own complications.
We do not discuss its definition here, since we reuse the relation
defined by \citet{Castagna2011}.
The use of semantic subtyping makes the definition of a typing algorithm
challenging: \citet{Castagna2014,Castagna2015} show how to define one
in an explicitly-typed setting.
Conversely, we study here an implicitly-typed language
and hence study the problem of type reconstruction (in the next section).

While this system is based on that described
by \citet{Castagna2014,Castagna2015},
there are significant differences which we discuss in Section~\ref{related}.
Notably, intersection types play a more limited role in our system
(no rule allows the derivation of an intersection of arrow types
for a function), making our type reconstruction complete.



\subsection{Types and subtyping}
\label{sec:variants-s-types}

As before, we consider a set $ \TypeVars $ of \emph{type variables}
(ranged over by $ \alpha $, $ \beta $, $ \gamma $, \dots)
and the sets $ \Constants $, $ \Tags $, and $ \BasicTypes $
of \emph{language constants}, \emph{tags}, and \emph{basic types}
(ranged over by $ c $, $ \Tagg $, and $ b $ respectively).

\begin{Definition}[Types]
\label{def:variants-s-types}
    A \emph{type} $ t $ is a term coinductively produced
    by the following grammar:
    \[
        \Bnf{ t
            \= \alpha \| b \| c
            \| t \to t \| t \times t \| \Tagg(t)
            \| t \lor t \| \lnot t \| \Empty }
    \]
    which satisfies two additional constraints:
    \begin{itemize}
        \item (\emph{regularity})
            the term must have a finite number of different sub-terms;
        \item (\emph{contractivity})
            every infinite branch must contain an infinite number
            of occurrences of atoms
            (i.e., a type variable or the immediate application
            of a type constructor: basic, constant, arrow, product, or variant).
    \end{itemize}
\end{Definition}

We introduce the following abbreviations:
\[
    t_1 \land t_2 \eqdef \lnot (\lnot t_1 \lor \lnot t_2) \qquad
    t_1 \setminus t_2 \eqdef t_1 \land (\lnot t_2) \qquad
    \Any \eqdef \lnot \Empty
    \: .
\]

With respect to the types in Definition~\ref{def:variants-k-types},
we add several new forms.
We introduce set-theoretic connectives
(union, intersection, and negation),
as well as bottom (the empty type $ \Empty $) and top ($ \Any $) types.
We add general (uniform) recursive types by interpreting the grammar \emph{coinductively},
while \VariantsK{} introduces recursion via kinds.
Contractivity is imposed to bar out ill-formed types such as those fulfilling the equation $ t = t \lor t $
(which does not give any information on the set of values it represents)
or $ t = \lnot t $ (which cannot represent any set of values).

We introduce explicit types for variants.
These types have the form $ \Tagg(t) $:
the type of variant expressions with tag $ \Tagg $
and an argument of type $ t $.%
    \footnote{\label{encoding}%
        We could encode $ \Tagg(t) $ by the product $ \Tagg \times t $.
        Although we have preferred to add explicit variant types,
        we still use this encoding to derive their subtyping properties:
        see~\citet{Petrucciani2015} for a detailed explanation.}
Type connectives allow us to represent all variant types of \VariantsK{}
by combining types of this form, as we describe in detail below.
Finally, we add singleton types for constants
(e.g., a type \Keyw{true} which is a subtype of \Keyw{bool}),
which we use to type pattern matching precisely.

\paragraph{Variant types and bounded quantification.}

\VariantsK{} uses constrained variables to type variants;
when these variables are quantified in a type scheme,
their kind constrains the possible instantiations of the scheme.
This is essentially a form of bounded quantification:
a variable of kind $ (L, U, T) $
may only be instantiated by other variables which fall within the bounds---the
lower bound being determined by $ L $ and $ T $,
the upper one by $ U $ and $ T $.

In \VariantsS, we can represent these bounds
as unions of variant types $ \Tagg(t) $.
For instance,
consider in \VariantsK{} a constrained variable $ \alpha $ of kind $
    (\Set{\TagLit{A}},
    \Set{\TagLit{A}, \TagLit{B}},
    \Set{\TagLit{A}\colon \Keyw{bool}, \TagLit{B}\colon \Keyw{int}})
$.
If we quantify $ \alpha $, we can then instantiate it
with variables whose kinds entail that of $ \alpha $.
Using our variant types and unions,
we write the lower bound as $ t_{\text{L}} = \TagLit{A}(\Keyw{bool}) $
and the upper one as $
    t_{\text{U}} = \TagLit{A}(\Keyw{bool}) \lor \TagLit{B}(\Keyw{int})
$.
In our system, $ \alpha $ should be a variable with bounded quantification,
which can only be instantiated by types $ t $
such that $ t_{\text{L}} \leq t \leq t_{\text{U}} $.

However, we do not need to introduce bounded quantification
as a feature of our language:
we can use type connectives to encode it
as proposed by~\citet[cf. Footnote~4 therein]{Castagna2011}.
The possible instantiations of $ \alpha $ (with the bounds above)
and the possible instantiations of $
    (t_{\text{L}} \lor \beta) \land t_{\text{U}}
$, with no bound on $ \beta $, are equivalent.
We use the latter form:
we internalize the bounds in the type itself by union and intersection.
In this way, we need no system of constraints extraneous to types.

\paragraph{Subtyping.}
\label{ssec:variant-s-types-subtyping}

There exists a \emph{subtyping} relation between types.
We write $ t_1 \leq t_2 $ when $ t_1 $ is a subtype of $ t_2 $;
we write $ t_1 \simeq t_2 $
when $ t_1 $ and $ t_2 $ are equivalent with respect to subtyping,
that is, when $ t_1 \leq t_2 $ and $ t_2 \leq t_1 $.
The definition and properties of this relation are studied
in \citet{Castagna2011},
except for variant types which, for this purpose, we encode as pairs (cf. Footnote~\ref{encoding}).

\begin{figure*}

    \begin{mathpar}
        \MyInfer[Ts-Var]
            { t \in \Inst(\Gamma(x)) }
            { \STyping{ \Gamma |- x: t } }
            {}
        \and
        \MyInfer[Ts-Const]
            { }
            { \STyping{ \Gamma |- c: c } }
            {}
        \and
        \MyInfer[Ts-Abstr]
            { \STyping{ \Gamma, \Set{x\colon t_1} |- e: t_2 } }
            { \STyping{ \Gamma |- \Abstr{x. e}: t_1 \to t_2 } }
            {}
        \and
        \MyInfer[Ts-Appl]
            { \STyping{ \Gamma |- e_1: t' \to t } \\
              \STyping{ \Gamma |- e_2: t' } }
            { \STyping{ \Gamma |- \Appl{e_1}{e_2}: t } }
            {}
        \and
        \MyInfer[Ts-Pair]
            { \STyping{ \Gamma |- e_1: t_1 } \\
              \STyping{ \Gamma |- e_2: t_2 } }
            { \STyping{ \Gamma |- (e_1, e_2): t_1 \times t_2 } }
            {}
        \and
        \MyInfer[Ts-Tag]
            { \STyping{ \Gamma |- e: t } }
            { \STyping{ \Gamma |- \Tagg(e): \Tagg(t) } }
            {}
        \and
        \MyInfer[Ts-Match]
            { \STyping{ \Gamma |- e_0: t_0 } \\
              t_0 \leq \textstyle\bigvee_{i \in I} \SAcc{p_i} \\
              t_i = (t_0 \setminus \textstyle\bigvee_{j < i} \SAcc{p_j})
                \land \SAcc{p_i} \\\\
              \forall i \in I \\
              \STyping{ \Gamma, \Gen_\Gamma(\SPatternEnv{ t_i // p_i }) |-
                e_i: t_i' } }
            { \STyping{ \Gamma |- \Match{ e_0 with (p_i \to e_i)_{i \in I} }:
                \textstyle\bigvee_{i \in I} t_i' } }
            {}
        \and
        \MyInfer[Ts-Subsum]
            { \STyping{ \Gamma |- e: t' } \\
              t' \leq t }
            { \STyping{ \Gamma |- e: t } }
            {}
    \end{mathpar}

    \caption{Typing relation of the \VariantsS{} type system.}
    \label{fig:variants-s-typing}
\end{figure*}

In brief, subtyping is given a semantic definition, in the sense that
$ t_1 \leq t_2 $ holds if and only if
$ \SInterpr{t_1} \subseteq \SInterpr{t_2} $,
where $ \SInterpr{\cdot} $ is an interpretation function
mapping types to sets of elements from some domain
(intuitively, the set of values of the language).
The interpretation is ``set-theoretic'' as it interprets union types as unions,
negation as complementation, and products as Cartesian products.

In general, in the semantic-subtyping approach,
we consider a type to denote the set of all values that have that type
(we will say that some type ``is'' the set of values of that type).
In particular, for arrow types, the type $ t_1 \to t_2 $ is that of function values (i.e., $\lambda$-abstractions) which,
if they are given an argument in $ \SInterpr{t_1} $ and they do not diverge,
yield a result in $ \SInterpr{t_2} $.
Hence, all types of the form $ \Empty \to t $, for any $ t $, are equivalent
(as only diverging expressions can have type $ \Empty $):
any of them is the type of all functions.
Conversely, $ \Any \to \Empty $ is the type of functions that
(provably) diverge on all inputs:
a function of this type should yield a value in the empty type
whenever it terminates, and that is impossible.

The presence of variables complicates the definition of semantic subtyping.
Here, we just recall from \citet{Castagna2011}
that subtyping is preserved by type substitutions:
$ t_1 \leq t_2 $ implies $ t_1 \theta \leq t_2 \theta $
for every type substitution $ \theta $.

\subsection{Type system}
\label{sec:variants-s-type-system}

We present \VariantsS{} focusing on the differences with respect
to the system of OCaml (i.e., \VariantsK); full definitions are in the Appendix.
Unlike in \VariantsK, type schemes here are defined just as in ML
as we no longer need kinding constraints.

\begin{Definition}[Type schemes]
\label{def:variants-s-type-schemes}
    A \emph{type scheme} $ s $ is of the form $ \Scheme{A. t} $,
    where $ A $ is a finite set $ \Set{\alpha_1, \dots, \alpha_n} $
    of type variables.
\end{Definition}

As in \VariantsK,
we identify a type scheme $ \Scheme{\varnothing. t} $ with the type $ t $
itself, we consider type schemes up to renaming of the variables they bind,
and we disregard useless quantification.

We write $ \Var(t) $ for the set of type variables occurring in a type $ t $;
we say they are the \emph{free variables} of $ t $,
and we say that $ t $ is \emph{ground} or \emph{closed}
if and only if $ \Var(t) $ is empty.
The (coinductive) definition of $ \Var $ can be found in \citet[Definition A.2]{Castagna2014}.

Unlike in ML, types in our system can contain variables
which are irrelevant to the meaning of the type.
For instance, $ \alpha \times \Empty $ is equivalent to $ \Empty $
(with respect to subtyping),
as we interpret product types into Cartesian products.
Thus, $ \alpha $ is irrelevant in $ \alpha \times \EmptyType $.
To capture this concept, we introduce the notion of \emph{meaningful variables}
in a type $t$. We define these to be the set
\[
    \Mvar(t) =
    \SetC{\alpha \in \Var(t) | t \SubstSingle[\Empty/\alpha] \nsimeq t}
    \: ,
\]
where the choice of $ \Empty $ to replace $ \alpha $ is arbitrary
(any other closed type yields the same definition).
Equivalent types have exactly the same meaningful variables.
To define generalization, we allow quantifying variables
which are free in the type environment but are meaningless in it
(intuitively, we act as if types were in a canonical form
without irrelevant variables).

We extend $ \Var $ to type schemes as
$ \Var(\Scheme{A. t}) = \Var(t) \setminus A $,
and do likewise for $ \Mvar $.

Type substitutions are defined in a standard way by coinduction;
there being no kinding system, we do not need the admissibility condition
of \VariantsK.

We define type environments $ \Gamma $ as usual.
The operations of generalization of types and instantiation of type schemes,
instead, must account for the presence of irrelevant variables
and of subtyping.

Generalization with respect to $ \Gamma $
quantifies all variables in a type
except for those that are free \emph{and meaningful} in $ \Gamma $:
\[
    \Gen_\Gamma(t) = \Scheme{A. t} \, ,
    \quad \text{where }
    A = \Var(t) \setminus \Mvar(\Gamma)
    \: .
\]
We extend $ \Gen $ pointwise to sets of bindings
$ \Set{x_1\colon t_1, \dots, x_n\colon t_n} $.

The set of instances of a type scheme is given by
\[
    \Inst(\Scheme{A. t}) = \SetC{t \theta | \Dom(\theta) \subseteq A}
    \: ,
\]
and we say that a type scheme $ s_1 $ is \emph{more general}
than a type scheme $ s_2 $---written $ \SMoreGen{ s_1 < s_2 } $---if
\begin{equation}\label{general}
    \forall t_2 \in \Inst(s_2) . \:
    \exists t_1 \in \Inst(s_1) . \: t_1 \leq t_2
    \: .
\end{equation}

Notice that the use of subtyping in the definition above generalizes the corresponding definition of ML (which uses equality) and subsumes the notion of ``admissibility'' of \VariantsK{} by a far simpler and more natural relation (cf.\ Definitions~\ref{def:variants-k-type-substitutions2} and~\ref{def:variants-k-instances}).

Figure~\ref{fig:variants-s-typing} defines the typing relation
$ \STyping{ \Gamma |- e: t } $ of the \VariantsS{} type system
(we use the \VariantsS{} subscript in the turnstile symbol
to distinguish this relation from that for \VariantsK).
All rules except that for pattern matching are straightforward.
Note that \RuleRef{Ts-Const} is more precise than in \VariantsK{}
since we have singleton types,
and that \RuleRef{Ts-Tag} uses the types we have introduced for variants.

The rule \RuleRef{Ts-Match} involves two new concepts
that we present below.
We start by typing the expression to be matched, $ e_0 $,
with some type $ t_0 $.
We also require every branch $ e_i $ to be well-typed with some type $ t_i' $:
the type of the whole \Keyw{match} expression is the union of all $ t_i' $.
We type each branch in an environment expanded
with types for the capture variables of $ p_i $:
this environment is generated by the function $ \SPatternEnv{ t_i // p_i } $ (described below)
and is generalized.

The advantage of our richer types here is that,
given any pattern,
the set of values it accepts
is always described precisely by a type.

\begin{Definition}[Accepted type]
    The \emph{accepted type} $ \SAcc{p} $ of a pattern $ p $
    is defined inductively as:
    \begin{align*}
        \SAcc{\Wildcard} = \SAcc{x} & = \Any &
        \SAcc{c} & = c \\
        \SAcc{(p_1, p_2)} & = \SAcc{p_1} \times \SAcc{p_2} &
        \SAcc{\Tagg(p)} & = \Tagg(\SAcc{p}) \\
        \SAcc{\PAnd{p_1 \& p_2}} & = \SAcc{p_1} \land \SAcc{p_2} &
        \SAcc{\POr{p_1 | p_2}} & = \SAcc{p_1} \lor \SAcc{p_2}
        \: .
    \end{align*}
\end{Definition}

\begin{figure*}

\begin{align*}
  \STransl_K[\alpha]
            & =
            \begin{cases}
                \alpha
                    & \text{if } K(\alpha) = \KUnkind \\
                (\SLow_K(L, T) \lor \alpha) \land \SUpp_K(U, T)
                    & \text{if } K(\alpha) = (L, U, T)
            \end{cases}
        \\
        \STransl_K[b]
            & =
            b \\
        \STransl_K[\tau_1 \to \tau_2]
            & =
            \STransl_K[\tau_1] \to \STransl_K[\tau_2] \\
        \STransl_K[\tau_1 \times \tau_2]
            & =
            \STransl_K[\tau_1] \times \STransl_K[\tau_2]
    \end{align*}
    \begin{align*}
       \text{where:}\qquad
    \SLow_K(L, T)
            & =
            \textstyle\bigvee_{\Tagg \in L}
                \TTagg(\textstyle\bigwedge_{\Tagg\colon \tau \in T}
                \STransl_K[\tau])
        \\
        \SUpp_K(U, T)
            & =
            \begin{cases}
                \textstyle\bigvee_{\Tagg \in U}
                    \TTagg(\textstyle\bigwedge_{\Tagg\colon \tau \in T}
                    \STransl_K[\tau])
                    & \text{if } U \neq \Tags
                \\[5pt]
                \textstyle\bigvee_{\Tagg \in \Domain(T)}
                    \TTagg(\textstyle\bigwedge_{\Tagg\colon \tau \in T}
                    \STransl_K[\tau])
                    \: \lor \:
                    (\AnyVariantType \setminus
                    \textstyle\bigvee_{\Tagg \in \Domain(T)}
                    \TTagg(\AnyType))
                    & \text{if } U = \Tags
            \end{cases}
    \end{align*}

    \caption{Translation of \Kin-types to \Sem-types.}
    \label{fig:variants-s-translation}
\end{figure*}

For well-typed values $ v $,
we have $
    \Matching{ v / p } \neq \MatchFail \iff
        \STyping{ \varnothing |- v: \SAcc{p} }
$.
We use accepted types to express the condition of \emph{exhaustiveness}:
$ t_0 \leq \bigvee_{i \in I} \SAcc{p_i} $
ensures that every value $ e_0 $ can reduce to (i.e., every value in $t_0$)
will match at least one pattern (i.e., is in the accepted type of some pattern).
We also use them to compute precisely the subtypes of $ t_0 $
corresponding to the values which will trigger each branch.
In the rule,
$ t_i $ is the type of all values which will be selected by the $ i $-th branch:
those in $ t_0 $ (i.e., generated by $ e_0 $),
not in any $ \SAcc{p_j} $ for $ j < i $
(i.e., not captured by any previous pattern),
and in $ \SAcc{p_i} $ (i.e., accepted by $ p_i $).
These types $ t_i $ allow us to express \emph{non-redundancy} checks:
if $ t_i \leq \Empty $ for some $ i $,
then the corresponding pattern will never be selected
(which likely means the programmer has made some mistake
and should receive a warning).%
\ifdirtytricks\else
\footnote{%
        We can also exploit redundancy information to exclude certain branches
        from typing (see Section~\ref{extensions-overloading}),
        though it is not always possible during type reconstruction.}
\fi

The last element we must describe is the generation of types
for the capture variables of each pattern
by the $ \SPatternEnv{ t_i // p_i } $ function.
Here, our use of $ t_i $ means we exploit
the shape of the pattern $ p_i $
and of the previous ones to generate more precise types;
environment generation in \VariantsK{} essentially uses only $ t_0 $
and is therefore less precise.

Environment generation relies on two functions $ \SPiFst $ and $ \SPiSnd $
which extract the first and second component
of a type $ t \leq \Any \times \Any $.
For instance, if $
    t = (\alpha \times \beta) \lor (\Keyw{bool} \times \Keyw{int})
$, we have $ \SPiFst(t) = \alpha \lor \Keyw{bool} $
and $ \SPiSnd(t) = \beta \lor \Keyw{int} $.
Given any tag $ \Tagg $,
$ \SPiTagg $ does likewise for variant types with that tag.
See \citet[Appendix C.2.1]{Castagna2014} and \citet{Petrucciani2015}
for the full details.

\begin{Definition}[Pattern environment generation]
    Given a pattern $ p $ and a type $ t \leq \SAcc{p} $,
    the type environment $ \SPatternEnv{ t // p } $
    generated by pattern matching
    is defined inductively as:

\smallskip
\noindent$\begin{array}{r@{{{}={}}}l@{\qquad}r@{{}={}}l}
        \SPatternEnv{ t // \Wildcard }
            &  \varnothing & \SPatternEnv{ t // (p_1, p_2) }
            &
            \SPatternEnv{ \SPiFst(t) // p_1 } \: \cup \:
            \SPatternEnv{ \SPiSnd(t) // p_2 }
            \\[1mm]
        \SPatternEnv{ t // x }
            &  \Set{x\colon t} &        \SPatternEnv{ t // \Tagg(p) }
            &  \SPatternEnv{ \SPiTagg(t) // p } \\[1mm]
        \SPatternEnv{ t // c }
            &  \varnothing
            &
        \SPatternEnv{ t // \PAnd{p_1 \& p_2} }
            &
            \SPatternEnv{ t // p_1 } \: \cup \: \SPatternEnv{ t // p_2 } \\[1mm]
\SPatternEnv{ t // \POr{p_1 | p_2} }
            &
            \makebox[1cm][l]{$\SPatternEnv{ (t \land \SAcc{p_1}) // p_1 }
            \: \SPatternEnvOr \:
            \SPatternEnv{ (t \setminus \SAcc{p_1}) // p_2 $}
        \:}\\[2mm]
    \end{array}$
\noindent
where $
        (\Gamma \SPatternEnvOr \Gamma')(x) = \Gamma(x) \lor \Gamma'(x)
    $.
\end{Definition}
\noindent
The \VariantsS{} type system is sound, as stated by the following properties.

\begin{Theorem}[Progress]
\label{thm:variants-s-progress}
    Let $ e $ be a well-typed, closed expression
    (i.e., $ \STyping{ \varnothing |- e: t } $ holds for some $ t $).
    Then, either $ e $ is a value
    or there exists an expression $ e' $ such that $ \Smallstep{ e ~> e' } $.
\end{Theorem}

\begin{Theorem}[Subject reduction]
\label{thm:variants-s-subject-reduction}
    Let $ e $ be an expression
    and $ t $ a type such that $ \STyping{ \Gamma |- e: t } $.
    If $ \Smallstep{ e ~> e' } $, then $ \STyping{ \Gamma |- e': t } $.
\end{Theorem}

\begin{Corollary}[Type soundness]
\label{cor:variants-s-soundness}
    Let $ e $ be a well-typed, closed expression,
    that is, such that $ \STyping{ \varnothing |- e: t } $ holds for some $ t $.
    Then, either $ e $ diverges or it reduces to a value $ v $
    such that $ \STyping{ \varnothing |- v: t } $.
\end{Corollary}

\subsection{Comparison with \VariantsK}
\label{sec:variants-s-comparisons}

Our type system \VariantsS{} extends  \VariantsK{} in the sense that
every well-typed program of \VariantsK{}
is also well-typed in \VariantsS:
we say that \VariantsS{} is \emph{complete} with respect to \VariantsK{}.

To show completeness,
we define a translation $ \STransl_K[\cdot] $
which maps \Kin-types (i.e., types of \VariantsK)
to \Sem-types (types of \VariantsS).
The translation is parameterized by a kinding environment
to make sense of type variables.

\begin{Definition}[Translation of types]
    Given a \Kin-type $ \tau $ in a non-recursive kinding environment $ K $,
    its \emph{translation} is the \Sem-type $ \STransl_K[\tau] $
    defined inductively by the rules in Figure~\ref{fig:variants-s-translation}.

    We define the translation of type schemes as
    $ \STransl_K[\KScheme{A. K' > \tau}] = \Scheme{A. \STransl_K,K'[\tau]} $
    and that of type environments by translating each type scheme pointwise.
\end{Definition}

The only complex case is the translation of a constrained variable.
We translate it to the same variable, in union with its lower bound
and in intersection with its upper bound.
Lower bounds and finite upper ones (i.e., those where $ U \neq \Tags $)
are represented by a union of variant types.
In \VariantsK, a tag in $ U $ may be associated
with more than one argument type,
in which case its argument should have all these types.
This is a somewhat surprising feature of the type system in OCaml---for
details, see \citet{Garrigue2002,Garrigue2015}---but
here we can simply take the intersection of all argument types.
For instance, the OCaml type \Caml![< `A of int | `B of unit > `A ] as 'a!, represented in \VariantsK{} by the type variable $\alpha$ with kind $
    (\Set{\TagLit{A}}, \Set{\TagLit{A}, \TagLit{B}},
        \Set{\TagLit{A}\colon \Keyw{int},
        \TagLit{B}\colon \Keyw{unit}})
$, is translated into $(\TagLit{A}(\Keyw{int})\lor\alpha)\land(\TagLit{A}(\Keyw{int})\lor\TagLit{B}(\Keyw{unit}))$.

\begin{figure*}

    \begin{mathpar}
        \MyInfer[TRs-Var]
            {  }
            { \SRConstrGen{ x: t => \Set{ \SRSub{x < t} } } }
            {}
        \and
        \MyInfer[TRs-Const]
            {  }
            { \SRConstrGen{ c: t => \Set{ \SRSub{c < t} } } }
            {}
        \and
        \MyInfer[TRs-Abstr]
            { \SRConstrGen{ e: \beta => C } }
            { \SRConstrGen{ \Abstr{x. e}: t => \Set{
                \SRDef{ \Set{x\colon \alpha} | C }, \:
                \SRSub{\alpha \to \beta < t} } } }
            {}
        \and
        \MyInfer[TRs-Appl]
            { \SRConstrGen{ e_1: \alpha \to \beta => C_1 } \\
              \SRConstrGen{ e_2: \alpha           => C_2 } }
            { \SRConstrGen{ \Appl{e_1}{e_2}: t    =>
                C_1 \cup C_2 \cup \Set{\SRSub{\beta < t}} } }
            {}
        \and
        \MyInfer[TRs-Pair]
            { \SRConstrGen{ e_1: \alpha_1 => C_1 } \\
              \SRConstrGen{ e_2: \alpha_2 => C_2 } }
            { \SRConstrGen{ (e_1, e_2): t =>
                C_1 \cup C_2 \cup
                \Set{ \SRSub{ \alpha_1 \times \alpha_2 < t} }
              } }
            {}
        \and
        \MyInfer[TRs-Tag]
            { \SRConstrGen{ e: \alpha => C } }
            { \SRConstrGen{ \Tagg(e): t =>
                C \cup \Set{ \SRSub{\TTagg(\alpha) < t} } } }
            {}
        \and
        \MyInfer[TRs-Match]
            { \SRConstrGen{ e_0: \alpha => C_0 } \\
              t_i = (\alpha \setminus \textstyle\bigvee_{j < i} \SAcc{p_j})
                \land \SAcc{p_i} \\\\
              \forall i \in I \\
              \SRPatternEnv{ t_i /// p_i => \Gamma_i, C_i } \\
              \SRConstrGen{ e_i: \beta => C_i' } \\\\
              C_0' = C_0 \cup (\textstyle\bigcup_{i \in I} C_i) \cup
                \Set{ \SRSub{\alpha < \textstyle\bigvee_{i \in I} \SAcc{p_i}} } 
            }
            { \SRConstrGen{
                \Match{e_0 with (p_i \to e_i)_{i \in I}}: t => \Set{
                \SRLet{ C_0' | (\Gamma_i \In C_i')_{i \in I}}, \:
                \SRSub{\beta < t} } }
            }
            {}
    \end{mathpar}

    \caption{Constraint generation rules.}
    \label{fig:reconstruction-constraint-generation-mono-noa}
\end{figure*}

The translation of an upper bound $ U = \Tags $ is more involved.
Ideally, we need the type
\[
    \textstyle\bigvee_{\Tagg \in \Domain(T)}
    \Tagg(\textstyle\bigwedge_{\Tagg\colon \tau \in T} \STransl_K[\tau])
    \enspace \lor \enspace
    \textstyle\bigvee_{\Tagg \notin \Domain(T)}
    \Tagg(\Any)
\]
which states that tags mentioned in $ T $ can only appear
with arguments of the proper type,
whereas tags not in $ T $ can appear with any argument.
However, the union on the right is infinite
and cannot be represented in our system;
hence, in the definition in Figure~\ref{fig:variants-s-translation}
we use its complement with respect to the top type of variants
$ \AnyVariantType $.%
    \footnote{\label{fn:anyvariant}%
        The type $ \AnyVariantType $ can itself be defined by complementation
        as
        \[
            \lnot \big( (\textstyle\bigvee_{b \in \BasicTypes} b)
                \lor (\Empty \to \Any) \lor (\Any \times \Any) \big) :
        \]
        the type of values which are not constants, nor abstractions,
        nor pairs.}

In practice, a type $ (t_{\text{L}} \lor \alpha) \land t_{\text{U}} $
can be replaced by its lower (respectively, upper) bound
if $ \alpha $ only appears in covariant (resp., contravariant) position.

We state the completeness property as follows.

\begin{Theorem}[Preservation of typing]
\label{thm:variants-s-conservativity-preservation}
    Let $ e $ be an expression,
    $ K $ a non-recursive kinding environment,
    $ \Gamma $ a \Kin-type environment,
    and $ \tau $ a \Kin-type.
    If $ \KTyping{ K; \Gamma |- e: \tau } $,
    then $ \STyping{ \STransl_K[\Gamma] |- e: \STransl_K[\tau] } $.
\end{Theorem}

Notice that we have defined $ \STransl_K[\cdot] $ by induction.
Therefore, strictly speaking, we have only proved that \VariantsS{} deduces all the judgments provable for non-recursive types in \VariantsK.
Indeed, in the statement we require the kinding environment $ K $
to be non-recursive%
    \footnote{
        We say $ K $ is non-recursive if it does not contain any cycle
        $ \alpha, \alpha_1, \dots, $ $ \alpha_n, \alpha $
        such that the kind of each variable $ \alpha_i $ contains
        $ \alpha_{i + 1} $.}.
We conjecture that the result holds also with recursive kindings and that it can be proven by coinductive techniques.

\section{Type reconstruction}\label{recon}

In this section, we study type reconstruction for the \VariantsS{} type system.
We build on the work of \citet{Castagna2015}, who study local type inference
and type reconstruction for the polymorphic version of \CDuce.
In particular, we reuse their work on the resolution
of the \emph{tallying problem}, which plays in our system
the same role as unification in ML.

Our contribution is threefold:
$(i)$ we prove type reconstruction for our system to be both sound and complete,
while in \citet{Castagna2015} it is only proven to be sound for \CDuce{} (indeed, we rely
on the restricted role of intersection types in our system to obtain this result);
$(ii)$ we describe reconstruction with let-polymorphism
and use structured constraints to separate constraint generation
from constraint solving;
$(iii)$ we define reconstruction for full pattern matching.
Both let-polymorphism and pattern matching
are omitted in \citet{Castagna2015}.

Type reconstruction for a program (a closed expression) $ e $
consists in finding a type $ t $
such that $ \STyping{ \varnothing |- e: t } $ can be derived:
we see it as finding a type substitution $ \theta $
such that $ \STyping{ \varnothing |- e: \alpha \theta } $ holds for some fresh variable $\alpha$.
We generalize this to non-closed expressions
and to reconstruction of types that are partially known.
Thus, we say that type reconstruction consists---given an expression $ e $,
a type environment $ \Gamma $, and a type $ t $---in computing
a type substitution $ \theta $ such that
$ \STyping{ \Gamma \theta |- e: t \theta } $ holds,
if any such $ \theta $ exists.

Reconstruction in our system proceeds in two main phases.
In the first, \emph{constraint generation} (Section~\ref{congen}),
we generate from an expression $ e $ and a type $ t $
a set of constraints that record the conditions
under which $ e $ may be given type $ t $.
In the second phase, \emph{constraint solving} (Sections~\ref{conrew}--\ref{typeconsol}), we solve (if possible) these constraints
to obtain a type substitution $ \theta $.

We keep these two phases separate following an approach inspired by
presentations of HM($X$) \citep{Pottier2005}:
we use structured constraints which contain expression variables,
so that constraint generation does not depend on the type environment $ \Gamma $
that $ e $ is to be typed in.
$ \Gamma $ is used later for constraint solving.

Constraint solving is itself made up of two steps:
\emph{constraint rewriting} (Section~\ref{conrew}) and \emph{type-constraint solving} (Section~\ref{typeconsol}).
In the former, we convert a set of structured constraints
into a simpler set of subtyping constraints.
In the latter, we solve this set of subtyping constraints
to obtain a set of type substitutions;
this latter step is analogous to unification in ML
and is computed using the tallying algorithm of \citet{Castagna2015}.
Constraint rewriting also uses type-constraint solving internally;
hence, these two steps are actually intertwined in practice.

\begin{figure*}

    \begin{mathpar}
        \MyInfer
            { \forall i \in I \\
              \SRConstrRewr{ \Gamma |- c_i => D_i } }
            { \SRConstrRewr{ \Gamma |- \SetC{c_i | i \in I} =>
                \textstyle\bigcup_{i \in I} D_i } }
            {}
        \and
        \MyInfer
            { }
            { \SRConstrRewr{ \Gamma |- \SRSub{t < t'} => \Set{\SRSub{t < t'}} }
            }
            {} 
        \and
        \MyInfer
            { \Gamma(x) = \Scheme{ \Set{\alpha_1, \dots, \alpha_n}. t_x } }
            { \SRConstrRewr{ \Gamma |- \SRSub{x < t} => \Set{
                \SRSub{t_x \SubstDots[\beta_1/\alpha_1 ... \beta_n/\alpha_n] <
                    t} } }
            }
            {}
        \and
        \MyInfer
            { \SRConstrRewr{ \Gamma, \Gamma' |- C => D } }
            { \SRConstrRewr{ \Gamma |- \SRDef{ \Gamma' | C } => D } }
            {}
        \and
        \MyInfer
            { \SRConstrRewr{ \Gamma |- C_0 => D_0 } \\
              \theta_0 \in \Tally(D_0) \\\\
              \forall i \in I \\
              \SRConstrRewr{
                \Gamma, \Gen_{\Gamma \theta_0}(\Gamma_i \theta_0) |-
                C_i => D_i }
            }
            { \SRConstrRewr{ \Gamma |-
                \SRLet{ C_0 | (\Gamma_i \In C_i)_{i \in I} } =>
                    \SREq(\theta_0) \cup \textstyle\bigcup_{i \in I} D_i }
            }
            {}
    \end{mathpar}

    \caption{Constraint rewriting rules.}
    \label{fig:reconstruction-constraint-rewriting-poly-noa}
\end{figure*}

\subsection{Constraint generation}\label{congen}

Given an expression $ e $ and a type $ t $,
constraint generation computes a finite set of constraints
of the form defined below.

\begin{Definition}[Constraints]
\label{def:reconstruction-constraints}
    A \emph{constraint} $ c $ is a term
    inductively generated by the following grammar:
    \[
      \Bnf{ c
        \= \SRSub{t < t}
        \| \SRSub{x < t}
        \| \SRDef{ \Gamma | C }
        \| \SRLet{ C | (\Gamma_i \In C_i)_{i \in I} } }
    \]
    where $ C $ ranges over \emph{constraint sets},
    that is, finite sets of constraints, and
    where the range of every type environment $ \Gamma $ in constraints
    of the form \Keyw{def} or \Keyw{let}
    only contains types (i.e., trivial type schemes).
\end{Definition}

A constraint of the form $ \SRSub{t < t'} $
requires $ t \theta \leq t' \theta $ to hold
for the final substitution $ \theta $.
One of the form $ \SRSub{x < t} $ constrains the type of $ x $
(actually, an instantiation of its type scheme with fresh variables)
in the same way.
A definition constraint $ \SRDef{ \Gamma | C } $
introduces new expression variables, as we do in abstractions;
these variables may then occur in $ C $.
We use \Keyw{def} constraints to introduce monomorphic bindings
(environments with types and not type schemes).

Finally, \Keyw{let} constraints introduce polymorphic bindings.
We use them for pattern matching:
hence, we define them with multiple branches (the constraint sets $C_i$'s),
each with its own environment
(binding the capture variables of each pattern to types).
To solve a constraint $ \SRLet{ C_0 | (\Gamma_i \In C_i)_{i \in I} } $,
we first solve $ C_0 $ to obtain a substitution $ \theta $;
then, we apply $ \theta $ to all types in each $ \Gamma_i $
and we generalize the resulting types;
finally, we solve each $ C_i $
(in an environment expanded with the generalization of $ \Gamma_i \theta $).

We define constraint generation as a relation $ \SRConstrGen{ e: t => C } $,
given by the rules in
Figure~\ref{fig:reconstruction-constraint-generation-mono-noa}.
We assume all variables introduced by the rules to be fresh (see the Appendix for the formal treatment of freshness: cf.\ Definition~\ref{def:a-sr-fresh} and Figures~\ref{fig:reconstruction-constraint-generation-matching-a} and \ref{fig:reconstruction-constraint-generation-a}).
Constraint generation for variables and constants (rules \Rule{TRs-Var} and \Rule{TRs-Const})
just yields a subtyping constraint.
For an abstraction $ \Abstr{x. e} $ (rule \Rule{TRs-Abstr}),
we generate constraints for the body and wrap them
into a definition constraint binding $ x $ to a fresh variable $ \alpha $;
we add a subtyping constraint
to ensure that $ \Abstr{x. e} $ has type $ t $ by subsumption.
The rules for applications, pairs, and tags are similar.

For pattern-matching expressions (rule \Rule{TRs-Match}),
we use an auxiliary relation $ \SRPatternEnv{ t /// p => \Gamma, C } $
to generate the pattern type environment $ \Gamma $,
together with a set of constraints $ C $
in case the environment contains new type variables.
The full definition is in the Appendix;
as an excerpt, consider the rules for variable and tag patterns.\vspace{-3.3mm}
\begin{mathpar}
  \MyInfer
    { }
    { \SRPatternEnv{ t /// x => \Set{x\colon t}, \varnothing } }
    {}
  \quad
  \MyInfer
    { \SRPatternEnv{ \alpha /// p => \Gamma, C } }
    { \SRPatternEnv{ t /// \Tagg(p) =>
        \Gamma, C \cup \Set{ \SRSub{t < \Tagg(\alpha)} } } }
    {}
\end{mathpar}
The rule for variable patterns produces no constraints (and the empty environment).
Conversely, the rule for tags must introduce a new variable $\alpha$
to stand for the argument type:
the constraint produced mirrors
the use of the projection operator $ \SPiTagg $ in the deductive system.
To generate constraints for a pattern-matching expression,
we generate them for the expression to be matched
and for each branch separately.
All these are combined in a \Keyw{let} constraint,
together with the constraints generated by patterns and with $
  \SRSub{\alpha < \bigvee_{i \in I} {\SAcc{p_i}}}
$, which ensures exhaustiveness.

\subsection{Constraint rewriting}\label{conrew}

The first step of constraint solving consists in rewriting the constraint set
into a simpler form that contains only subtyping constraints, that is, into a set of the form
$ \Set{\SRSub{t_1 < t_1'}, \dots, \SRSub{t_n < t_n'}} $ (i.e., no \texttt{let}, \texttt{def}, or expression variables). We call such sets \emph{type-constraint sets} (ranged over by $ D $).

Constraint rewriting is defined as a relation
$ \SRConstrRewr{ \Gamma |- C => D } $:
between type environments, constraints or constraint sets,
and type-constraint sets.
It is given by the rules in
Figure~\ref{fig:reconstruction-constraint-rewriting-poly-noa}.

We rewrite constraint sets pointwise.
We leave subtyping constraints unchanged.
In variable type constraints, we replace the variable $ x $
with an instantiation of the type scheme $ \Gamma(x) $
with the variables $ \beta_1 $, \dots, $ \beta_n $,
which we assume to be fresh.
We rewrite \Keyw{def} constraints by expanding the environment
and rewriting the inner constraint set.

The complex case is that of \Keyw{let} constraints,
which is where rewriting already performs type-constraint solving.
We first rewrite the constraint set $ C_0 $.
Then we extract a solution $ \theta_0 $---if any exists---by the $ \Tally $ algorithm
(described below).
The algorithm can produce multiple alternative solutions:
hence, this step is non-deterministic.
Finally, we rewrite each of the $ C_i $ in an expanded environment.
We perform generalization, so \Keyw{let} constraints
may introduce polymorphic bindings.
The resulting type-constraint set is the union
of the type-constraint sets obtained for each branch
plus $ \SREq(\theta_0) $, which is defined as
\[
    \SREq(\theta_0)
        =
        \textstyle\bigcup_{\alpha \in \Dom(\theta_0)}
    \Set{\SRSub{\alpha < \alpha \theta_0}, \,
      \SRSub{\alpha \theta_0 < \alpha}}
    \: .
\]

We add the constraints of $ \SREq(\theta_0) $ because tallying might generate
multiple incompatible solutions for the constraints in $ D_0 $.
The choice of $ \theta_0 $ is arbitrary,
but we must force subsequent steps of constraint solving to abide by it.
Adding $ \SREq(\theta_0) $ ensures that every solution $ \theta $
to the resulting type-constraint set will satisfy
$ \alpha \theta \simeq \alpha \theta_0 \theta $ for every $ \alpha $,
and hence will not contradict our choice.

\subsection{Type-constraint solving}\label{typeconsol}

\citet{Castagna2015} define the \emph{tallying problem} as the problem---in our
terminology---of finding a substitution that satisfies a given
type-constraint set.

\begin{Definition}
\label{def:reconstruction-tallying}
  We say that a type substitution $ \theta $ satisfies a type-constraint set
  $ D $, written $ \SRSat{ \theta |- D } $, if
  $ t \theta \leq t' \theta $ holds for every $ \SRSub{t < t'} $ in $ D $.
  When $ \theta $ satisfies $ D $, we say it is a solution
  to the \emph{tallying problem} of $ D $.
\end{Definition}

The tallying problem is the analogue in our system
of the unification problem in ML.
However, there is a very significant difference:
while unification admits principal solutions, tallying does not.
Indeed, the algorithm to solve the tallying problem for a type-constraint set
produces a finite set of type substitutions. The algorithm is sound
in that all substitutions it generates are solutions.
It is complete in the sense that any other solution is less general than one
of those in the set:
we have a finite number of solutions which are principal when taken together,
but not necessarily a single solution that is principal on its own.

This is a consequence of our semantic definition of subtyping.
As an example, consider subtyping for product types:
with a straightforward syntactic definition,
a constraint $ t_1 \times t_1' \leq t_2 \times t_2' $
would simplify to the conjunction of two constraints
$ t_1 \leq t_2 $ and $ t_1' \leq t_2' $.
With semantic subtyping---where products are seen as Cartesian products---that
simplification is sound, but it is not the only possible choice:
either $ t_1 \leq \Empty $ or $ t_1' \leq \Empty $ is also enough
to ensure $ t_1 \times t_1' \leq t_2 \times t_2' $,
since both ensure $ t_1 \times t_1' \simeq \Empty $.
The three possible choices can produce incomparable solutions.

\citet[Section 3.2 and Appendix C.1]{Castagna2015}
define a sound, complete, and terminating algorithm to solve the tallying problem,
which can be adapted to our types by encoding variants as pairs.
We refer to this algorithm here as $ \Tally $
(it is $ \CDuceSol_\varnothing $ in the referenced work)
and state its properties.

\begin{Property}[Tallying algorithm]
    There exists a terminating algorithm $ \Tally $ such that,
    for any type-constraint set $ D $,
    $ \Tally(D) $ is a finite, possibly empty, set of type substitutions.
\end{Property}

\begin{Theorem}[Soundness and completeness of $ \Tally $]
    Let $ D $ be a type-constraint set. For any type substitution $ \theta $:
    \begin{compactitemize}
        \item if $ \theta \in \Tally(D) $, then $ \SRSat{\theta |- D} $;
        \item if $ \SRSat{ \theta |- D } $, then $
            \Exists \theta' \in \Tally(D), \theta'' .
            \Forall \alpha \in \Dom(\theta) .
            \alpha \theta \simeq \alpha \theta' \theta''
          $.
    \end{compactitemize}
\end{Theorem}

Hence, given a type-constraint set, we can use $ \Tally $
to either find a set of solutions or determine it has no solution:
$ \Tally(D) = \varnothing $ occurs if and only if
there exists no $ \theta $ such that $ \SRSat{ \theta |- D } $.

\subsubsection{Properties of type reconstruction}

Type reconstruction as a whole consists in generating a constraint set $ C $
from an expression, rewriting this set into a type-constraint set $ D $
(which can require solving intermediate type-constraint sets)
and finally solving $ D $ by the $ \Tally $ algorithm.
Type reconstruction is both sound and complete
with respect to the deductive type system \VariantsS.
We state these properties in terms of constraint rewriting.

\begin{Theorem}[Soundness of constraint generation and rewriting]
    Let $ e $ be an expression, $ t $ a type,
    and $ \Gamma $ a type environment.
    If $ \SRConstrGen{ e: t => C } $, $ \SRConstrRewr{ \Gamma |- C => D } $,
    and $ \SRSat{ \theta |- D } $,
    then $ \STyping{ \Gamma \theta |- e: t \theta } $.
\end{Theorem}

\begin{Theorem}[Completeness of constraint generation and rewriting]
    Let $ e $ be an expression, $ t $ a type,
    and $ \Gamma $ a type environment.
    Let $ \theta $ be a type substitution
    such that $ \STyping{ \Gamma \theta |- e: t \theta } $.

    Let $ \SRConstrGen{ e: t => C } $.
    There exist a type-constraint set $ D $ and a type substitution $ \theta' $,
    with $ \Domain(\theta) \cap \Domain(\theta') = \varnothing $,
    such that $ \SRConstrRewr{ \Gamma |- C => D } $
    and $ \SRSat{ (\theta \cup \theta') |- D } $.
\end{Theorem}
These theorems and the properties above
express soundness and completeness for the reconstruction system. Decidability is a direct consequence of the termination of the tallying algorithm.

\subsubsection{Practical issues}

As compared to reconstruction in ML,
our system has the disadvantage of being non-deterministic:
in practice, an implementation should check every solution that tallying generates
at each step of type-constraint solving
until it finds a choice of solution which makes the whole program well-typed.
This should be done at every step of generalization
(that is, for every \Keyw{match} expression)
and might cripple efficiency.
Whether this is significant in practice or not is a question that
requires further study and experimentation.
 Testing multiple solutions cannot be avoided
since our system does not admit principal types. For instance the function
\begin{CamlBlock}
let f(x,y) = (function (`A,`A)|(`B,`B)->`C)(x,y)
\end{CamlBlock}
has both type \Caml!(`A,`A)->`C!
and type \Caml!(`B,`B)->`C! (and neither is better than the other) but it is not possible to deduce for it their least upper
bound \Caml!(`A,`A)$\lor$(`B,`B)->`C! (which would be principal).

Multiple solutions often arise by instantiating some type variables by the empty type. Such solutions are in many cases subsumed by other more general solutions, but not always. For instance, consider the
\Caml!'a list! data-type (encoded as the recursive type
\Caml!X = ('a,X)$\lor$[]!) together with the
classic \Caml!map! function over lists (the type of which is
 \Caml!('a -> 'b)$\,$->$\,$'a$\,$list$\,$->$\,$'b$\,$list!). The application of
 \Caml!map! to the successor
function \Caml!succ$\,$:$\,$int$\,$->$\,$int! has type
 \Caml!int$\,$list$\,$->$\,$int$\,$list!,
but also type \Caml![]$\,$->$\,$[]! (obtained by instantiating all the
variables of the type of \Caml!map! by the empty type). The latter
type is correct, cannot be derived (by instantiation and/or subtyping) from the former, but it is seldom useful (it just states that
\Caml!map(succ)! maps the empty list into the empty list).  As such,
it should be possible to define some preferred choice of solution (i.e., the
solution that does not involve empty types) which is likely to be the
most useful in practice.   As it happens, we would like to try to restrict the system so that it
only considers solutions without empty types.  While it would make us
lose completeness with respect to \VariantsS, it would be interesting
to compare the restricted system with ML (with respect to which it
could still be complete).

\section{Extensions}\label{extensions}
In this section, we present three extensions or modifications
to the \VariantsS{} type system; the presentation is just sketched for space reasons:
the details of all three can be found in the Appendix.

The first is the introduction of overloaded functions
typed via intersection types, as done in \CDuce.
The second is a refinement of the typing of pattern matching,
which we have shown as part of Example 2
(the function \Keyw{g} and our definition of \Keyw{map}).
Finally, the third is a restriction of our system
to adapt it to the semantics of the OCaml implementation
which, unlike our calculus,
cannot compare safely untagged values of different types at runtime.

\subsection{Overloaded functions}
\label{extensions-overloading}

\CDuce{} allows the use of intersection types
to type overloaded functions precisely:
for example, it can type the negation function\\[1mm]
\centerline{$
  \Keyw{not} \eqdef
  \Abstr{x. \Match{x with \Keyw{true} \to \Keyw{false} \|
                            \Keyw{false} \to \Keyw{true}}}
$}\medskip

\noindent
with the type
$ (\Keyw{true} \to \Keyw{false}) \land (\Keyw{false} \to \Keyw{true}) $,
which is more precise than $ \Keyw{bool} \to \Keyw{bool} $.
We can add this feature by changing the rule
to type \textlambda-abstractions to
\begin{mathpar}
  \MyInfer[]
    { \forall j \in J .
        \enspace \Typing{ \Gamma, \Set{x\colon t_j'} |- e: t_j } }
    { \Typing{ \Gamma |- \Abstr{x. e}:
        \textstyle\bigwedge_{j \in J} t_j' \to t_j } }
    {}
\end{mathpar}
which types the abstraction with an intersection of arrow types,
provided each of them can be derived for it. The rule above roughly
corresponds to the one introduced by Reynolds for the language
Forsythe~\cite{Reynolds1997}. With this rule alone, however, one has
only the so-called \emph{coherent overloading}~\cite{PierceThesis},
that is, the possibility of assigning different types to the same
piece of code, yielding an intersection type.  In full-fledged
overloading, instead, different pieces of code are executed for
different types of the input. This possibility was first introduced
by \CDuce~\cite{FCB02,Benzaken2003} and it is obtained by typing
pattern matching without taking into account the type of the branches
that cannot be selected for a given input type.  Indeed, the
function ``\Keyw{not}'' above cannot be given the type we want if we just
add the rule above: it can neither be typed as
$ \Keyw{true} \to \Keyw{false} $ nor as $ \Keyw{false} \to \Keyw{true}
$.

To use intersections effectively for pattern matching,
we need to exclude redundant patterns from typing.
We do so by changing the rule \RuleRef{Ts-Match}
(in Figure~\ref{fig:variants-s-typing}):
when for some branch $ i $ we have $ t_i \leq \Empty $, 
we do not type that branch at all, and we do not consider it in the result type
(that is, we set $ t_i' = \Empty $).
In this way, if we take $ t_j' = \Keyw{true} $,
we can derive $ t_j = \Keyw{false} $ (and vice versa).
Indeed, if we assume that the argument is $ \Keyw{true} $,
the second branch will never be selected:
it is therefore sound not to type it at all.
This typing technique is peculiar to \CDuce's overloading.
However, functions in \CDuce{} are explicitly typed.
As type reconstruction is undecidable
for unrestricted intersection type systems,
this extension would make annotations necessary in our system as well.
We plan to study the extension of our system
with intersection types for functions
and to adapt reconstruction to also consider explicit annotations.

\subsection{Refining the type of expressions in pattern matching}\label{refinedpatterns}

Two of our motivating examples concerning pattern matching
(from Section~\ref{intro}, Example 2)
involved a refinement of the typing of pattern matching
that we have not described yet, but which can be added
as a small extension of our \VariantsS{} system.

Recall the function \Keyw{g} defined as
\(
  \Abstr{x.
    \Match{x with \TagLit{A} \to \Appl{\Keyw{id2}}{x} \|
      \Wildcard \to x}}
\),
where \Keyw{id2} has domain $ \TagLit{A} \lor \TagLit{B} $.
Like OCaml, \VariantsS{} requires the type of $ x $
to be a subtype of $ \TagLit{A} \lor \TagLit{B} $,
but this constraint is unnecessary
because $ \Appl{\Keyw{id2}}{x} $ is only computed when $ x = \TagLit{A} $.
To capture this, we need pattern matching to introduce more precise types
for variables in the matched expression;
this is a form of \emph{occurrence typing} \citep{Tobin-Hochstadt2010}
or \emph{flow typing} \citep{Pearce2013}.

We first consider pattern matching on a variable.
In an expression $ \Match{x with (p_i \to e_i)_{i \in I}} $
we can obtain this increased precision by using the type $ t_i $---actually,
its generalization---for $ x $ while typing the $ i $-th branch.
In the case of \Keyw{g}, the first branch is typed assuming $ x $
has type $ t_0 \land \TagLit{A} $,
where $ t_0 $ is the type we have derived for $ x $.
As a result, the constraint $
  t_0 \land \TagLit{A} \leq \TagLit{A} \lor \TagLit{B}
$ does not restrict $ t_0 $.
\ifdirtytricks\else

\fi
We can express this so as to reuse pattern environment generation.
Let $ \SExprPattern{\cdot}\colon \Expressions \to \Patterns $
be a function
such that $ \SExprPattern{x} = x $
and $ \SExprPattern{e} = \Wildcard $ when $ e $ is not a variable.
Then, we obtain the typing above if we use
\ifdirtytricks\vspace{-.8mm}\fi
\[
  \Gamma, \Gen_\Gamma(\SPatternEnv{t_i // \SMExprPattern{e_0}}),
  \Gen_\Gamma(\SPatternEnv{t_i // p_i})
\]
as the type environment in which we type the $ i $-th branch, rather than
$
  \Gamma, \Gen_\Gamma(\SPatternEnv{t_i // p_i})
$.

We generalize this approach to refine types also for variables
occurring inside pairs and variants.
To do so, we redefine $ \SExprPattern{\cdot} $.
On variants, we let $ \SExprPattern{\Tagg(e)} = \Tagg(\SExprPattern{e}) $.
On pairs, ideally we want $
  \SExprPattern{(e_1, e_2)} = (\SExprPattern{e_1}, \SExprPattern{e_2})
$: however, pair patterns cannot have repeated variables,
while $ (e_1, e_2) $ might.
We therefore introduce a new form of pair pattern $ \SMPatternPair<p_1, p_2> $
(only for internal use) which admits repeated variables:
environment generation for such patterns intersects the types it obtains
for each occurrence of a variable.

\subsection{Applicability to OCaml}

A thesis of this work is that the type system of OCaml---specifically,
the part dealing with polymorphic variants and pattern matching---could
be profitably replaced by an alternative, set-theoretic system.
Of course, we need the set-theoretic system to be still type safe.

In Section~\ref{setvariants}, we stated that \VariantsS{} is sound
with respect to the semantics we gave in Section~\ref{language}.
However, this semantics is not precise enough,
as it does not correspond to the behaviour of the OCaml implementation
on ill-typed terms.%
  \footnote{%
    We can observe this if we bypass type-checking,
    for instance by using \Keyw{Obj.magic} for unsafe type conversions.}

Notably, OCaml does not record type information at runtime:
values of different types cannot be compared safely
and constants of different basic types might have the same representation
(as, for instance, \Keyw{1} and \Keyw{true}).
Consider as an example the two functions
\begin{gather*}
  \Abstr{x.
    \Match{x with \Keyw{true} \to \Keyw{true} \| \Wildcard \to \Keyw{false}}}
  \\
  \Abstr{x.
    \Match{x with (\Keyw{true}, \Keyw{true}) \to \Keyw{true}
      \| \Wildcard \to \Keyw{false}}}
  \: .
\end{gather*}
Both can be given the type $ \Any \to \Keyw{bool} $ in \VariantsS,
which is indeed safe in our semantics.
Hence, we can apply both of them to \Keyw{1}, and both return \Keyw{false}.
In OCaml, conversely, the first would return \Keyw{true}
and the second would cause a crash.
The types $
  \Keyw{bool} \to \Keyw{bool}
$ and $
  \Keyw{bool} \times \Keyw{bool} \to \Keyw{bool}
$, respectively, would be safe for these functions in OCaml.
\ifdirtytricks\else

\fi
To model OCaml more faithfully, we define an alternative semantics
where matching a value $ v $ against a pattern $ p $ can have three outcomes
rather than two:
it can succeed ($ \Matching{v / p} = \varsigma $),
fail ($ \Matching{v / p} = \MatchFail $),
or be undefined ($ \Matching{v / p} = \MatchUndef $).
Matching is undefined whenever it is unsafe in OCaml:
for instance, $
  \Matching{ \Keyw{1} / \Keyw{true} } =
  \Matching{ \Keyw{1} / (\Keyw{true}, \Keyw{true}) } = \MatchUndef
$ (see Appendix~\ref{ocaml} for the full definition).

We use the same definition as before for reduction
(see Section~\ref{sec:variants-k-semantics}).
Note that a \Keyw{match} expression on a value reduces to the first branch
for which matching is successful
\emph{if the result is $ \MatchFail $ for all previous branches}.
If matching for a branch is undefined, no branch after it can be selected;
hence, there are fewer possible reductions with this semantics.

Adapting the type system requires us to restrict
the typing of pattern matching so that undefined results cannot arise.
We define the \emph{compatible type} $ \SCompat{p} $ of a pattern $ p $
as the type of values $ v $ which can be safely matched with it:
those for which $ \Matching{v / p} \neq \MatchUndef $.
For instance, $ \SCompat{\Keyw{1}} = \Keyw{int} $.
The rule for pattern matching should require
that the type $ t_0 $ of the matched expression
be a subtype of all $ \SCompat{p_i} $.

Note that this restricts the use of union types in the system.
For instance, if we have a value of type $ \Keyw{bool} \lor \Keyw{int} $,
we can no longer use pattern matching to discriminate
between the two cases.
This is to be expected in a language without runtime type tagging:
indeed, union types are primarily used for variants,
which reintroduce tagging explicitly.
Nevertheless, having unions of non-variant types in the system is still useful,
both internally (to type pattern matching)
and externally (see Example 3 in Section~\ref{intro}, for instance).

\section{Related work}\label{related}
We discuss here the differences between our system
and other formalizations of variants in ML.
We also compare our work with the work on \CDuce{} and other union/intersection type systems.

\subsection{Variants in ML: formal models and OCaml}
\label{ssec:related-variants}

\VariantsK{} is based on the framework of \emph{structural polymorphism}
and more specifically on the presentations by
\citet{Garrigue2002,Garrigue2015}.
There exist several other systems with structural polymorphism:
for instance, the earlier one by \citet{Garrigue1998}
and more expressive constraint-based frameworks,
like the presentation of HM($ X $) by \citet{Pottier2005}.
We have chosen as a starting point the system which corresponds most closely
to the actual implementation in OCaml.

With respect to the system in \citet{Garrigue2002,Garrigue2015},
\VariantsK{} differs mainly in three respects.
First, Garrigue's system describes constraints more abstractly
and can accommodate different forms of polymorphic typing
of variants and of records.
We only consider variants and, as a result, give a more concrete presentation.
Second, we model full pattern matching instead of ``shallow'' case analysis.
To our knowledge, pattern matching on polymorphic variants in OCaml
is only treated in \citet{Garrigue2004a} and only as concerns some problems
with type reconstruction.
We have chosen to formalize it to compare \VariantsK{}
to our set-theoretic type system \VariantsS{},
which admits a simpler formalization and more precise typing.
However, we have omitted a feature of OCaml that allows refinement
of variant types in alias patterns and which is modeled in \citet{Garrigue2002}
by a \Keyw{split} construct.
While this feature makes OCaml more precise than \VariantsK,
it is subsumed in \VariantsS{} by the precise typing of capture variables.
Third, we did not study type inference for \VariantsK{}.
Since \VariantsS{} is more expressive than \VariantsK{}
and since we describe complete reconstruction for it,
extending Garrigue's inference system to pattern matching
was unnecessary for the goals of this work.

As compared to OCaml itself (or, more precisely, to the fragment we consider)
our formalization is different because it requires exhaustiveness;
this might not always by practical in \VariantsK,
but non-exhaustive pattern matching is no longer useful
once we introduce more precise types, as in \VariantsS.
Other differences include not considering variant refinement in alias patterns,
as noted above, and the handling of conjunctive types,
where OCaml is more restrictive than we are in order to infer more intuitive types
\citep[as discussed in][Section 4.1]{Garrigue2004a}.

\subsection{\VariantsS{} and the \CDuce{} calculus}
\label{ssec:related-cduce}

\VariantsS{} reuses the subtyping relation defined by \citet{Castagna2011}
and some of the work described in \citet{Castagna2014,Castagna2015}
(notably, the encoding of bounded polymorphism via type connectives
and the algorithm to solve the tallying problem).
Here, we explore the application of these elements
to a markedly different language.

\citet{Castagna2014,Castagna2015} study polymorphic typing for the \CDuce{}
language, which features type-cases. Significantly, such type-cases
can discriminate between functions of different types;
pattern matching in ML cannot
(indeed, it cannot distinguish between functions and non-functional values).
As a result, the runtime semantics of \CDuce{} is quite involved and, unlike
ours, not type-erasing;
our setting has allowed us to simplify the type system too.
Moreover, most of the work in \citet{Castagna2014,Castagna2015}
studies an explicitly-typed language
(where functions can be typed with intersection types).
In contrast, our language is implicitly typed.
We focus our attention on type reconstruction and prove it sound and complete,
thanks to the limited use we make of intersections.
We have also introduced differences in presentation to conform our system
to standard descriptions of the Hindley-Milner system.

\subsection{Union types and pattern matching}

The use of union and intersection types in ML has been studied
in the literature of \emph{refinement type} systems.
For example, the theses of \citet{Davies2005} and \citet{Dunfield2007}
describe systems where declared datatypes
(such as the ordinary variants of OCaml)
are refined by finite discriminated unions.
Here we study a very different setting, because we consider
\emph{polymorphic} variants and, above all, we focus on providing complete type reconstruction,
while the cited works describe forms of bidirectional type checking
which require type annotations.
Conversely, our system makes a more limited use of intersection types,
since it does not allow the derivation of intersection types for functions.
Refinement type systems are closer in spirit to the work on \CDuce{} which is why we refer the reader to Section 7 on related work in \citet{Castagna2014} for a comprehensive comparison.

For what concerns programming languages we are not aware of any implicitly-typed language
with full-fledged union types. The closest match to our work is probably Typed
Racket~\citep{Tobin-Hochstadt2008,Tobin-Hochstadt2010} which
represents datatypes as unions of tagged types, as we do. However it does not perform type reconstruction: it
is an explicitly-typed language with local type inference, that is,
the very same setting studied for \CDuce{} in \citet{Castagna2015} whose Section 6
contains  a thorough comparison with the type
system of Typed Racket.\footnote{Actually, \CDuce{} local type inference is more general than the one in Typed Racket, insofar as in an application it locally infers the instantiation for both the function and the argument while Typed Racket does only the former.}  Typed Racket also features \emph{occurrence
typing}, which refines the types of variables according to the results
of tests (combinations of predicates on base types and selectors) to
give a form of flow sensitivity.  We introduced a similar feature in
Section~\ref{refinedpatterns}: we use pattern matching and hence
consider tests which are as expressive as theirs, but we do not allow
them to be abstracted out as functions.

\section{Conclusion}\label{conclusion}
This work shows how to add general union, intersection and difference types in  implicitly-typed languages that traditionally use the HM type system. Specifically, we showed how to improve the current OCaml type system of
polymorphic variants in four different aspects: its formalization, its
meta-theoretic properties, the expressiveness of the system, and its
practical ramifications.  These improvements are obtained by a drastic departure from
the current unification-based approach and by the injection in the system of
set-theoretic types and semantic subtyping.

Our approach arguably improves the formalization of polymorphic variants: in our system we directly
encode all meta-theoretic notions in a core---albeit rich---type theory, while
the current OCaml system must introduce sophisticated ``ad hoc'' constructions
(e.g., the definition of constrained kind, cf. Definition~\ref{def:variants-k-kinds}) to
simulate subtyping. This is why, in our approach, bounded polymorphism can be encoded
in terms of union and intersection types, and meta-theoretic properties such as exhaustiveness and redundancy in pattern matching
can be internalized and
expressed in terms of types and subtyping. Likewise, the most pleasant surprise of our formalization is the definition of
the generality relation $ \sqsubseteq $ on type schemes (cf. equation~\eqref{general}):
the current OCaml formalization requires complicated definitions such as the admissibility
of type substitutions, while in our system it turns out to be the
straightforward and natural generalization to subtyping of the usual relation of
ML. A similar consideration can be done for unification, which is here generalized by the notion of tallying.

In the end we obtain a type system which is very natural: if we
abstract the technicalities of the rule for pattern matching, the type system really
is what one expects it to be: all (and only) the classic typing rules
plus a subsumption rule. And even the rule \Rule{Ts-Match}, the most complicated
one, is at the end what one should expect it to be: (1) type the
matched expression $e_0$, (2) check whether the patterns are
exhaustive, (3) for each branch (3.i) compute the set of the
results of $e_0$ that are captured by the pattern of the branch, (3.ii) use them to deduce
the type of the capture variables of the pattern (3.iii) generalize
the types of these variables in order to type the body of the branch, and (4) return the union of the types of the branches.

The advantages of our approach are not limited to the formalization. The
resulting system is more expressive---it types more programs while
preserving static type safety---and natural, insofar as it removes the
pathological behaviours we outlined in the introduction as well as problems found in real life \citep[e.g.,][]{nicollet11,taw06}. The solution can be even more satisfactory if we extend the current syntax of OCaml types. For instance, \citet{nicollet11} shows the OCaml function \Caml:function `A ->$$ `B | x ->$$ x: which transforms \Caml:`A: into \Caml:`B: and leaves any other constructor unchanged. OCaml gives to this function the somewhat nonsensical type \Caml:([> `A | `B ] as 'a) ->$$ 'a:. Our reconstruction algorithm deduces instead the type \Caml:'a ->$$ (`B | ('a\`A)):: it correctly deduces that the result can be either  \Caml:`B: or the variant in input, but can never be  \Caml:`A: \citep[for further examples of the use of difference types see][]{caml-difference,caml-difference2}. If we want to preserve the current syntax of OCaml types, this type should be approximated as  \Caml:([> `B ] as 'a) ->$$ 'a:;
however, if we extend the syntax with differences (that in our system come for free),
we gain the expressiveness that the kinding approach can only achieve with explicit row variables and that is needed, for instance, to encode exceptions~\citep{Blume2008}.
But we can do more:  by allowing also intersections in the syntax of OCaml types we could type Nicollet's function by the type
\Caml:(`A ->$$ `B) & (('a\`A) ->$$ ('a\`A)):, which is exact since it states that the function maps  \Caml:`A: to \Caml:`B: and leaves any argument other than  \Caml:`A: unchanged. As an aside, notice that types of this form provide an exact typing of exception handlers as intended by \citet{Blume2008} (Nicollet's function can be seen as a handler that catches the exception \Caml:`A: yielding \Caml:`B: and lets all other values pass through).


Finally, our work improves some aspects of the theory of semantic subtyping
as well: our type reconstruction copes with let-polymorphism
and pattern matching and it is proven to be not only sound but also complete,
all properties that the system in \citet{Castagna2015} does not possess.
Furthermore, the refinement we proposed in Section~\ref{refinedpatterns}
applies to \CDuce{} patterns as well, and it has already been
implemented in the development version of \CDuce.

This work  is just the first step of a long-term research. Our short-term
plan is to finish an ongoing implementation and test it,
especially as concerns messages to show to the programmer.
We also need to extend the subtyping relation used here to cope with types containing cyclic values (e.g., along the lines of the work of \citet{BonsangueEtAl14}): the subtyping relation of~\citet{Castagna2011} assumes that types contain only finite values, but cyclic values can be defined in OCaml.

The interest of this work is not limited to polymorphic
variants. In the long term we plan to check whether building on
this work it is possible to extend the syntax of OCaml patterns and
types, so as to encode XML document types and provide the OCaml programmer
with processing capabilities for XML documents like those that can be
found in XML-centred programming languages such as \CDuce. Likewise we
want to explore the addition of intersection types to OCaml (or
Haskell) in order to allow the programmer to define refinement types
and check how such an integration blends with existing features,
notably GADTs.

\paragraph{Acknowledgments.} We are very grateful to Jacques Garrigue who gave invaluable feedback on an early version of this work. This work was partially supported by the Oracle Labs External Research Project ``Deep Integration of Programming Languages in Databases with Truffle''.



\bibliography{main}
\bibliographystyle{abbrvnat}

\SetupAppendix
\appendix

\section{Appendix}

In this Appendix, we present full definitions of the language
and type systems we have described, together with complete proofs
of all results.

\subsection{The language of polymorphic variants}

\subsubsection{Syntax}

We assume that there exist
a countable set $ \ExprVars $ of \emph{expression variables},
ranged over by $ x $, $ y $, $ z $, \dots,
a set $ \Constants $ of constants, ranged over by $ c $,
and a set $ \Tags $ of tags, ranged over by $ \Tagg $.

\begin{Definition}[Expressions]
\label{def:a-lang-expressions}
    An \emph{expression} $ e $ is a term inductively generated
    by the following grammar:
    \[
        \Bnf{ e
            \= x \| c \| \Abstr{x. e} \| \Appl{e}{e} \| (e, e) \| \Tagg(e)
            \| \Match{e with (p_i \to e_i)_{i \in I}} }
    \]
    where $ p $ ranges over the set $ \Patterns $ of patterns, defined below.
    We write $ \Expressions $ to denote the set of all expressions.
\end{Definition}

We define $ \Fvar(e) $ to be the set of expression variables
occurring free in the expression $ e $,
and we say that $ e $ is \emph{closed} if and only if $ \Fvar(e) $ is empty.

As customary, we consider expressions up to \textalpha-renaming
of the variables bound by abstractions and by patterns.

\begin{Definition}[Patterns]
\label{def:a-lang-patterns}
    A \emph{pattern} $ p $ is a term
    inductively generated by the following grammar:
    \[
        \Bnf{ p
            \= \Wildcard \| x \| c \| (p, p) \| \Tagg(p)
            \| \PAnd{p \& p} \| \POr{p | p} }
    \]
    such that
    \begin{itemize}
        \item in a pair pattern $ (p_1, p_2) $
            or an intersection pattern $ \PAnd{p_1 \& p_2} $,
            $ \Capt(p_1) \cap \Capt(p_2) = \varnothing $;
        \item in a union pattern $ \POr{p_1 | p_2} $,
            $ \Capt(p_1) = \Capt(p_2) $,
    \end{itemize}
    where $ \Capt(p) $ denotes the set of expression variables
    occurring as sub-terms in a pattern $ p $
    (called the \emph{capture variables} of $ p $).

    We write $ \Patterns $ to denote the set of all patterns.
\end{Definition}

\subsubsection{Semantics}

\begin{Definition}[Values]
\label{def:a-lang-values}
    A \emph{value} $ v $ is a closed expression
    inductively generated by the following grammar.
    \[
        \Bnf{ v \= c \| \Abstr{x. e} \| (v, v) \| \Tagg(v) }
    \]
\end{Definition}

\begin{Definition}[Expression substitution]
\label{def:a-lang-substitution}
    An \emph{expression substitution} $ \varsigma $
    is a partial mapping of expression variables to values.
    We write $ \SubstC[v_i/x_i | i \in I] $ for the substitution
    which replaces free occurrences of $ x_i $ with $ v_i $,
    for each $ i \in I $.
    We write $ e \varsigma $ for the application of the substitution
    to an expression $ e $. We write $ \varsigma_1 \cup \varsigma_2 $
    for the union of disjoint substitutions.
\end{Definition}

\begin{Definition}[Semantics of pattern matching]
\label{def:a-lang-matching}
    We write $ \Matching{v/p} $ for the result of matching
    a value $ v $ against a pattern $ p $.
    We have either $ \Matching{v/p} = \varsigma $,
    where $ \varsigma $ is a substitution defined
    on the variables in $ \Capt(p) $,
    or $ \Matching{v/p} = \MatchFail $.
    In the former case, we say that $ v $ matches $ p $
    (or that $ p $ accepts $ v $); in the latter, we say that matching fails.

    The definition of $ \Matching{v/p} $ is given inductively
    in Figure~\ref{fig:variants-semantics-matching-app}.
\end{Definition}

\begin{figure}

    \begin{align*}
        \Matching{v / \Wildcard}
            & = \EmptySubst
            \\
        \Matching{v / x}
            & = \SubstSingle[v / x]
            \\
        \Matching{v / c}
            & =
            \begin{cases}
                \EmptySubst & \text{if } v = c \\
                \MatchFail  & \text{otherwise}
            \end{cases}
            \\
        \Matching{v / (p_1, p_2)}
            & =
            \begin{cases}
                \varsigma_1 \cup \varsigma_2 &
                    \text{if } v = (v_1, v_2)
                    \text{ and } \forall i . \:
                    \Matching{v_i / p_i} = \varsigma_i \\
                \MatchFail & \text{otherwise}
            \end{cases}
            \\
        \Matching{v / \Tagg(p_1)}
            & =
            \begin{cases}
                \varsigma_1 &
                    \text{if } v = \Tagg(v_1)
                    \text{ and } \Matching{v_1 / p_1} = \varsigma_1 \\
                \MatchFail & \text{otherwise}
            \end{cases}
            \\
        \Matching{v / \PAnd{p_1 \& p_2}}
            & =
            \begin{cases}
                \varsigma_1 \cup \varsigma_2 &
                    \text{if } \forall i . \: \Matching{v / p_i} = \varsigma_i
                    \\
                \MatchFail & \text{otherwise}
            \end{cases}
            \\
        \Matching{v / \POr{p_1 | p_2}}
            & =
            \begin{cases}
                \Matching{v / p_1} &
                    \text{if } \Matching{v / p_1} \neq \MatchFail \\
                \Matching{v / p_2} & \text{otherwise}
            \end{cases}
    \end{align*}

    \caption{Semantics of pattern matching.}
    \label{fig:variants-semantics-matching-app}
\end{figure}

\begin{Definition}[Evaluation contexts]
\label{def:a-lang-contexts}
    Let the symbol $ \Hole $ denote a hole.
    An \emph{evaluation context} $ E $ is a term
    inductively generated by the following grammar.
    \[
        \Bnf{ E
              \= \Hole \| \Appl{E}{e} \| \Appl{v}{E}
              \| (E, e) \| (v, E) \| \Tagg(E)
              \| \Match{E with (p_i \to e_i)_{i \in I}} }
    \]

    We write $ \CtxE{e} $ for the expression obtained by replacing
    the hole in $ E $ with the expression $ e $.
\end{Definition}

\begin{Definition}[Reduction]
\label{def:a-lang-reduction}
    The reduction relation $ \SmallstepArrow $ between expressions
    is given by the rules in Figure~\ref{fig:variants-semantics-reduction}.
\end{Definition}

\begin{figure}

    \begin{mathpar}
        \MyInfer[R-Appl]
            { }
            { \Smallstep{
                \Appl{ (\Abstr{x. e}) }{ v } ~>
                e \SubstSingle[v/x] } }
            {}
        \and
        \MyInfer[R-Match]
            { \Matching{ v / p_j } = \varsigma \\
              \forall i < j . \: \Matching{ v / p_i } = \MatchFail }
            { \Smallstep{
                \Match{ v with (p_i \to e_i)_{i \in I} } ~>
                e_j \varsigma } }
            { j \in I }
        \and
        \MyInfer[R-Ctx]
            { \Smallstep{ e ~> e' } }
            { \Smallstep{ \CtxE{e} ~> \CtxE{e'} } }
            {}
    \end{mathpar}
    
    \caption{Small-step reduction relation.}
    \label{fig:variants-semantics-reduction}
\end{figure}

\subsection{Typing variants with kinding constraints}

\subsubsection{Definition of the \VariantsK{} type system}

We assume that there exists a countable set $ \TypeVars $
of \emph{type variables},
ranged over by $ \alpha $, $ \beta $, $ \gamma $, \dots.
We also consider a finite set $ \BasicTypes $ of \emph{basic types},
ranged over by $ b $,
and a function $ b_{(\Placeholder)} $ from constants to basic types.

\begin{Definition}[Types]
\label{def:a-k-types}
    A \emph{type} $ \tau $ is a term
    inductively generated by the following grammar.
    \[
        \Bnf{ \tau \= \alpha \| b \| \tau \to \tau \| \tau \times \tau }
    \]
\end{Definition}

\begin{Definition}[Kinds]
\label{def:a-k-kinds}
    A \emph{kind} $ \kappa $ is
    either the \emph{unconstrained kind} ``$ \KUnkind $''
    or a \emph{constrained kind}, that is, a triple $ (L, U, T) $ where:
    \begin{itemize}
        \item $ L $ is a finite set of tags $ \Set{\Tagg_1, \dots, \Tagg_n} $;
        \item $ U $ is either a finite set of tags
            or the set $ \Tags $ of all tags;
        \item $ T $ is a finite set of pairs of a tag and a type, written
            $ \{ \Tagg_1\colon \tau_1, $ $ \dots, \Tagg_n\colon \tau_n \} $
            (its domain $ \Dom(T) $ is the set of tags occurring in it);
    \end{itemize}
    and where the following conditions hold:
    \begin{itemize}
        \item $ L \subseteq U $, $ L \subseteq \Dom(T) $, and,
            if $ U \neq \Tags $, $ U \subseteq \Dom(T) $;
        \item tags in $ L $ have a single type in $ T $, that is,
            if $ \Tagg \in L $, whenever both $ \Tagg\colon \tau_1 \in T $
            and $ \Tagg\colon \tau_2 \in T $, we have $ \tau_1 = \tau_2 $.
    \end{itemize}
\end{Definition}

\begin{Definition}[Kind entailment]
\label{def:a-k-entailment}
    The entailment relation $ \KKindEntail{\cdot |= \cdot} $
    between constrained kinds is defined as
    \[
        \KKindEntail{(L, U, T) |= (L', U', T')}
            \iff L \supseteq L' \land U \subseteq U' \land T \supseteq T'
        \: .
    \]
\end{Definition}

\begin{Definition}[Kinding environments]
\label{def:a-k-kindingenv}
    A \emph{kinding environment} $ K $ is a partial mapping
    from type variables to kinds.
    We write kinding environments as $
        K = \Set{\KKinding{\alpha_1 :: \kappa_1}, \dots,
                \KKinding{\alpha_n :: \kappa_n}}
    $.
    We write $ K, K' $ for the updating of the kinding environment $ K $
    with the new bindings in $ K' $. It is defined as follows.
    \[
        (K, K')(\alpha) =
            \begin{cases}
                K'(\alpha) & \text{if } \alpha \in \Dom(K') \\
                K(\alpha) & \text{otherwise}
            \end{cases}
    \]

    We say that a kinding environment is \emph{closed}
    if all the type variables that appear in the types in its range
    also appear in its domain.
    We say it is \emph{canonical} if it is infinite
    and contains infinitely many variables of every kind.
\end{Definition}

\begin{Definition}[Type schemes]
\label{def:a-k-schemes}
    A \emph{type scheme} $ \sigma $ is of the form
    $ \KScheme{A. K > \tau} $, where:
    \begin{itemize}
        \item $ A $ is a finite set $ \Set{\alpha_1, \dots, \alpha_n} $
            of type variables;
        \item $ K $ is a kinding environment such that $ \Dom(K) = A $.
    \end{itemize}
\end{Definition}

We identify a type scheme $ \KScheme{\varnothing. \varnothing > \tau} $,
which quantifies no variable, with the type $ \tau $ itself.
We consider type schemes up to renaming of the variables they bind
and disregard useless quantification
(i.e., quantification of variables that do not occur in the type).

\begin{Definition}[Free variables]
\label{def:a-k-freevars}
    The set of \emph{free variables} $ \Var_K(\sigma) $
    of a type scheme $ \sigma $ with respect to a kinding environment $ K $
    is the minimum set satisfying the following equations.
    \begin{align*}
        \Var_K(\KScheme{A. K' > \tau})
            & = \Var_{K, K'}(\tau) \setminus A \\
        \Var_K(\alpha) & =
            \begin{cases}
                \Set{\alpha} \cup \bigcup_{\Tagg\colon \tau \in T} \Var_K(\tau)
                    & \text{if } K(\alpha) = (L, U, T) \\
                \Set{\alpha} & \text{otherwise}
            \end{cases} \\
        \Var_K(b) & = \varnothing \\
        \Var_K(\tau_1 \to \tau_2)
            & = \Var_K(\tau_1) \cup \Var_K(\tau_2) \\
        \Var_K(\tau_1 \times \tau_2)
            & = \Var_K(\tau_1) \cup \Var_K(\tau_2)
    \end{align*}

    We say that a type $ \tau $ is \emph{ground} or \emph{closed}
    if and only if $ \Var_\varnothing(\tau) $ is empty.
    We say that a type or a type scheme is
    \emph{closed in a kinding environment $ K $}
    if all its free variables are in the domain of $ K $.
\end{Definition}

\begin{Definition}[Type substitutions]
\label{def:a-k-typesubst}
    A \emph{type substitution} $ \theta $ is a finite mapping
    of type variables to types.
    We write $ \SubstC[\tau_i/\alpha_i | i \in I] $ for the type substitution
    which simultaneously replaces $ \alpha_i $ with $ \tau_i $,
    for each $ i \in I $.
    We write $ \tau \theta $ for the application of the substitution $ \theta $
    to the type $ \tau $, which is defined as follows.
    \begin{align*}
        \alpha \theta & =
            \begin{cases}
                \tau' & \text{if } \SubstF{\tau'/\alpha} \in \theta \\
                \alpha & \text{otherwise}
            \end{cases} \\
        b \theta & = b \\
        (\tau_1 \to \tau_2) \theta & = (\tau_1 \theta) \to (\tau_2 \theta) \\
        (\tau_1 \times \tau_2) \theta & = (\tau_1 \theta) \times (\tau_2 \theta)
    \end{align*}

    We extend the $ \Var $ operation to substitutions as
    \[
        \Var_K(\theta) = \bigcup_{\alpha \in \Dom(\theta)} \Var_K(\alpha \theta)
        \: .
    \]

    We extend application of substitutions to the typing component
    of a constrained kind $ (L, U, T) $: $ T \theta $ is given by the pointwise
    application of $ \theta $ to all types in $ T $.
    We extend it to kinding environments: $ K \theta $ is given by the pointwise
    application of $ \theta $ to the typing component of every constrained kind
    in the range of $ K $.
    We extend it to type schemes $ \KScheme{A. K > \tau} $:
    by renaming quantified variables, we assume $
        A \cap (\Dom(\theta) \cup \Var_\varnothing(\theta)) = \varnothing
    $, and we have $
        (\KScheme{A. K > \tau}) \theta = \KScheme{A. K \theta > \tau \theta}
    $.

    We write $ \theta_1 \cup \theta_2 $ for the union of disjoint substitutions
    and $ \theta_1 \circ \theta_2 $ for the composition of substitutions.
\end{Definition}

\begin{Definition}[Admissibility of a type substitution]
\label{def:a-k-admissibility}
    A type substitution $ \theta $ is
    \emph{admissible} between two kinding environments $ K $ and $ K' $,
    written $ \KAdmissible{K |- \theta: K'} $,
    if and only if,
    for every type variable $ \alpha $ such that $ K(\alpha) = (L, U, T) $,
    $ \alpha \theta $ is a type variable
    such that $ K'(\alpha \theta) = (L', U', T') $
    and $ \KKindEntail{(L', U', T') |= (L, U, T \theta)} $.
\end{Definition}

\begin{Definition}[Type environments]
\label{def:a-k-typeenv}
    A \emph{type environment} $ \Gamma $ is a partial mapping
    from expression variables to type schemes.
    We write type environments as $
        \Gamma = \Set{x_1\colon \sigma_1, \dots, x_n\colon \sigma_n} $.

    We write $ \Gamma, \Gamma' $ for the updating
    of the type environment $ \Gamma $ with the new bindings in $ \Gamma' $.
    It is defined as follows.
    \[
        (\Gamma, \Gamma')(x) =
            \begin{cases}
                \Gamma'(x) & \text{if $ x \in \Dom(\Gamma') $} \\
                \Gamma(x) & \text{otherwise}
            \end{cases}
    \]

    We extend the $ \Var $ operation to type environments as
    \[
        \Var_K(\Gamma) = \bigcup_{\sigma \in \Range(\Gamma)} \Var_K(\sigma)
        \: .
    \]
\end{Definition}

\begin{Definition}[Generalization]
\label{def:a-k-gen}
    We define the \emph{generalization} of a type $ \tau $
    with respect to a kinding environment $ K $
    and a type environment $ \Gamma $
    as the type scheme
    \[
        \Gen_{K; \Gamma}(\tau) = \KScheme{A. K' > \tau}
    \]
    where $ A = \Var_K(\tau) \setminus \Var_K(\Gamma) $
    and $ K' = \SetC{\KKinding{\alpha :: K(\alpha)} | \alpha \in A} $.

    We extend this definition to type environments which only contain types
    (i.e., trivial type schemes) as
    \[
        \Gen_{K; \Gamma}(\SetC{x_i\colon \tau_i | i \in I}) =
            \SetC{x_i\colon \Gen_{K; \Gamma}(\tau_i) | i \in I}
        \: .
    \]
\end{Definition}

\begin{Definition}[Instances of a type scheme]
\label{def:a-k-inst}
    The set of \emph{instances} of a type scheme $ \KScheme{A. K' > \tau} $
    in a kinding environment $ K $ is defined as
    \[
        \Inst_K(\KScheme{A. K' > \tau}) =
            \SetC{\tau \theta |
                \Dom(\theta) \subseteq A \: \land \:
                \KAdmissible{K, K' |- \theta: K}}
        \: .
    \]

    We say that a type scheme $ \sigma_1 $ is \emph{more general}
    than a type scheme $ \sigma_2 $ in $ K $,
    and we write $ \KMoreGen{K |- \sigma_1 < \sigma_2} $,
    if $ \Inst_K(\sigma_1) \supseteq \Inst_K(\sigma_2) $.

    We extend this notion to type environments as
    \[
        \KMoreGen{K |- \Gamma_1 < \Gamma_2} \iff
            \Dom(\Gamma_1) = \Dom(\Gamma_2) \land
            \forall x \in \Dom(\Gamma_1) . \:
                \KMoreGen{K |- \Gamma_1(x) < \Gamma_2(x)}
        \: .
    \]
\end{Definition}

\begin{Definition}[Pattern environment generation]
\label{def:a-k-patternenv}
    The environment generated by pattern matching is given by the relation
    $ \KPatternTyping{K |- p: \tau => \Gamma} $
    (\emph{the pattern $ p $ can match type $ \tau $ in $ K $,
    producing the bindings in $ \Gamma $}),
    defined by the rules in Figure~\ref{fig:variants-k-pattern-environment}.
\end{Definition}

\begin{figure*}

    \begin{mathpar}
        \MyInfer[TPk-Wildcard]
            { }
            { \KPatternTyping{ K |- \Wildcard: \tau => \varnothing } }
            {}
        \and
        \MyInfer[TPk-Var]
            { }
            { \KPatternTyping{ K |- x: \tau => \Set{x\colon \tau} } }
            {}
        \and
        \MyInfer[TPk-Const]
            { }
            { \KPatternTyping{ K |- c: b_c => \varnothing } }
            {}
        \and
        \MyInfer[TPk-Pair]
            { \KPatternTyping{ K |- p_1: \tau_1 => \Gamma_1 } \\
              \KPatternTyping{ K |- p_2: \tau_2 => \Gamma_2 } }
            { \KPatternTyping{ K |- (p_1, p_2): \tau_1 \times \tau_2 =>
                \Gamma_1 \cup \Gamma_2 } }
            {}
        \and
        \MyInfer[TPk-Tag]
            { \KPatternTyping{ K |- p: \tau => \Gamma } \\
              K(\alpha) = (L, U, T) \\
              (\Tagg \in U \text{ implies } \Tagg \colon \tau \in T) }
            { \KPatternTyping{ K |- \Tagg(p): \alpha => \Gamma } }
            {}
        \and
        \MyInfer[TPk-And]
            { \KPatternTyping{ K |- p_1: \tau => \Gamma_1 } \\
              \KPatternTyping{ K |- p_2: \tau => \Gamma_2 } }
            { \KPatternTyping{ K |- \PAnd{p_1 \& p_2}: \tau =>
                \Gamma_1 \cup \Gamma_2 } }
            {}
        \and
        \MyInfer[TPk-Or]
            { \KPatternTyping{ K |- p_1: \tau => \Gamma } \\
              \KPatternTyping{ K |- p_2: \tau => \Gamma } }
            { \KPatternTyping{ K |- \POr{p_1 | p_2}: \tau => \Gamma } }
            {}
    \end{mathpar}

    \caption{Pattern environment generation for \VariantsK.}
    \label{fig:variants-k-pattern-environment}
\end{figure*}

\begin{Definition}[Exhaustiveness]
\label{def:a-k-exhaustive}
    We say that a set of patterns $ P $ is \emph{exhaustive}
    with respect to a type $ \tau $ in a kinding environment $ K $,
    and we write $ \KExhaustive{K |- \tau < P} $, when
    \[
        \forall K', \theta, v . \:
            (\KAdmissible{ K |- \theta: K' } \, \land \,
            \KTyping{ K'; \varnothing |- v: \tau \theta }) \implies
            \exists p \in P, \varsigma . \: \Matching{ v / p } = \varsigma
        \: .
    \]
\end{Definition}

\begin{Definition}[Typing relation]
\label{def:a-k-typing}
    The typing relation $ \KTyping{K; \Gamma |- e: \tau} $
    (\emph{$ e $ is given type $ \tau $ in the kinding environment $ K $
    and the type environment $ \Gamma $})
    is defined by the rules in Figure~\ref{fig:variants-k-typing},
    where we require $ K $ to be closed
    and $ \Gamma $ and $ \tau $ to be closed with respect to $ K $.
    We also assume that $ K $ is canonical.
\end{Definition}

\begin{figure*}

    \begin{mathpar}
        \MyInfer[Tk-Var]
            { \tau \in \Inst_K(\Gamma(x)) }
            { \KTyping{ K; \Gamma |- x: \tau } }
            {}
        \and
        \MyInfer[Tk-Const]
            { }
            { \KTyping{ K; \Gamma |- c: b_c } }
            {}
        \and
        \MyInfer[Tk-Abstr]
            { \KTyping{ K; \Gamma, \Set{x\colon \tau_1} |- e: \tau_2 } }
            { \KTyping{ K; \Gamma |- \Abstr{x. e}: \tau_1 \to \tau_2 } }
            {}
        \and
        \MyInfer[Tk-Appl]
            { \KTyping{ K; \Gamma |- e_1: \tau' \to \tau } \\
              \KTyping{ K; \Gamma |- e_2: \tau' } }
            { \KTyping{ K; \Gamma |- \Appl{e_1}{e_2}: \tau } }
            {}
        \and
        \MyInfer[Tk-Pair]
            { \KTyping{ K; \Gamma |- e_1: \tau_1 } \\
              \KTyping{ K; \Gamma |- e_2: \tau_2 } }
            { \KTyping{ K; \Gamma |- (e_1, e_2): \tau_1 \times \tau_2 } }
            {}
        \and
        \MyInfer[Tk-Tag]
            { \KTyping{ K; \Gamma |- e: \tau } \\
              \KKindEntail{ K(\alpha) |=
                (\Set{\Tagg}, \Tags, \Set{\Tagg\colon \tau}) } }
            { \KTyping{ K; \Gamma |- \Tagg(e): \alpha } }
            {}
        \and
        \MyInfer[Tk-Match]
            { \KTyping{ K; \Gamma |- e_0: \tau_0 } \\
              \KExhaustive{ K |- \tau_0 < \SetC{p_i | i \in I} } \\\\
              \forall i \in I \\
              \KPatternTyping{ K |- p_i: \tau_0 => \Gamma_i } \\
              \KTyping{ K; \Gamma, \Gen_{K;\Gamma}(\Gamma_i) |- e_i: \tau } }
            { \KTyping{ K; \Gamma |-
                \Match{e_0 with (p_i \to e_i)_{i \in I}}: \tau } }
            {}
    \end{mathpar}

    \caption{Typing relation for \VariantsK.}
    \label{fig:variants-k-typing}
\end{figure*}

\subsubsection{Properties of the \VariantsK{} type system}

\begin{Lemma}[Generation for values]
\label{lem:a-k-generation}
    Let $ v $ be a value. Then:
    \begin{itemize}
        \item if $ \KTyping{ K; \Gamma |- v: b } $,
            then $ v = c $ for some constant $ c $ such that $ b_c = b $;
        \item if $ \KTyping{ K; \Gamma |- v: \tau_1 \to \tau_2 } $,
            then $ v $ is of the form $ \Abstr{x. e} $
            and $ \KTyping{ K; \Gamma, \Set{x\colon \tau_1} |- e: \tau_2 } $;
        \item if $ \KTyping{ K; \Gamma |- v: \tau_1 \times \tau_2 } $,
            then $ v $ is of the form $ (v_1, v_2) $,
            $\KTyping{ K; \Gamma |- v_1: \tau_1 } $,
            and $ \KTyping{ K; \Gamma |- v_2: \tau_2 } $;
        \item if $ \KTyping{ K; \Gamma |- v: \alpha } $,
            then $ v $ is of the form $ \Tagg(v_1) $,
            $ K(\alpha) = (L, U, T) $, $ \Tagg \in L $,
            and $ \KTyping{ K; \Gamma |- v_1: \tau_1 } $
            for the only type $ \tau_1 $ such that $ \Tagg\colon \tau_1 \in T $.
    \end{itemize}
\end{Lemma}

\begin{Proof}
    The typing rules are syntax-directed,
    so the last rule applied to type a value is fixed by its form.
    All these rules derive types of different forms, thus the form of the type
    assigned to a value determines the last rule used.
    In each case the premises of the rule entail the consequences above.
\end{Proof}

\begin{Lemma}[Correctness of environment generation]
\label{lem:a-k-patterns-correct}
    Let $ p $ be a pattern and $ v $ a value
    such that $ \Matching{v / p} = \varsigma $.
    If $ \KTyping{ K; \Gamma |- v: \tau } $
    and $ \KPatternTyping{ K |- p: \tau => \Gamma' } $,
    then, for all $ x \in \Capt(p) $,
    $ \KTyping{ K; \Gamma |- x \varsigma: \Gamma'(x) } $.
\end{Lemma}

\begin{Proof}
    By induction on the derivation of $
        \KPatternTyping{K |- p: \tau => \Gamma'}
    $. We reason by cases on the last applied rule.

    \begin{LCases}
    \Cases[\Rule{TPk-Wildcard} and \Rule{TPk-Const}]
        There is nothing to prove since $ \Capt(p) = \varnothing $.

    \Case[\Rule{TPk-Var}]
        We have
        \[
            \Matching{v / x} = \SubstSingle[v/x] \qquad
            \KPatternTyping{K |- x: \tau => \Set{x\colon \tau}}
        \]
        and must prove $
            \KTyping{K; \Gamma |- x \SubstSingle[v/x]: \Set{x\colon \tau}(x)}
        $, which we know by hypothesis.

    \Case[\Rule{TPk-Pair}]
        We have
        \[
            \KPatternTyping{K |- (p_1, p_2): \tau_1 \times \tau_2 =>
                \Gamma'_1 \cup \Gamma'_2}
            \qquad
            \KPatternTyping{K |- p_1: \tau_1 => \Gamma'_1} \qquad
            \KPatternTyping{K |- p_2: \tau_2 => \Gamma'_2}
            \: .
        \]

        By Lemma~\ref{lem:a-k-generation}, $
            \KTyping{K; \Gamma |- v: \tau_1 \times \tau_2}
        $ implies $ v = (v_1, v_2) $ and $
            \KTyping{K; \Gamma |- v_i: \tau_i}
        $ for both $ i $. Furthermore, $
            \Matching{(v_1, v_2) / (p_1, p_2)} = \varsigma =
            \varsigma_1 \cup \varsigma_2
        $, and $ \Matching{v_i / p_i} = \varsigma_i $ for both $ i $.
        For each capture variable $ x $,
        we can apply the induction hypothesis to the sub-pattern
        which contains $ x $ and conclude.

    \Case[\Rule{TPk-Tag}]
        We have
        \begin{gather*}
            \KPatternTyping{K |- \Tagg(p_1): \alpha => \Gamma'} \qquad
            \KPatternTyping{K |- p_1: \tau_1 => \Gamma'} \\
            K(\alpha) = (L, U, T) \qquad
            (\Tagg \in U \text{ implies } \Tagg\colon \tau_1 \in T)
            \: .
        \end{gather*}

        Since $ \Matching{v / \Tagg(p_1)} = \varsigma $,
        we know $ v = \Tagg(v_1) $.
        Hence, by Lemma~\ref{lem:a-k-generation}, we have $ \Tagg \in L $
        and $ \KTyping{K; \Gamma |- v_1: \tau_1'} $
        with $ \Tagg\colon \tau_1' \in T $.
        Since $ \Tagg \in U $, we also have $ \Tagg\colon \tau_1 \in T $
        and hence $ \tau_1 = \tau_1' $
        (as $ \Tagg $ is also in $ L $
        and can only have a single type in $ T $).

        We therefore know $ \KPatternTyping{K |- p_1: \tau_1 => \Gamma'} $
        and $ \KTyping{K; \Gamma |- v_1: \tau_1} $,
        as well as $ \Matching{v_1 / p_1} = \varsigma $.
        We can apply the induction hypothesis to conclude.

    \Cases[\Rule{TPk-And} and \Rule{TPk-Or}]
        Straightforward application of the induction hypothesis,
        to both sub-patterns for intersections
        and to the one that is actually selected for unions.
        \qedhere
    \end{LCases}
\end{Proof}

\begin{Lemma}[Stability of environment generation under type substitutions]
\label{lem:a-k-patterns-typesubst}
    If $ \KPatternTyping{ K |- p: \tau => \Gamma } $,
    then $ \KPatternTyping{ K' |- p: \tau \theta => \Gamma \theta } $
    for every type substitution $ \theta $
    such that $ \KAdmissible{K |- \theta: K'} $.
\end{Lemma}

\begin{Proof}
    By induction on the derivation of $
        \KPatternTyping{K |- p: \tau => \Gamma}
    $. We reason by cases on the last applied rule.

    \begin{LCases}
    \Cases[\Rule{TPk-Wildcard}, \Rule{TPk-Var}, and \Rule{TPk-Const}]
        Straightforward.

    \Case[\Rule{TPk-Pair}]
        We have
        \[
            \KPatternTyping{K |- (p_1, p_2): \tau_1 \times \tau_2 =>
                \Gamma_1 \cup \Gamma_2}
            \qquad
            \KPatternTyping{K |- p_1: \tau_1 => \Gamma_1} \qquad
            \KPatternTyping{K |- p_2: \tau_2 => \Gamma_2}
            \: .
        \]

        By the induction hypothesis we derive both $
            \KPatternTyping{K' |- p_1: \tau_1 \theta => \Gamma_1 \theta}
        $ and $
            \KPatternTyping{K' |- p_2: \tau_2 \theta => \Gamma_2 \theta}
        $, then we apply \Rule{TPk-Pair} again to conclude.

    \Case[\Rule{TPk-Tag}]
        We have
        \begin{gather*}
            \KPatternTyping{K |- \Tagg(p_1): \alpha => \Gamma} \qquad
            \KPatternTyping{K |- p_1: \tau_1 => \Gamma} \\
            K(\alpha) = (L, U, T) \qquad
            (\Tagg \in U \text{ implies } \Tagg\colon \tau_1 \in T)
            \: .
        \end{gather*}

        By the induction hypothesis we derive $
            \KPatternTyping{K' |- p_1: \tau_1 \theta => \Gamma \theta}
        $. Since $ \KAdmissible{K |- \theta: K'} $,
        $ \alpha \theta $ must be a variable $ \beta $ such that $
            K'(\beta) = (L', U', T')
        $. To apply \Rule{TPk-Tag} and conclude, we must establish that,
        if $ \Tagg \in U' $, then $ \Tagg\colon \tau_1 \theta \in T' $.
        Since admissibility also implies $
            \KKindEntail{(L', U', T') |= (L, U, T \theta)}
        $, we have $ U' \subseteq U $ and $ T \theta \subseteq T' $.
        Hence, if $ \Tagg \in U' $, then $ \Tagg \in U $,
        in which case $ \Tagg\colon \tau_1 \in T $
        and therefore $ \Tagg\colon \tau_1 \theta \in T \theta $,
        and $ \Tagg\colon \tau_1 \theta \in T' $.

    \Cases[\Rule{TPk-And} and \Rule{TPk-Or}]
        Straightforward application of the induction hypothesis,
        analogously to the case of pair patterns.
        \qedhere
    \end{LCases}
\end{Proof}

\begin{Lemma}[Stability of exhaustiveness under type substitutions]
\label{lem:a-k-exhaust-typesubst}
    If $ \KExhaustive{K |- \tau < P} $,
    then $ \KExhaustive{K' |- \tau \theta < P} $
    for any type substitution $ \theta $
    such that $ \KAdmissible{K |- \theta: K'} $.
\end{Lemma}

\begin{Proof}
    We must prove, for every $ K'' $, $ \theta' $ such that $
        \KAdmissible{K' |- \theta': K''}
    $ and every $ v $ such that $
        \KTyping{K''; \EmptyEnv |- v: \tau \theta \theta'}
    $, that there exists a $ p \in P $ which accepts $ v $.
    This holds because $ \theta' \circ \theta $ is such that $
        \KAdmissible{K |- \theta' \circ \theta: K''}
    $: for any $ \alpha $ such that $ K(\alpha) = (L, U, T) $, we have $
        K'(\alpha \theta) = (L', U', T')
    $ and hence $
        K''(\alpha \theta \theta') = (L'', U'', T'')
    $; we have $
        \KKindEntail{(L', U', T') |= (L, U, T \theta)}
    $ and $
        \KKindEntail{(L'', U'', T'') |= (L', U', T' \theta')}
    $ and therefore $
        \KKindEntail{(L'', U'', T'') |= (L, U, T \theta \theta')}
    $.
    The conclusion follows by the definition of $ \KExhaustive{K |- \tau < P} $.
\end{Proof}

\begin{Lemma}
\label{lem:a-k-generalization}
    If $ \Var_K(\Gamma_1) \subseteq \Var_K(\Gamma_2) $,
    then, for every type $ \tau $, $
        \KMoreGen{ K |-
            \Gen_{K;\Gamma_1}(\tau) < \Gen_{K;\Gamma_2}(\tau) }
    $.
\end{Lemma}

\begin{Proof}
    An instance of $ \Gen_{K;\Gamma_2}(\tau) $ is a type $ \tau \theta $
    such that $ \Dom(\theta) \subseteq \Var_K(\tau) \setminus \Var_K(\Gamma_2) $
    and $ \KAdmissible{K |- \theta: K} $.
    It is also an instance of $ \Gen_{K;\Gamma_1}(\tau) $,
    with the same $ \theta $, since $
        \Var_K(\tau) \setminus \Var_K(\Gamma_2) \subseteq
        \Var_K(\tau) \setminus \Var_K(\Gamma_1)
    $.
\end{Proof}

\begin{Lemma}[Weakening]
\label{lem:a-k-weakening}
    Let $ K $ be a kinding environment
    and $ \Gamma_1 $, $ \Gamma_2 $ two type environments
    such that $ \KMoreGen{K |- \Gamma_1 < \Gamma_2} $
    and $ \Var_K(\Gamma_1) \subseteq \Var_K(\Gamma_2) $.
    If $ \KTyping{ K; \Gamma_2 |- e: \tau } $,
    then $ \KTyping{ K; \Gamma_1 |- e: \tau } $.
\end{Lemma}

\begin{Proof}
    By induction on the derivation of $ \KTyping{K; \Gamma_2 |- e: \tau} $.
    We reason by cases on the last applied rule.

    \begin{LCases}
    \Case[\Rule{Tk-Var}]
        We have:
        \[
            \KTyping{K; \Gamma_2 |- x: \tau} \qquad
            \tau \in \Inst_K(\Gamma_2(x))
        \]
        and hence, since $ \KMoreGen{ K |- \Gamma_1 < \Gamma_2 } $, we have
        $ \tau \in \Inst_K(\Gamma_1(x)) $ and apply \Rule{Tk-Var} to conclude.

    \Case[\Rule{Tk-Const}]
        Straightforward.

    \Case[\Rule{Tk-Abstr}]
        We have:
        \[
            \KTyping{K; \Gamma_2 |- \Abstr{x. e_1}: \tau_1 \to \tau_2}
            \qquad
            \KTyping{K; \Gamma_2, \Set{x\colon \tau_1} |- e_1: \tau_2}
            \: .
        \]

        Since $ \KMoreGen{ K |- \Gamma_1 < \Gamma_2 } $, we have $
            \KMoreGen{ K |- \Gamma_1, \Set{x\colon \tau_1} <
                \Gamma_2, \Set{x\colon \tau_1} }
        $, and, since $ \Var_K(\Gamma_1) \subseteq \Var_K(\Gamma_2) $,
        we have $
            \Var_K(\Gamma_1, \Set{x\colon \tau_1}) \subseteq
            \Var_K(\Gamma_2, \Set{x\colon \tau_1})
        $. Thus we may derive $
            \KTyping{K; \Gamma_1, \Set{x\colon \tau_1} |- e_1: \tau_2}
        $ by the induction hypothesis and apply \Rule{Tk-Abstr} to conclude.

    \Cases[\Rule{Tk-Appl}, \Rule{Tk-Pair}, and \Rule{Tk-Tag}]
        Straightforward application of the induction hypothesis.

    \Case[\Rule{Tk-Match}]
        We have
        \begin{gather*}
            \KTyping{K; \Gamma_2 |- \Match{e_0 with (p_i \to e_i)_{i \in I}}:
                \tau} \\
            \KTyping{K; \Gamma_2 |- e_0: \tau_0} \qquad
            \KExhaustive{K |- \tau_0 < \SetC{p_i | i \in I}} \\
            \forall i \in I . \enspace
            \KPatternTyping{K |- p_i: \tau_0 => \Gamma_i} \qquad
            \KTyping{K; \Gamma_2, \Gen_{K;\Gamma_2}(\Gamma_i) |- e_i: \tau}
            \enspace .
        \end{gather*}

        By the induction hypothesis, we derive $
            \KTyping{K; \Gamma_1 |- e_0: \tau_0}
        $.

        For every branch, note that by Lemma~\ref{lem:a-k-generalization} $
            \Var_K(\Gamma_1) \subseteq \Var_K(\Gamma_2)
        $ implies $
            \KMoreGen{ K |- \Gen_{K;\Gamma_1}(\tau) < \Gen_{K;\Gamma_2}(\tau)}
        $ for any $ \tau $. Hence, we have $
            \KMoreGen{K |- \Gamma_1, \Gen_{K;\Gamma_1}(\Gamma_i) <
                \Gamma_2, \Gen_{K;\Gamma_2}(\Gamma_i)}
        $. Additionally, since $
            \Var_K(\Gen_{K;\Gamma_1}(\Gamma_i)) \subseteq \Var_K(\Gamma_1)
        $, we have $
            \Var_K(\Gamma_1, \Gen_{K;\Gamma_1}(\Gamma_i)) \subseteq
            \Var_K(\Gamma_2, \Gen_{K;\Gamma_2}(\Gamma_i))
        $.

        Hence we may apply the induction hypothesis for all $ i $ to derive $
            \KTyping{K; \Gamma_1, \Gen_{K;\Gamma_1}(\Gamma_i) |- e_i: \tau}
        $ and then apply \Rule{Tk-Match} to conclude.
        \qedhere
    \end{LCases}
\end{Proof}

\begin{Lemma}[Stability of typing under type substitutions]
\label{lem:a-k-typesubst}
    Let $ K $, $ K' $ be two closed, canonical kinding environments
    and $ \theta $ a type substitution
    such that $ \KAdmissible{K |- \theta: K'} $.
    If $ \KTyping{ K; \Gamma |- e: \tau } $,
    then $ \KTyping{ K'; \Gamma \theta |- e: \tau \theta } $.
\end{Lemma}

\begin{Proof}
    By induction on the derivation of $ \KTyping{K; \Gamma |- e: \tau} $.
    We reason by cases on the last applied rule.

    \begin{LCases}
    \Case[\Rule{Tk-Var}]
        We have
        \begin{gather*}
            \KTyping{K; \Gamma |- x: \tau} \qquad
            \tau \in \Inst_K(\Gamma(x)) \\
            \Gamma(x) = \KScheme{A. K_x > \tau_x} \qquad
            \tau = \tau_x \theta_x \qquad
            \Dom(\theta_x) \subseteq A \qquad
            \KAdmissible{K, K_x |- \theta_x: K}
        \end{gather*}
        and must show
        \[
            \KTyping{K'; \Gamma \theta |- x: \tau \theta}
            \: .
        \]

        By \textalpha-renaming we can assume that $ \theta $
        does not involve $ A $, that is, $ A \cap \Dom(\theta) = \varnothing $
        and $ A \cap \Var_\varnothing(\theta) = \varnothing $,
        and also that $
            A \cap (\Dom(K') \cup \Var_\varnothing(K')) = \varnothing
        $, that is, that the variables in $ A $ are not assigned a kind
        in $ K' $ nor do they appear in the types in the typing component
        of the kinds in $ K' $.

        Under these assumptions, $
            (\Gamma \theta)(x) = \KScheme{A. K_x \theta > \tau_x \theta}
        $. We must show that $ \tau \theta = \tau_x \theta \theta_x' $
        for a substitution $ \theta_x' $ such that $
            \Dom(\theta_x') \subseteq A
        $ and $ \KAdmissible{K', K_x \theta |- \theta_x': K'} $.

        Let $
            \theta_x' = \SubstC[\alpha \theta_x \theta / \alpha | \alpha \in A]
        $. First, we show that $
            \tau_x \theta \theta_x' = \tau_x \theta_x \theta = \tau \theta
        $, by showing that, for any $ \alpha $, $
            \alpha \theta \theta_x' = \alpha \theta_x \theta
        $. If $ \alpha \in A $, then $
            \alpha \theta \theta_x' = \alpha \theta_x' = \alpha \theta_x \theta
        $ ($ \theta $ is not defined on the variables in $ A $).
        If $ \alpha \notin A $, then $ \alpha \theta \theta_x' = \alpha \theta $
        ($ \theta $ never produces any variable in $ A $)
        and $ \alpha \theta_x \theta = \alpha \theta $
        as $ \alpha \notin \Dom(\theta_x) $.

        Since $ \Dom(\theta_x') \subseteq A $ holds, we only need to establish
        that $ \KAdmissible{K', K_x \theta |- \theta_x': K'} $.
        This requires proving, for each $ \alpha $ such that $
            (K', K_x \theta)(\alpha) = (L, U, T)
        $, that $ \alpha \theta_x' $ is a type variable such that $
            K'(\alpha \theta_x') = (L', U', T')
        $ and $ \KKindEntail{(L', U', T') |= (L, U, T \theta_x')} $.

        Such an $ \alpha $ can either be in the domain of $ K_x \theta $
        (if and only if it is in $ A $) or in the domain of $ K' $.
        In the latter case, we have $ \alpha \theta_x' = \alpha $,
        since $ \alpha \notin A $, and hence its kind in $ K' $
        is the same as in $ K', K_x \theta $.
        We must prove $ \KKindEntail{(L, U, T) |= (L, U, T \theta_x')} $,
        which holds because the variables in $ A $ do not appear in $ T $
        since $ (L, U, T) \in \Range(K') $.

        In the former case, we have $
            (K_x \theta)(\alpha) = (L, U, T)
        $ and hence $ K_x(\alpha) = (L, U, T_1) $,
        with $ T = T_1 \theta $.
        Also, $ \alpha \theta_x' = \alpha \theta_x \theta $.
        Since $ \KAdmissible{K, K_x |- \theta_x: K} $,
        $ K(\alpha \theta_x) = (L_2, U_2, T_2) $.
        Then, since $ \KAdmissible{K |- \theta: K'} $,
        $ K'(\alpha \theta_x \theta) = (L', U', T') $.
        We know $ \KKindEntail{(L_2, U_2, T_2) |= (L, U, T_1 \theta_x)} $
        and $ \KKindEntail{(L', U', T') |= (L_2, U_2, T_2 \theta)} $.
        Both $ L' \supseteq L $ and $ U' \subseteq U $ hold by transitivity.
        We show $ T' \supseteq T \theta_x' $ holds as well.
        If $ \Tagg\colon \tau \in T \theta_x' $, since $ T = T_1 \theta $,
        then $ \Tagg\colon \tau_1 \in T_1 $
        and $ \tau = \tau_1 \theta \theta_x' = \tau_1 \theta_x \theta $.
        We thus have $ \Tagg\colon \tau_1 \theta_x \in T_1 \theta_x $
        and therefore $ \Tagg\colon \tau_1 \theta_x \in T_2 $
        and $ \Tagg\colon \tau_1 \theta_x \theta \in T' $.

    \Case[\Rule{Tk-Const}]
        Straightforward.

    \Case[\Rule{Tk-Abstr}]
        We have:
        \[
            \KTyping{K; \Gamma |- \Abstr{x. e_1}: \tau_1 \to \tau_2}
            \qquad
            \KTyping{K; \Gamma, \Set{x\colon \tau_1} |- e_1: \tau_2}
            \: .
        \]

        By the induction hypothesis we have $
            \KTyping{K'; \Gamma \theta, \Set{x\colon \tau_1 \theta} |-
                e_1: \tau_2 \theta}
        $. Then by \Rule{Tk-Abstr} we derive $
            \KTyping{K'; \Gamma \theta |- \Abstr{x. e_1}:
                (\tau_1 \to \tau_2) \theta}
        $, since $
            (\tau_1 \to \tau_2) \theta = (\tau_1 \theta) \to (\tau_2 \theta)
        $.

    \Cases[\Rule{Tk-Appl} and \Rule{Tk-Pair}]
        Straightforward application of the induction hypothesis.

    \Case[\Rule{Tk-Match}]
        For the sake of clarity, we first prove the simpler case corresponding
        to (the encoding of) \Keyw{let}, where---simplifying environment
        generation---we have
        \begin{gather*}
            \KTyping{K; \Gamma |- \Match{e_0 with x \to e_1}: \tau} \qquad
            \KTyping{K; \Gamma |- e_0: \tau_0} \qquad
            \KTyping{K; \Gamma, \Gen_{K; \Gamma}(\Set{x\colon \tau_0}) |-
                e_1: \tau}
        \end{gather*}
        and must show
        \[
            \KTyping{K'; \Gamma \theta |- \Match{e_0 with x \to e_1}:
                \tau \theta}
        \]
        which we prove by establishing, for some type $ \hat{\tau_0} $, that
        \[
            \KTyping{K'; \Gamma \theta |- e_0: \hat{\tau_0}} \qquad
            \KTyping{K'; \Gamma \theta, \Gen_{K'; \Gamma \theta}(
                \Set{x\colon \hat{\tau_0}}) |- e_1: \tau \theta}
            \: .
        \]

        Let $
            A = \Set{\alpha_1, \dots, \alpha_n} =
            \Var_K(\tau_0) \setminus \Var_K(\Gamma)
        $. We assume that the variables in $ A $ do not appear in the kinds
        of variables not in $ A $, that is, that if $
            K(\alpha) = (L, U, T)
        $ and $ \alpha \notin A $, then $
            \Var_K(T) \cap A = \varnothing $.

        This assumption is justified by the following observations.
        The variables in $ A $ only appear quantified in the environment used
        for the typing derivation for $ e_1 $. Therefore we may assume that
        they do not appear in $ \tau $: if they do, it is because they have
        been chosen when instantiating some type scheme and, since $ K $ is
        canonical, we might have chosen some other variable of the same kind.
        As for the occurrences of the variables in $ A $ in the derivation for
        $ e_0 $, a similar reasoning applies. These variables do not appear
        free in the environment (neither directly in a type in $ \Gamma $,
        nor in the kinds of variables which appear free in $ \Gamma $).
        Therefore, if they occur in $ \tau_0 $ it is because they have been
        chosen either during instantiation of a type scheme or when typing
        an abstraction, and in both cases we might have chosen a different
        variable.

        Now we rename these variables so that $ \theta $ will not have effect
        on them. Let $ B = \Set{\beta_1, \dots, \beta_n} $ be a set of type
        variables such that $
            B \cap (\Dom(\theta) \cup \Var_\varnothing(\theta)) = \varnothing
        $ and $ B \cap \Var_\varnothing(\Gamma) = \varnothing $.
        Let $ \theta_0 = \SubstDots[\beta_1/\alpha_1 ... \beta_n/\alpha_n] $
        and $ \theta' = \theta \circ \theta_0 $. Since $ K' $ is canonical,
        we can choose each $ \beta_i $ so that, if $
            K(\alpha_i) = \KUnkind
        $, then $ K'(\beta_i) = \KUnkind $,
        and if $ K(\alpha_i) = (L, U, T) $,
        then $ K(\beta_i) = (L, U, T \theta') $.
        As for $ A $, we choose $ B $ so that the kinds in $ K' $
        for variables not in $ B $ do not contain variables of $ B $.

        We show $ \KAdmissible{K |- \theta': K'} $.
        For each $ \alpha $ such that $ K(\alpha) = (L, U, T) $,
        if $ \alpha \in A $ then $ \alpha = \alpha_i $ for some $ i $,
        $ \alpha \theta' = \beta_i $ and kind entailment holds
        straightforwardly by our choice of $ \beta_i $.
        If $ \alpha \notin A $, then $ \alpha \theta' = \alpha \theta $
        and the admissibility of $ \theta $ implies $
            K'(\alpha \theta) = (L', U', T')
        $ and $ \KKindEntail{(L', U', T') |= (L, U, T \theta)} $.
        We have $ T \theta = T \theta' $ because of our assumption on $ A $.

        Since $ \theta' $ is admissibile,
        by the induction hypothesis applied to $ \theta' $, we derive $
            \KTyping{K; \Gamma \theta' |- e_0: \tau_0 \theta'}
        $. Since the variables in $ A $ do not appear in $ \Gamma $,
        we have $ \Gamma \theta' = \Gamma \theta $.
        We choose $ \hat{\tau_0} $ to be $ \tau_0 \theta' $.

        We apply the induction hypothesis to the derivation for $ e_1 $,
        this time using $ \theta $ as the substitution. Now we have:
        \[
            \KTyping{K'; \Gamma \theta |- e_0: \tau_0 \theta'} \qquad
            \KTyping{K'; \Gamma \theta,
                (\Gen_{K; \Gamma}(\Set{x\colon \tau_0})) \theta |-
                e_1: \tau \theta}
            \: .
        \]

        We apply weakening (Lemma~\ref{lem:a-k-weakening})
        to derive from the latter the typing we need, that is,
        \[
            \KTyping{K'; \Gamma \theta,
                \Gen_{K'; \Gamma \theta}(\Set{x: \tau_0 \theta'}) |- e_1:
                \tau \theta}
            \: .
        \]

        To do so we must show
        \begin{gather*}
            \KMoreGen{K' |-
                \Gamma \theta,
                \Gen_{K'; \Gamma \theta}(\Set{x\colon \tau_0 \theta'}) <
                \Gamma \theta,
                (\Gen_{K; \Gamma}(\Set{x\colon \tau_0)}) \theta}
            \\
            \Var_{K'}(\Gamma \theta,
                \Gen_{K'; \Gamma \theta}(\Set{x\colon \tau_0 \theta'}))
                \subseteq
            \Var_{K'}(\Gamma \theta,
                (\Gen_{K; \Gamma}(\Set{x\colon \tau_0)}) \theta)
            \: .
        \end{gather*}

        The latter holds because $
            \Var_{K'}(\Gamma \theta,
                \Gen_{K'; \Gamma \theta}(\Set{x\colon \tau_0 \theta'}))
            \subseteq
            \Var_{K'}(\Gamma \theta) $.

        As for the former, we prove $
            \KMoreGen{K' |-
                \Gen_{K'; \Gamma \theta}(\Set{x\colon \tau_0 \theta'}) <
                (\Gen_{K; \Gamma}(\Set{x\colon \tau_0)}) \theta}
        $. We have
        \begin{gather*}
            \Gen_{K; \Gamma}(\Set{x\colon \tau_0}) =
                \KScheme{A. K_x > \tau_0} \qquad
            K_x =
                \SetC{\KKinding{\alpha :: K(\alpha)} | \alpha \in A}
            \: .
        \end{gather*}
        By \textalpha-renaming of the quantified variables we can write
        \begin{gather*}
            \Gen_{K; \Gamma}(\Set{x\colon \tau_0}) =
                \KScheme{B. K^\star_x > \tau_0 \theta_0} \\
            K^\star_x =
                \SetC{\KKinding{\beta_i :: \KUnkind} |
                    \KKinding{\alpha_i :: \KUnkind} \in K_x}
                \cup \SetC{\KKinding{\beta_i :: (L, U, T \theta_0)} |
                    \KKinding{\alpha_i :: (L, U, T)} \in A}
        \end{gather*}
        and, since $ \theta $ does not involve $ B $,
        \begin{gather*}
            (\Gen_{K; \Gamma}(\Set{x\colon \tau_0})) \theta =
            \KScheme{B. K^\star_x \theta > \tau_0 \theta_0 \theta} =
            \KScheme{B. K'_x > \tau_0 \theta'} \\
            K'_x = \SetC{\KKinding{\beta :: K'(\beta)} | \beta \in B}
            \: .
        \end{gather*}

        The other type scheme is
        \begin{gather*}
            \Gen_{K';\Gamma \theta}(\tau_0 \theta') =
                \KScheme{C. K'_C > \tau_0 \theta'}
            \\
            C = \Var_{K'}(\tau_0 \theta') \setminus \Var_{K'}(\Gamma \theta)
            \qquad
            K'_C = \SetC{\KKinding{\beta :: K'(\beta)} | \beta \in C}
            \: .
        \end{gather*}

        We show $ B \subseteq C $, which concludes the proof
        (because the kinding environments are both restrictions of $ K' $).
        Consider $ \beta_i \in B $. We have $
            \alpha_i \in \Var_K(\tau_0) \setminus \Var_K(\Gamma)
        $. Then $ \beta_i = \alpha_i \theta' \in \Var_{K'}(\tau_0 \theta') $.
        Furthermore $ \beta_i \notin \Var_{K'}(\Gamma \theta) $ holds
        because $ \Gamma \theta $ does not contain variables in $ B $
        ($ \Gamma $ does not contain them and $ \theta $ does not
        introduce them) and variables in $ B $ do not appear in the kinds
        of other variables which are not themselves in $ B $.

        We now consider the rule \Rule{Tk-Match} in its generality. We have
        \begin{gather*}
            \KTyping{K; \Gamma |- \Match{e_0 with (p_i \to e_i)_{i \in I}}:
                \tau} \\ \qquad
            \KTyping{K; \Gamma |- e_0: \tau_0} \qquad
            \KExhaustive{K |- \tau_0 < \SetC{p_i | i \in I}} \\ \qquad
            \forall i \in I . \enspace
            \KPatternTyping{K |- p_i: \tau_0 => \Gamma_i} \qquad
            \KTyping{K; \Gamma, \Gen_{K;\Gamma}(\Gamma_i) |- e_i: \tau}
        \end{gather*}
        and must show
        \begin{gather*}
            \KTyping{K'; \Gamma \theta |-
                \Match{e_0 with (p_i \to e_i)_{i \in I}}: \tau \theta}
        \end{gather*}
        which we prove by establishing, for some $ \hat{\tau_0} $ and
        $ \SetC{\hat{\Gamma_i} | i \in I} $, that
        \begin{gather*}
            \KTyping{K'; \Gamma \theta |- e_0: \hat{\tau_0}} \qquad
            \KExhaustive{K' |- \hat{\tau_0} < \SetC{p_i | i \in I}} \\
            \forall i \in I . \enspace
            \KPatternTyping{K' |- p_i: \hat{\tau_0} => \hat{\Gamma_i}} \qquad
            \KTyping{K'; \Gamma \theta, \Gen_{K';\Gamma \theta}(\hat{\Gamma_i})
                |- e_i: \tau \theta}
            \enspace .
        \end{gather*}

        For the derivation for $ e_0 $ we proceed as above and have
        $ \hat{\tau_0} = \tau_0 \theta' $.
        By Lemma~\ref{lem:a-k-exhaust-typesubst} we have $
            \KExhaustive{K' |- \tau_0 \theta' < \SetC{p_i | i \in I}}
        $. By Lemma~\ref{lem:a-k-patterns-typesubst}, we have $
            \KPatternTyping{K' |- p_i: \tau_0 \theta' => \Gamma_i \theta'}
        $ and thus take $ \hat{\Gamma_i} = \Gamma_i \theta' $.

        We proceed as before also for the derivations for each branch.
        The difference is that, to apply weakening, we must prove the two
        premises for the environments and not for $ \tau_0 $ alone.
        The condition on variables is straightforward, as before.
        For the other we prove, for each $ x \in \Capt(p_i) $ and assuming
        $ \Gamma_i(x) = \tau_x $,
        \[
            \KMoreGen{ K' |-
                \Gamma \theta,
                \Gen_{K'; \Gamma \theta}(\tau_x \theta') <
                \Gamma \theta,
                (\Gen_{K; \Gamma}(\tau_x)) \theta }
            \: .
        \]

        We show it as for $ \tau_0 $ above:
        $ \Var_K(\tau_x) $ is always a subset of $ \Var_K(\tau_0) $
        because environment generation does not introduce new variables.
    \qedhere
    \end{LCases}
\end{Proof}

\begin{Lemma}[Expression substitution]
\label{lem:a-k-exprsubst}
    Let $ x_1 $, \dots, $ x_n $ be distinct variables
    and $ v_1 $, \dots, $ v_n $ values.
    Let $ \Gamma' = \Set{x_1\colon \sigma_1, \dots, x_n\colon \sigma_n} $
    and $ \varsigma = \SubstDots[v_1/x_1 ... v_n/x_n] $.

    If $ \KTyping{ K; \Gamma, \Gamma' |- e: \tau } $
    and, for all $ k \in \Set{1, \dots, n} $
    and for all $ \tau_k \in \Inst_K(\sigma_k) $,
    $ \KTyping{ K; \Gamma |- v_k: \tau_k } $,
    then $ \KTyping{ K; \Gamma |- e \varsigma: \tau } $.
\end{Lemma}

\begin{Proof}
    By induction on the derivation of $
        \KTyping{K; \Gamma, \Gamma' |- e: \tau}
    $. We reason by cases on the last applied rule.

    \begin{LCases}
    \Case[\Rule{Tk-Var}]
        We have
        \[
            \KTyping{K; \Gamma, \Gamma' |- x: \tau} \qquad
            \tau \in \Inst_K((\Gamma, \Gamma')(x))
            \: .
        \]

        Either $ x = x_k $ for some $ k $ or not.
        In the latter case, $ x \varsigma = x $, $ x \notin \Dom(\Gamma') $
        and hence $ (\Gamma, \Gamma')(x) = \Gamma(x) $.
        Then, since $ \tau \in \Inst_K((\Gamma, \Gamma')(x)) $,
        $ \tau \in \Inst_K(\Gamma(x)) $ and \Rule{Tk-Var} can be applied.

        If $ x = x_k $, then $ (\Gamma, \Gamma')(x) = \Gamma'(x) = \sigma_k $.
        We must then prove $ \KTyping{K; \Gamma |- v_k: \tau} $,
        which we know by hypothesis since $ \tau \in \Inst_K(\sigma_k) $.

    \Case[\Rule{Tk-Const}]
        Straightforward.

    \Case[\Rule{Tk-Abstr}]
        We have
        \[
            \KTyping{K; \Gamma, \Gamma' |- \Abstr{x. e_1}: \tau_1 \to \tau_2}
            \qquad
            \KTyping{K; \Gamma, \Gamma', \Set{x\colon \tau_1} |- e_1: \tau_2}
            \: .
        \]

        By \textalpha-renaming we can assume $ x \notin \Dom(\Gamma') $;
        then $ (\Abstr{x. e_1}) \varsigma = \Abstr{x. (e_1 \varsigma)} $
        and $
            \Gamma, \Gamma', \Set{x\colon \tau_1} =
            \Gamma, \Set{x\colon \tau_1}, \Gamma'
        $. Therefore we have $
            \KTyping{K; \Gamma, \Set{x\colon \tau_1}, \Gamma' |- e_1: \tau_2}
        $ and, by the induction hypothesis, $
            \KTyping{K; \Gamma, \Set{x\colon \tau_1} |- e_1 \varsigma: \tau_2}
        $. We apply \Rule{Tk-Abstr} to conclude.

    \Cases[\Rule{Tk-Appl}, \Rule{Tk-Pair}, and \Rule{Tk-Tag}]
        Straightforward application of the induction hypothesis.

    \Case[\Rule{Tk-Match}]
        We have
        \begin{gather*}
            \KTyping{K; \Gamma, \Gamma' |-
                \Match{e_0 with (p_i \to e_i)_{i \in I}}: \tau} \\
            \KTyping{K; \Gamma, \Gamma' |- e_0: \tau_0} \qquad
            \KExhaustive{K |- \tau_0 < \SetC{p_i | i \in I}} \\
            \forall i \in I . \enspace
            \KPatternTyping{K |- p_i: \tau_0 => \Gamma_i} \qquad
            \KTyping{K; \Gamma, \Gamma', \Gen_{K;\Gamma,\Gamma'}(\Gamma_i) |-
                e_i: \tau}
            \enspace .
        \end{gather*}

        We assume by \textalpha-renaming that no capture variable
        of any pattern is in the domain of $ \Gamma' $.
        Then, $
            (\Match{e_0 with (p_i \to e_i)_{i \in I}}) \varsigma =
            \Match{e_0 \varsigma{} with (p_i \to e_i \varsigma)_{i \in I}}
        $ and $
            \Gamma, \Gamma', \Gen_{K;\Gamma,\Gamma'}(\Gamma_i)  =
            \Gamma, \Gen_{K;\Gamma,\Gamma'}(\Gamma_i), \Gamma'
        $ for any $ i $.

        By the induction hypothesis, we derive $
            \KTyping{K; \Gamma |- e_0 \varsigma: \tau_0}
        $ and $
            \KTyping{K; \Gamma, \Gen_{K;\Gamma,\Gamma'}(\Gamma_i) |-
                e_i \varsigma: \tau}
        $ for all $ i $. From the latter, we prove $
            \KTyping{K; \Gamma, \Gen_{K;\Gamma}(\Gamma_i) |-
                e_i \varsigma: \tau}
        $ by weakening (Lemma~\ref{lem:a-k-weakening}): we have $
            \KMoreGen{K |- \Gen_{K;\Gamma}(\Gamma_i) <
                \Gen_{K;\Gamma,\Gamma'}(\Gamma_i)}
        $ by Lemma~\ref{lem:a-k-generalization}---since $
            \Var_K(\Gamma) \subseteq \Var_K(\Gamma,\Gamma')
        $---and clearly we have $
            \Var_K(\Gamma, \Gen_{K;\Gamma}(\Gamma_i)) \subseteq
            \Var_K(\Gamma, \Gen_{K;\Gamma,\Gamma'}(\Gamma_i))
        $ since $ \Var_K(\Gen_{K;\Gamma}(\Gamma_i)) \subseteq \Var_K(\Gamma) $.
        \qedhere
    \end{LCases}
\end{Proof}

\begin{Theorem}[Progress]
\label{thm:a-k-progress}
    Let $ e $ be a well-typed, closed expression.
    Then, either $ e $ is a value
    or there exists an expression $ e' $ such that $ \Smallstep{ e ~> e' } $.
\end{Theorem}

\begin{Proof}
    By hypothesis we have $ \KTyping{K; \EmptyEnv |- e: \tau} $.
    The proof is by induction on its derivation;
    we reason by cases on the last applied rule.

    \begin{LCases}
        \Case[\Rule{Tk-Var}]
            This case does not occur because variables are not closed.

        \Case[\Rule{Tk-Const}]
            In this case $ e $ is a constant $ c $ and therefore a value.

        \Case[\Rule{Tk-Abstr}]
            In this case $ e $ is an abstraction $ \Abstr{x. e_1} $.
            Since it is also closed, it is a value.

        \Case[\Rule{Tk-Appl}]
            We have
            \[
                \KTyping{K; \EmptyEnv |- \Appl{e_1}{e_2}: \tau} \qquad
                \KTyping{K; \EmptyEnv |- e_1: \tau' \to \tau} \qquad
                \KTyping{K; \EmptyEnv |- e_2: \tau'}
                \: .
            \]

            By the induction hypothesis,
            each of $ e_1 $ and $ e_2 $ either is a value or may reduce.
            If $ \Smallstep{e_1 ~> e_1'} $, then $
                \Smallstep{\Appl{e_1}{e_2} ~> \Appl{e_1'}{e_2}}
            $. If $ e_1 $ is a value and $ \Smallstep{e_2 ~> e_2'} $,
            then $ \Smallstep{\Appl{e_1}{e_2} ~> \Appl{e_1}{e_2'}} $.

            If both are values then, by Lemma~\ref{lem:a-k-generation},
            $ e_1 $ has the form $ \Abstr{x. e_3} $ for some $ e_3 $.
            Then, we can apply \Rule{R-Appl}
            and $ \Smallstep{\Appl{e_1}{e_2} ~> e_3 \SubstSingle[e_2/x]} $.

        \Case[\Rule{Tk-Pair}]
            We have
            \[
                \KTyping{K; \EmptyEnv |- (e_1, e_2): \tau_1 \times \tau_2}
                    \qquad
                \KTyping{K; \EmptyEnv |- e_1: \tau_1} \qquad
                \KTyping{K; \EmptyEnv |- e_2: \tau_2}
                \: .
            \]

            By the induction hypothesis, each of $ e_1 $ and $ e_2 $
            either is a value or may reduce.
            If $ \Smallstep{e_1 ~> e_1'} $,
            then $ \Smallstep{(e_1, e_2) ~> (e_1', e_2)} $.
            If $ e_1 $ is a value and $ \Smallstep{e_2 ~> e_2'} $, then $
                \Smallstep{(e_1, e_2) ~> (e_1, e_2')}
            $. If both are values, then $ (e_1, e_2) $ is also a value.

        \Case[\Rule{Tk-Tag}]
            We have
            \[
                \KTyping{K; \EmptyEnv |- \Tagg(e_1): \alpha} \qquad
                \KTyping{K; \EmptyEnv |- e_1: \tau_1}
                \: .
            \]

            Analogously to the previous case,
            by the induction hypothesis we have that either $ e_1 $ is a value
            or $ \Smallstep{e_1 ~> e_1'} $.
            In the former case, $ \Tagg(e_1) $ is a value as well.
            In the latter, we have $ \Smallstep{\Tagg(e_1) ~> \Tagg(e_1')} $.

        \Case[\Rule{Tk-Match}]
            We have
            \[
                \KTyping{K; \EmptyEnv |-
                         \Match{e_0 with (p_i \to e_i)_{i \in I}}: \tau} \qquad
                \KTyping{K; \EmptyEnv |- e_0: \tau_0} \qquad
                \KExhaustive{K |- \tau_0 < \SetC{p_i | i \in I}}
                \: .
            \]

            By the inductive hypothesis, either $ e_0 $ is a value
            or it may reduce.
            In the latter case, if $ \Smallstep{e_0 ~> e_0'} $,
            then $
                \Smallstep{ \Match{e_0 with (p_i \to e_i)_{i \in I}} ~>
                    \Match{e_0' with (p_i \to e_i)_{i \in I}} }
            $.

            If $ e_0 $ is a value, on the other hand,
            the expression may reduce by application of \Rule{R-Match}.
            Since $ \KExhaustive{K |- \tau_0 < \SetC{p_i | i \in I}} $
            and $ e_0 $ is a value of type $ \tau_0 $
            (and therefore satisfies the premises of the definition
            of exhaustiveness, with $ \theta = \EmptySubst $ and $ K = K' $),
            there exists at least an $ i \in I $ such that $
                \Matching{e_0 / p_i} = \varsigma
            $ for some substitution $ \varsigma $.
            Let $ j $ be the least of these $ i $
            and $ \varsigma_j $ the corresponding substitution; then $
                \Smallstep{ \Match{e_0 with (p_i \to e_i)_{i \in I}} ~>
                    e_j \varsigma_j }
            $.
            \qedhere
    \end{LCases}
\end{Proof}

\begin{Theorem}[Subject reduction]
\label{thm:a-k-subject}
    Let $ e $ be an expression and $ \tau $ a type
    such that $ \KTyping{ K; \Gamma |- e: \tau } $.
    If $ \Smallstep{ e ~> e' } $, then $ \KTyping{ K; \Gamma |- e': \tau } $.
\end{Theorem}

\begin{Proof}
    By induction on the derivation of $ \KTyping{K; \Gamma |- e: \tau} $.
    We reason by cases on the last applied rule.

    \begin{LCases}
    \Cases[\Rule{Tk-Var}, \Rule{Tk-Const}, and \Rule{Tk-Abstr}]
        These cases may not occur: variables, constants, and abstractions
        never reduce.

    \Case[\Rule{Tk-Appl}]
        We have
        \[
            \KTyping{K; \Gamma |- \Appl{e_1}{e_2}: \tau} \qquad
            \KTyping{K; \Gamma |- e_1: \tau' \to \tau} \qquad
            \KTyping{K; \Gamma |- e_2: \tau'}
            \: .
        \]

        $ \Smallstep{\Appl{e_1}{e_2} ~> e'} $ occurs in any of three ways:
        $(i)$ $ \Smallstep{e_1 ~> e_1'} $ and $ e' = \Appl{e_1'}{e_2} $;
        $(ii)$ $ e_1 $ is a value, $ \Smallstep{e_2 ~> e_2'} $ and
        $ e' = \Appl{e_1}{e_2'} $;
        $(iii)$ both $ e_1 $ and $ e_2 $ are values, $ e_1 $ is of the form $ \Abstr{x. e_3} $, and $ e' = e_3 \SubstSingle[e_2/x] $.

        In the first case, we derive by the induction hypothesis that $
            \KTyping{K; \Gamma |- e_1': \tau' \to \tau}
        $ and conclude by applying \Rule{Tk-Appl} again.
        The second case is analogous.

        In the third case, we know by Lemma~\ref{lem:a-k-generation} that $
            \KTyping{K; \Gamma, \Set{x\colon \tau'} |- e_3: \tau}
        $. We also know that $ e_2 $ is a value such that $
            \KTyping{K; \Gamma |- e_2: \tau'}
        $. Then, by Lemma~\ref{lem:a-k-exprsubst}, $
            \KTyping{K; \Gamma |- e_3 \SubstSingle[e_2/x]: \tau}
        $.

    \Case[\Rule{Tk-Pair}]
        We have
        \[
            \KTyping{K; \Gamma |- (e_1, e_2): \tau_1 \times \tau_2} \qquad
            \KTyping{K; \Gamma |- e_1: \tau_1} \qquad
            \KTyping{K; \Gamma |- e_2: \tau_2}
            \: .
        \]

        $ \Smallstep{(e_1, e_2) ~> e'} $ occurs either because
        $ \Smallstep{e_1 ~> e_1'} $ and $ e' = (e_1', e_2) $,
        or because $ e_1 $ is a value, $ \Smallstep{e_2 ~> e_2'} $,
        and $ e' = (e_1, e_2') $.
        In either case, the induction hypothesis allows us to derive
        that the type of the component that reduces is preserved;
        therefore, we can apply \Rule{Tk-Pair} again to conclude.

    \Case[\Rule{Tk-Tag}]
        Analogously to the previous case,
        a variant expression only reduces if its argument does,
        so we apply the induction hypothesis and \Rule{Tk-Tag} to conclude.

    \Case[\Rule{Tk-Match}]
        We have
        \begin{gather*}
            \KTyping{K; \Gamma |- \Match{e_0 with (p_i \to e_i)_{i \in I}}:
                \tau} \\
            \KTyping{K; \Gamma |- e_0: \tau_0} \qquad
            \forall i \in I . \enspace
            \KPatternTyping{K |- p_i: \tau_0 => \Gamma_i} \qquad
            \KTyping{K; \Gamma, \Gen_{K;\Gamma}(\Gamma_i) |- e_i: \tau}
            \enspace .
        \end{gather*}

        $ \Smallstep{\Match{e_0 with (p_i \to e_i)_{i \in I}} ~> e'} $ occurs
        either because $ \Smallstep{e_0 ~> e_0'} $ and $
            e' = \Match{e_0' with (p_i \to e_i)_{i \in I}}
        $ or because $ e_0 $ is a value and $ e' = e_j \varsigma $, where $
            \Matching{e_0 / p_j} = \varsigma
        $ and, for all $ i < j $, $ \Matching{e_0 / p_i} = \MatchFail $.
        In the former case, we apply the induction hypothesis
        and conclude by \Rule{Tk-Match}.

        In the latter case, $ \varsigma $ is a substitution
        from the capture variables of $ p_j $ to values, and we know
        by Lemma~\ref{lem:a-k-patterns-correct}
        that, for all $ x \in \Capt(p_j) $, $
            \KTyping{K; \Gamma |- x \varsigma: \Gamma_j(x)}
        $. We show that, additionally, $
            \KTyping{K; \Gamma |- x \varsigma: \tau_x}
        $ holds for every $ \tau_x \in \Inst_K(\Gen_{K;\Gamma}(\Gamma_j(x))) $.
        Every such $ \tau_x $ is equal to $ \Gamma_j(x) \theta $
        for a $ \theta $ such that $
            \Dom(\theta) \subseteq \Var_K(\Gamma_j(x)) \setminus \Var_K(\Gamma)
        $ and $ \KAdmissible{K |- \theta: K} $
        (the kinding environment captured by generalization is
        just a subset of $ K $). Then, $
            \KTyping{K; \Gamma |- x \varsigma: \Gamma_j(x) \theta}
        $ holds by Lemma~\ref{def:a-k-typesubst},
        since $ \Gamma \theta = \Gamma $
        (the substitution does not change any free variable of $ \Gamma $).

        From $ \KTyping{K; \Gamma, \Gen_{K;\Gamma}(\Gamma_j) |- e_j: \tau} $
        and from the fact that we have $
            \KTyping{K; \Gamma |- x \varsigma: \tau_x}
        $ for all $ x \in \Dom(\Gamma_j) $ and all $
            \tau_x \in \Inst_K(\Gen_{K;\Gamma}(\Gamma_j(x)))
        $, we derive $ \KTyping{K; \Gamma |- e_j \varsigma: \tau } $
        by Lemma~\ref{lem:a-k-exprsubst}.
    \qedhere
    \end{LCases}
\end{Proof}

\begin{Corollary}[Type soundness]
\label{cor:a-k-soundness}
    Let $ e $ be a well-typed, closed expression, that is, such that
    $ \KTyping{ K; \varnothing |- e: \tau } $ holds for some $ \tau $.
    Then, either $ e $ diverges or it reduces to a value $ v $
    such that $ \KTyping{ K; \EmptyEnv |- v: \tau } $.
\end{Corollary}

\begin{Proof}
    Consequence of Theorem~\ref{thm:a-k-progress}
    and Theorem~\ref{thm:a-k-subject}.
\end{Proof}

\subsection{Typing variants with set-theoretic types}

\subsubsection{Definition of the \VariantsS{} type system}

We consider a set $ \TypeVars $ of \emph{type variables}
(ranged over by $ \alpha $, $ \beta $, $ \gamma $, \dots)
and the sets $ \Constants $, $ \Tags $, and $ \BasicTypes $
of \emph{language constants}, \emph{tags}, and \emph{basic types}
(ranged over by $ c $, $ \Tagg $, and $ b $ respectively).

\begin{Definition}[Types]
\label{def:a-s-types}
    A \emph{type} $ t $ is a term coinductively produced
    by the following grammar:
    \[
        \Bnf{ t
            \= \alpha \| b \| c
            \| t \to t \| t \times t \| \Tagg(t)
            \| t \lor t \| \lnot t \| \Empty }
    \]
    which satisfies two additional constraints:
    \begin{itemize}
        \item (\emph{regularity})
            the term must have a finite number of different sub-terms;
        \item (\emph{contractivity})
            every infinite branch must contain an infinite number
            of occurrences of atoms
            (i.e., a type variable or the immediate application
            of a type constructor: basic, constant, arrow, product, or variant).
    \end{itemize}
\end{Definition}

We introduce the following abbreviations:
\[
     t_1 \land t_2 \eqdef \lnot (\lnot t_1 \lor \lnot t_2) \qquad
     t_1 \setminus t_2 \eqdef t_1 \land (\lnot t_2) \qquad
     \Any \eqdef \lnot \Empty
     \: .
\]

\begin{Definition}[Type schemes]
\label{def:a-s-schemes}
    A \emph{type scheme} $ s $ is of the form $ \Scheme{A. t} $,
    where $ A $ is a finite set $ \Set{\alpha_1, \dots, \alpha_n} $
    of type variables.
\end{Definition}

We identify a type scheme $ \Scheme{\varnothing. t} $ with the type $ t $
itself.
Furthermore, we consider type schemes up to renaming of the variables they bind,
and we disregard useless quantification.

\begin{Definition}[Free variables]
\label{def:a-s-freevars}
    We write $ \Var(t) $ for the set of type variables occurring
    in a type $ t $;
    we say they are the \emph{free variables} of $ t $,
    and we say that $ t $ is \emph{ground} or \emph{closed}
    if and only if $ \Var(t) $ is empty.

    We extend the definition to type schemes as $
        \Var(\Scheme{A. t}) = \Var(t) \setminus A
    $.
\end{Definition}

The (coinductive) definition of $ \Var $ can be found in \citet[Definition A.2]{Castagna2014}.

\begin{Definition}[Meaningful variables]
\label{def:a-s-mvar}
    We define the set $ \Mvar(t) $ of \emph{meaningful variables}
    of a type $ t $ as
    \[
        \Mvar(t) =
        \SetC{\alpha \in \Var(t) |
            t \SubstSingle[\EmptyType/\alpha] \nsimeq t}
        \: .
    \]

    We extend the definition to type schemes as $
        \Mvar(\Scheme{A. t}) = \Mvar(t) \setminus A
    $.
\end{Definition}

\begin{Definition}[Type substitutions]
\label{def:a-s-typesubst}
    A \emph{type substitution} $ \theta $ is a finite mapping
    of type variables to types.
    We write $ \SubstC[t_i/\alpha_i | i \in I] $ for the type substitution
    which simultaneously replaces $ \alpha_i $ with $ t_i $,
    for each $ i \in I $.
    We write $ t \theta $ for the application of the substitution $ \theta $
    to the type $ t $;
    application is defined coinductively by the following equations.
    \begin{align*}
        \alpha \theta & =
            \begin{cases}
                t' & \text{if $ \SubstF{t'/\alpha} \in \theta $} \\
                \alpha & \text{otherwise}
            \end{cases} \\
        b \theta & = b \\
        c \theta & = c \\
        (t_1 \to t_2) \theta & = (t_1 \theta) \to (t_2 \theta) \\
        (t_1 \times t_2) \theta & = (t_1 \theta) \times (t_2 \theta)
        \\
        (\Tagg(t)) \theta & = \Tagg(t \theta) \\
        (t_1 \lor t_2) \theta & = (t_1 \theta) \lor (t_2 \theta) \\
        (\lnot t) \theta & = \lnot(t \theta) \\
        \Empty \theta & = \Empty
    \end{align*}

    We extend the $ \Var $ operation to substitutions as
    \[
        \Var(\theta) = \bigcup_{\alpha \in \Dom(\theta)} \Var(\alpha \theta)
        \: .
    \]
    and we extend $ \Mvar $ likewise.

    We extend application of substitutions
    to type schemes $ \Scheme{A. t} $:
    by renaming quantified variables, we assume $
        A \cap (\Dom(\theta) \cup \Var(\theta)) = \varnothing
    $, and we have $
        (\Scheme{A. t}) \theta = \Scheme{A. t \theta}
    $.

    We write $ \theta_1 \cup \theta_2 $ for the union of disjoint substitutions
    and $ \theta_1 \circ \theta_2 $ for the composition of substitutions.
\end{Definition}

\begin{Definition}[Type environments]
\label{def:a-s-typeenv}
    A \emph{type environment} $ \Gamma $ is a partial mapping
    from expression variables to type schemes.
    We write type environments as $
        \Gamma = \Set{x_1\colon s_1, \dots, x_n\colon s_n} $.

    We write $ \Gamma, \Gamma' $ for the updating
    of the type environment $ \Gamma $ with the new bindings in $ \Gamma' $.
    It is defined as follows.
    \[
        (\Gamma, \Gamma')(x) =
            \begin{cases}
                \Gamma'(x) & \text{if $ x \in \Dom(\Gamma') $} \\
                \Gamma(x) & \text{otherwise}
            \end{cases}
    \]

    We extend the $ \Var $ operation to type environments as
    \[
        \Var(\Gamma) = \bigcup_{s \in \Range(\Gamma)} \Var(s)
        \: ,
    \]
    and we extend $ \Mvar $ likewise.
\end{Definition}

\begin{Definition}[Generalization]
\label{def:a-s-gen}
    We define the \emph{generalization} of a type $ t $
    with respect to the type environment $ \Gamma $
    as the type scheme
    \[
        \Gen_\Gamma(t) = \Scheme{A. t}
    \]
    where $ A = \Var(t) \setminus \Mvar(\Gamma) $.

    We extend this definition to type environments which only contain types
    (i.e., trivial type schemes) as
    \[
        \Gen_\Gamma(\SetC{x_i\colon t_i | i \in I}) =
            \SetC{x_i\colon \Gen_\Gamma(t_i) | i \in I}
        \: .
    \]
\end{Definition}

\begin{Definition}[Instances of a type scheme]
\label{def:a-s-inst}
    The set of \emph{instances} of a type scheme $ \Scheme{A. t} $
    is defined as
    \[
        \Inst(\Scheme{A. t}) =
            \SetC{t \theta | \Dom(\theta) \subseteq A}
        \: .
    \]

    We say that a type scheme $ s_1 $ is \emph{more general}
    than a type scheme $ s_2 $,
    and we write $ \SMoreGen{s_1 < s_2} $, if
    \[
        \forall t_2 \in \Inst(s_2) . \:
        \exists t_1 \in \Inst(s_1) . \: t_1 \leq t_2
        \: .
    \]

    We extend this notion to type environments as
    \[
        \SMoreGen{\Gamma_1 < \Gamma_2} \iff
            \Dom(\Gamma_1) = \Dom(\Gamma_2) \land
            \forall x \in \Dom(\Gamma_1) . \:
                \SMoreGen{\Gamma_1(x) < \Gamma_2(x)}
        \: .
    \]
\end{Definition}

\begin{Definition}[Accepted type]
\label{def:a-s-accepted}
    The \emph{accepted type} $ \SAcc{p} $ of a pattern $ p $
    is defined inductively as:
    \begin{align*}
        \SAcc{\Wildcard} = \SAcc{x} & = \AnyType &
        \SAcc{c} & = c \\
        \SAcc{(p_1, p_2)} & = \SAcc{p_1} \times \SAcc{p_2} &
        \SAcc{\Tagg(p)} & = \Tagg(\SAcc{p}) \\
        \SAcc{\PAnd{p_1 \& p_2}} & = \SAcc{p_1} \land \SAcc{p_2} &
        \SAcc{\POr{p_1 | p_2}} & = \SAcc{p_1} \lor \SAcc{p_2}
        \: .
    \end{align*}
\end{Definition}

The projection operators $ \SPiFst $ and $ \SPiSnd $ for product types
are defined by \citet[Appendix C.2.1]{Castagna2014}.
We do not repeat the definition,
but we state below the properties we need in the proofs.
The projection operators for variant types correspond to $ \SPiSnd $
if we encode variant types as pairs;
we therefore rephrase the same properties for them.

\begin{Property}[Projections of product types]
\label{pro:a-s-productproj}
    There exist two functions $ \SPiFst $ and $ \SPiSnd $ which,
    given a type $ t \leq \Any \times \Any $,
    yield types $ \SPiFst(t) $ and $ \SPiSnd(t) $ such that:
    \begin{itemize}
        \item $ t \leq \SPiFst(t) \times \SPiSnd(t) $;
        \item if $ t \leq t_1 \times t_2 $, then $ \SPiIth(t) \leq t_i $;
        \item if $ t \leq t' \leq \Any \times \Any $,
            then $ \SPiIth(t) \leq \SPiIth(t') $;
        \item for all type substitutions $ \theta $,
            $ \SPiIth(t \theta) \leq \SPiIth(t) \theta $.
    \end{itemize}
\end{Property}

\begin{Property}[Projections of variant arguments]
\label{pro:a-s-variantproj}
    For every tag $ \Tagg $ there exists a function $ \SPiTagg $ which,
    given a type $ t \leq \Tagg(\Any) $,
    yields a type $ \SPiTagg(t) $ such that:
    \begin{itemize}
        \item $ t \leq \Tagg(\SPiTagg(t)) $;
        \item if $ t \leq \Tagg(t') $, then $ \SPiTagg(t) \leq t' $;
        \item if $ t \leq t' \leq \Tagg(\Any) $,
            then $ \SPiTagg(t) \leq \SPiTagg(t') $;
        \item for all type substitutions $ \theta $,
            $ \SPiTagg(t \theta) \leq \SPiTagg(t) \theta $.
    \end{itemize}
\end{Property}

\begin{Definition}[Pattern environment generation]
\label{def:a-s-patternenv}
    Given a pattern $ p $ and a type $ t \leq \SAcc{p} $,
    the type environment $ \SPatternEnv{ t // p } $
    generated by pattern matching
    is defined inductively as:
    \begin{align*}
        \SPatternEnv{ t // \Wildcard }
            & = \varnothing \\
        \SPatternEnv{ t // x }
            & = \Set{x\colon t} \\
        \SPatternEnv{ t // c }
            & = \varnothing \\
        \SPatternEnv{ t // (p_1, p_2) }
            & =
            \SPatternEnv{ \SPiFst(t) // p_1 } \: \cup \:
            \SPatternEnv{ \SPiSnd(t) // p_2 } \\
        \SPatternEnv{ t // \Tagg(p) }
            & = \SPatternEnv{ \SPiTagg(t) // p } \\
        \SPatternEnv{ t // \PAnd{p_1 \& p_2} }
            & =
            \SPatternEnv{ t // p_1 } \: \cup \: \SPatternEnv{ t // p_2 } \\
        \SPatternEnv{ t // \POr{p_1 | p_2} }
            & =
            \SPatternEnv{ (t \land \SAcc{p_1}) // p_1 }
            \: \SPatternEnvOr \:
            \SPatternEnv{ (t \setminus \SAcc{p_1}) // p_2 }
        \: ,
    \end{align*}
    where $
        (\Gamma \SPatternEnvOr \Gamma')(x) = \Gamma(x) \lor \Gamma'(x)
    $.
\end{Definition}

\begin{Definition}[Typing relation]
\label{def:a-s-typing}
    The typing relation $ \STyping{ \Gamma |- e: t } $
    (\emph{$ e $ is given type $ t $ in the type environment $ \Gamma $})
    is defined by the rules in Figure~\ref{fig:variants-s-typing-app}.
\end{Definition}

\begin{figure*}

    \begin{mathpar}
        \MyInfer[Ts-Var]
            { t \in \Inst(\Gamma(x)) }
            { \STyping{ \Gamma |- x: t } }
            {}
        \and
        \MyInfer[Ts-Const]
            { }
            { \STyping{ \Gamma |- c: c } }
            {}
        \and
        \MyInfer[Ts-Abstr]
            { \STyping{ \Gamma, \Set{x\colon t_1} |- e: t_2 } }
            { \STyping{ \Gamma |- \Abstr{x. e}: t_1 \to t_2 } }
            {}
        \and
        \MyInfer[Ts-Appl]
            { \STyping{ \Gamma |- e_1: t' \to t } \\
              \STyping{ \Gamma |- e_2: t' } }
            { \STyping{ \Gamma |- \Appl{e_1}{e_2}: t } }
            {}
        \and
        \MyInfer[Ts-Pair]
            { \STyping{ \Gamma |- e_1: t_1 } \\
              \STyping{ \Gamma |- e_2: t_2 } }
            { \STyping{ \Gamma |- (e_1, e_2): t_1 \times t_2 } }
            {}
        \and
        \MyInfer[Ts-Tag]
            { \STyping{ \Gamma |- e: t } }
            { \STyping{ \Gamma |- \Tagg(e): \Tagg(t) } }
            {}
        \and
        \MyInfer[Ts-Match]
            { \STyping{ \Gamma |- e_0: t_0 } \\
              t_0 \leq \textstyle\bigvee_{i \in I} \SAcc{p_i} \\
              t_i = (t_0 \setminus \textstyle\bigvee_{j < i} \SAcc{p_j})
                \land \SAcc{p_i} \\\\
              \forall i \in I \\
              \STyping{ \Gamma, \Gen_\Gamma(\SPatternEnv{ t_i // p_i }) |-
                e_i: t_i' } }
            { \STyping{ \Gamma |- \Match{ e_0 with (p_i \to e_i)_{i \in I} }:
                \textstyle\bigvee_{i \in I} t_i' } }
            {}
        \and
        \MyInfer[Ts-Subsum]
            { \STyping{ \Gamma |- e: t' } \\
              t' \leq t }
            { \STyping{ \Gamma |- e: t } }
            {}
    \end{mathpar}

    \caption{Typing relation of the \VariantsS{} type system.}
    \label{fig:variants-s-typing-app}
\end{figure*}

\subsubsection{Properties of the \VariantsS{} type system}

\begin{Lemma}[Generation for values]
\label{lem:a-s-generation}
    Let $ v $ be a value. Then:
    \begin{itemize}
        \item if $ \STyping{ \Gamma |- v: c } $, then $ v = c $;
        \item if $ \STyping{ \Gamma |- v: b } $,
            then $ v = c $ for some $ c $ such that $ b_c \leq b $;
        \item if $ \STyping{ \Gamma |- v: t_1 \to t_2 } $,
            then $ v $ is of the form $ \Abstr{x. e} $
            and $ \STyping{ \Gamma, \Set{x\colon t_1} |- e: t_2 } $;
        \item if $ \STyping{ \Gamma |- v: t_1 \times t_2 } $,
            then $ v $ is of the form $ (v_1, v_2) $,
            $ \STyping{ \Gamma |- v_1: t_1 } $,
            and $ \STyping{ \Gamma |- v_2: t_2 } $;
        \item if $ \STyping{ \Gamma |- v: \Tagg(t_1) } $,
            then $ v $ is of the form $ \Tagg(v_1) $
            and $ \STyping{ \Gamma |- v_1: t_1 } $.
    \end{itemize}
\end{Lemma}

\begin{Proof}
    By induction on the typing derivation:
    values must be typed by an application of the rule
    corresponding to their form to appropriate premises,
    possibly followed by applications of \RuleRef{Ts-Subsum}.

    The base cases are straightforward.
    In the inductive step, we just apply the induction hypothesis;
    for abstractions,
    the result follows from the behaviour of subtyping on arrow types.
\end{Proof}

We state the next three lemmas without proof,
as they rely on the model of types which we have not discussed.
Details can be found
in \citet{Frisch2008} and \citet{Castagna2011},
as well as in Alain Frisch's PhD thesis.%
    \footnote{%
        A. Frisch. \emph{Th\'eorie, conception et r\'ealisation d'un langage
        de programmation adapt\'e à XML}. PhD thesis, Universit\'e Paris 7 --
        Denis Diderot, 2004.}

\begin{Lemma}
\label{lem:a-s-patterns-union}
    For each $ i \in I $, let $ p_i $ be a pattern.
    If $ \STyping{ \Gamma |- v: \bigvee_{i \in I} \SAcc{p_i} } $,
    then there exists an $ i \in I $
    such that $ \STyping{ \Gamma |- v: \SAcc{p_i} } $.
\end{Lemma}

\begin{Lemma}
\label{lem:a-s-patterns-intersection}
    Let $ t $ be a type.
    Let $ t' $ be a type such that either
    $ t' = \SAcc{p} $ or $ t' = \lnot \SAcc{p} $, for some pattern $ p $.
    If $ \STyping{ \Gamma |- v: t } $ and $ \STyping{ \Gamma |- v: t' } $,
    then $ \STyping{ \Gamma |- v: t \land t' } $.
\end{Lemma}

\begin{Lemma}
\label{lem:a-s-patterns-accepted}
    Let $ v $ be a well-typed value
    (i.e., $ \STyping{ \varnothing |- v: t } $ holds for some $ t $)
    and $ p $ a pattern. Then:
    \begin{itemize}
        \item $ \STyping{ \varnothing |- v: \SAcc{p} } $ holds
            if and only if $ \Matching{v / p} = \varsigma $
            for some substitution $ \varsigma $;
        \item $ \STyping{ \varnothing |- v: \lnot \SAcc{p} } $ holds
            if and only if $ \Matching{v / p} = \MatchFail $.
    \end{itemize}
\end{Lemma}

\begin{Lemma}
\label{lem:a-s-patterns-subtyping}
    Let $ p $ be a pattern and $ t $, $ t' $ two types.
    If $ t \leq t' \leq \SAcc{p} $, then, for all $ x \in \Capt(p) $,
    $ (\SPatternEnv{t // p})(x) \leq (\SPatternEnv{t' // p})(x) $.
\end{Lemma}

\begin{Proof}
    By structural induction on $ p $.

    \begin{LCases}
    \Cases[$ p = \Wildcard $ and $ p = c $]
        There is nothing to prove since $ \Capt(p) = \varnothing $.

    \Case[$ p = x $]
        We must prove $
            (\SPatternEnv{t // x})(x) \leq (\SPatternEnv{t' // x})(x)
        $, that is, $ t \leq t' $, which we know by hypothesis.

    \Case[$ p = (p_1, p_2) $]
        Each $ x \in \Capt(p) $ is either in $ \Capt(p_1) $ or in
        $ \Capt(p_2) $. Assume $ x \in \Capt(p_i) $; then, $
            (\SPatternEnv{t // p})(x) = (\SPatternEnv{\SPiIth(t) // p_i})(x)
        $ and $
            (\SPatternEnv{t' // p})(x) = (\SPatternEnv{\SPiIth(t') // p_i})(x)
        $. Since $ t \leq t' $ implies $ \SPiIth(t) \leq \SPiIth(t') $
        by Property~\ref{pro:a-s-productproj},
        we can apply the induction hypothesis to conclude.

    \Case[$ p = \Tagg(p) $]
        Analogous to the previous case,
        because $ t \leq t' $ implies $ \SPiTagg(t) \leq \SPiTagg(t') $
        by Property~\ref{pro:a-s-variantproj}.

    \Case[$ p = \PAnd{p_1 \& p_2} $]
        Each $ x \in \Capt(p) $ is either in $ \Capt(p_1) $ or in
        $ \Capt(p_2) $. Assume $ x \in \Capt(p_i) $; then, $
            (\SPatternEnv{t // p})(x) = (\SPatternEnv{t // p_i})(x)
        $ and $ (\SPatternEnv{t' // p})(x) = (\SPatternEnv{t' // p_i})(x) $.
        We apply the induction hypothesis to conclude.

    \Case[$ p = \POr{p_1 | p_2} $]
        Every $ x \in \Capt(p) $ is both in $ \Capt(p_1) $ and in
        $ \Capt(p_2) $. We have that $
            (\SPatternEnv{t // p})(x) =
                (\SPatternEnv{t \land \SAcc{p_1} // p_1})(x) \lor
                (\SPatternEnv{t \setminus \SAcc{p_1} // p_2})(x)
        $ and likewise for $ t' $. Since $
            t \land \SAcc{p_1} \leq t' \land \SAcc{p_1}
        $ and $
            t \setminus \SAcc{p_1} \leq t' \setminus \SAcc{p_1}
        $, we can apply the induction hypothesis to both sub-patterns
        to derive $
            (\SPatternEnv{t \land \SAcc{p_1} // p_1})(x) \leq
            (\SPatternEnv{t' \land \SAcc{p_1} // p_1})(x)
        $ and $
            (\SPatternEnv{t \setminus \SAcc{p_1} // p_2})(x) \leq
            (\SPatternEnv{t' \setminus \SAcc{p_1} // p_2})(x)
        $. Then we have $
            (\SPatternEnv{t \land \SAcc{p_1} // p_1})(x) \lor
                (\SPatternEnv{t \setminus \SAcc{p_1} // p_2})(x)
            \leq
            (\SPatternEnv{t' \land \SAcc{p_1} // p_1})(x) \lor
                (\SPatternEnv{t' \setminus \SAcc{p_1} // p_2})(x)
        $.
    \qedhere
    \end{LCases}
\end{Proof}

\begin{Lemma}[Correctness of environment generation]
\label{lem:a-s-patterns-correct}
    Let $ p $ be a pattern and $ v $ a value
    such that $ \STyping{ \Gamma |- v: t } $ for some $ t \leq \SAcc{p} $.
    Then, for all $ x \in \Capt(p) $,
    $ \STyping{ \Gamma |- x (\Matching{v/p}): (\SPatternEnv{t//p})(x) } $.
\end{Lemma}

\begin{Proof}
    By structural induction on $ p $.

    \begin{LCases}
    \Cases[$ p = \Wildcard $ and $ p = c $]
        There is nothing to prove since $ \Capt(p) = \varnothing $.

    \Case[$ p = x $]
        We must prove $
            \STyping{\Gamma |- x \SubstSingle[v/x]: (\SPatternEnv{t // x})(x)}
        $, which is the hypothesis $ \STyping{\Gamma |- v: t} $.

    \Case[$ p = (p_1, p_2) $]
        We have $ t \leq \Any \times \Any $, hence $
            t \leq \SPiFst(t) \times \SPiSnd(t)
        $; then, since $ \STyping{\Gamma |- v: \SPiFst(t) \times \SPiSnd(t)} $
        by subsumption, we have by Lemma~\ref{lem:a-s-generation}
        that $ v = (v_1, v_2) $ and that $ \STyping{\Gamma |- v_i: \SPiFst(t)} $
        for both $ i $. Moreover, $
            t \leq \SAcc{(p_1, p_2)} = \SAcc{p_1} \times \SAcc{p_2}
        $. Hence, by Property~\ref{pro:a-s-productproj},
        $ \SPiIth(t) \leq \SAcc{p_i} $ for both $ i $.

        Each $ x \in \Capt(p) $ is either in $ \Capt(p_1) $ or in
        $ \Capt(p_2) $. Assume $ x \in \Capt(p_i) $; then, $
            x (\Matching{v / p}) = x (\Matching{v_i / p_i})
        $ and $
            (\SPatternEnv{t // p})(x) = (\SPatternEnv{\SPiIth(t) // p_i})(x)
        $. We apply the induction hypothesis to conclude.

    \Case[$ p = \Tagg(p) $]
        Analogous to the previous case.

    \Case[$ p = \PAnd{p_1 \& p_2} $]
        Each $ x \in \Capt(p) $ is either in $ \Capt(p_1) $ or in
        $ \Capt(p_2) $. Assume $ x \in \Capt(p_i) $; then, we can directly
        apply the induction hypothesis since $
            t \leq \SAcc{\PAnd{p_1 \& p_2}}
        $ implies $ t \leq \SAcc{p_1} $ and $ t \leq \SAcc{p_2} $.

    \Case[$ p = \POr{p_1 | p_2} $]
        Either $ \Matching{v / p} = \Matching{v / p_1} $ or $
            \Matching{v / p} = \Matching{v / p_2} $
        (in which case $ \Matching{v / p_1} = \MatchFail $).

    \begin{LCases}
        \Case[$ \Matching{v / p} = \Matching{v / p_1} $]
            By Lemma~\ref{lem:a-s-patterns-accepted} we have $
                \STyping{\Gamma |- v: \SAcc{p_1}}
            $; by Lemma~\ref{lem:a-s-patterns-intersection} we have $
                \STyping{\Gamma |- v: t \land \SAcc{p_1}}
            $. Since $ t \land \SAcc{p_1} \leq \SAcc{p_1} $,
            by the induction hypothesis we have, for all $
                x \in \Capt(p_1) = \Capt(p)
            $, $
                \STyping{\Gamma |- x (\Matching{v / p}):
                    (\SPatternEnv{t \land \SAcc{p_1} // p_1})(x)}
            $ and, by subsumption, $
                \STyping{\Gamma |- x (\Matching{v / p}):
                    (\SPatternEnv{t \land \SAcc{p_1} // p_1})(x) \lor
                    (\SPatternEnv{t \setminus \SAcc{p_1} // p_2})(x)
                }
            $.

        \Case[$ \Matching{v / p} = \Matching{v / p_2} $]
            By Lemma~\ref{lem:a-s-patterns-accepted}
            and Lemma~\ref{lem:a-s-patterns-intersection}, we have $
                \STyping{\Gamma |- v: t \setminus \SAcc{p_1}}
            $. Additionally, $
                t \setminus \SAcc{p_1} \leq \SAcc{p_2}
            $ holds because it is equivalent to $
                t \leq \SAcc{p_1} \lor \SAcc{p_2}
            $. Therefore by the induction hypothesis we have, for all $
                x \in \Capt(p_1) = \Capt(p)
            $, $
                \STyping{\Gamma |- x (\Matching{v / p}):
                    (\SPatternEnv{t \setminus \SAcc{p_1} // p_2})(x)}
            $ and, by subsumption, $
                \STyping{\Gamma |- x (\Matching{v / p}):
                    (\SPatternEnv{t \land \SAcc{p_1} // p_1})(x) \lor
                    (\SPatternEnv{t \setminus \SAcc{p_1} // p_2})(x)}
            $.
    \qedhere
    \end{LCases}
    \end{LCases}
\end{Proof}

\begin{Lemma}
\label{lem:a-s-patterns-typesubst}
    Let $ p $ be a pattern, $ t $ a type such that $ t \leq \SAcc{p} $,
    and $ \theta $ a type substitution.
    Then, for all $ x \in \Capt(p) $, $
        (\SPatternEnv{t \theta // p})(x) \leq
        ((\SPatternEnv{t // p}) (x)) \theta
    $.
\end{Lemma}

\begin{Proof}
    By structural induction on $ p $.

    \begin{LCases}
    \Cases[$ p = \Wildcard $ and $ p = c $]
        There is nothing to prove since $ \Capt(p) = \varnothing $.

    \Case[$ p = x $]
        We must prove $
            (\SPatternEnv{t \theta // x})(x) \leq
            (\SPatternEnv{t // x})(x) \theta
        $, which is $ t \theta \leq t \theta $.

    \Case[$ p = (p_1, p_2) $]
        Each $ x \in \Capt(p) $ is either in $ \Capt(p_1) $ or in
        $ \Capt(p_2) $. Assume $ x \in \Capt(p_i) $; then, $
            (\SPatternEnv{t \theta // p})(x) =
            (\SPatternEnv{\SPiIth(t \theta) // p_i})(x)
        $ and $
            (\SPatternEnv{t // p})(x) \theta =
            (\SPatternEnv{\SPiIth(t) // p_i})(x) \theta
        $.

        Since $ \SPiIth(t \theta) \leq \SPiIth(t) \theta $,
        by Lemma~\ref{lem:a-s-patterns-subtyping} we have $
            (\SPatternEnv{\SPiIth(t \theta) // p_i})(x) \leq
            (\SPatternEnv{\SPiIth(t) \theta // p_i})(x)
        $. By the induction hypothesis we have $
            (\SPatternEnv{\SPiIth(t) \theta // p_i})(x) \leq
            (\SPatternEnv{\SPiIth(t) // p_i})(x) \theta
        $.

    \Case[$ p = \Tagg(p) $]
        Analogous to the previous case,
        since $ \SPiTagg(t \theta) \leq \SPiTagg(t) \theta $.

    \Case[$ p = \PAnd{p_1 \& p_2} $]
        Each $ x \in \Capt(p) $ is either in $ \Capt(p_1) $ or in
        $ \Capt(p_2) $. Assume $ x \in \Capt(p_i) $; then, $
            (\SPatternEnv{t \theta // p})(x) =
            (\SPatternEnv{t \theta // p_i})(x)
        $ and $
            (\SPatternEnv{t // p})(x) \theta =
            (\SPatternEnv{t // p_i})(x) \theta
        $. We conclude by the induction hypothesis.

    \Case[$ p = \POr{p_1 | p_2} $]
        Every $ x \in \Capt(p) $ is both in $ \Capt(p_1) $ and in
        $ \Capt(p_2) $. We have $
            (\SPatternEnv{t \theta // p})(x) =
                (\SPatternEnv{(t \land \SAcc{p_1}) \theta // p_1})(x) \lor
                (\SPatternEnv{(t \setminus \SAcc{p_1}) \theta // p_2})(x)
        $---pattern types are closed, so we can apply $ \theta $ to them
        too---and $
            (\SPatternEnv{t // p})(x) \theta =
                (\SPatternEnv{t \land \SAcc{p_1} // p_1})(x) \theta \lor
                (\SPatternEnv{t \setminus \SAcc{p_1} // p_2})(x) \theta
        $. We conclude by applying the induction hypothesis
        to both members of the union.
    \qedhere
    \end{LCases}
\end{Proof}

\begin{Lemma}
\label{lem:a-s-mvar}
    Let $ t_1 $ and $ t_2 $ be equivalent types ($ t_1 \simeq t_2 $).
    Then, $ \Mvar(t_1) = \Mvar(t_2) $.
\end{Lemma}

\begin{Proof}
    Since subtyping is preserved by type substitutions,
    for every $ \alpha $ we have $
        t_1 \SubstSingle[\Empty/\alpha] \simeq t_2 \SubstSingle[\Empty/\alpha]
    $.
    If $ \alpha \in \Mvar(t_1) $, we have $
        t_1 \SubstSingle[\Empty/\alpha] \nsimeq t_1
    $ by the definition of $ \Mvar $.
    This necessarily implies $
        t_2 \SubstSingle[\Empty/\alpha] \nsimeq t_2
    $, otherwise we would have $
        t_1 \SubstSingle[\Empty/\alpha] \simeq t_1
    $ by transitivity.
\end{Proof}

\begin{Lemma}
\label{lem:a-s-mvar2}
    Let $ t $ be a type and $ \theta $ a type substitution
    such that $ \Dom(\theta) \cap \Mvar(t) = \varnothing $.
    Then $ t \theta \simeq t $.
\end{Lemma}

\begin{Proof}
    Let $ t' = t \SubstDots[\Empty/\alpha_1 ... \Empty/\alpha_n] $
    where $ \Set{\alpha_1, \dots, \alpha_n} = \Var(t) \setminus \Mvar(t) $.
    We have $ t \simeq t' $ and $ \Var(t') = \Mvar(t) $.
    Since substitutions preserve subtyping (and hence equivalence),
    we have also $ t \theta \simeq t' \theta $.
    But $ t' \theta = t' \simeq t $;
    hence, we reach the conclusion by the transitivity of equivalence.
\end{Proof}

\begin{Lemma}
\label{lem:a-s-generalization}
    Let $ \Gamma_1 $, $ \Gamma_2 $ be two type environments
    such that $ \Mvar(\Gamma_1) \subseteq \Mvar(\Gamma_2) $
    and $ t_1 $, $ t_2 $ two types such that $ t_1 \leq t_2 $.
    Then, $ \SMoreGen{ \Gen_{\Gamma_1}(t_1) < \Gen_{\Gamma_2}(t_2) } $.
\end{Lemma}

\begin{Proof}
    An instance of $ \Gen_{\Gamma_2}(t_2) $ is a type $ t_2 \theta_2 $
    such that $ \Dom(\theta_2) \subseteq \Var(t_2) \setminus \Mvar(\Gamma_2) $.
    Let $ \theta_1 $ be the restriction of $ \theta_2 $
    to the variables in $ \Var(t_1) \setminus \Mvar(\Gamma_1) $.
    Then, $ t_1 \theta_1 $ is an instance of $ \Gen_{\Gamma_1}(t_1) $.

    We have $ t_1 \theta_1 = t_1 \theta_2 $
    because the two substitutions differ only on variables
    in $ \Var(t_2) \setminus \Var(t_1) $
    (which do not appear in $ t_1 $ at all)
    or in $ \Mvar(\Gamma_1) \setminus \Mvar(\Gamma_2) $ (which is empty).
    Finally, we have $ t_1 \theta_2 \leq t_2 \theta_2 $
    because subtyping is preserved by substitutions.
\end{Proof}

\begin{Lemma}[Weakening]
\label{lem:a-s-weakening}
    Let $ \Gamma_1 $, $ \Gamma_2 $ be two type environments
    such that $ \SMoreGen{ \Gamma_1 < \Gamma_2 } $
    and $ \Mvar(\Gamma_1) \subseteq \Mvar(\Gamma_2) $.
    If $ \STyping{ \Gamma_2 |- e: t } $,
    then $ \STyping{ \Gamma_1 |- e: t } $.
\end{Lemma}

\begin{Proof}
    By induction on the derivation of $ \STyping{\Gamma_2 |- e: t} $.
    We reason by cases on the last applied rule.

    \begin{LCases}
    \Case[\Rule{Ts-Var}]
        We have
        \[
            \STyping{\Gamma_2 |- x: t} \qquad
            t \in \Inst(\Gamma_2(x))
        \]
        and hence, since $ \SMoreGen{\Gamma_1 < \Gamma_2} $,
        there exists a $ t' \in \Inst(\Gamma_1(x)) $ such that $ t' \leq t $.
        We apply \Rule{Ts-Var} to derive $ \STyping{\Gamma_1 |- x: t'} $
        and \Rule{Ts-Subsum} to conclude.

    \Case[\Rule{Ts-Const}]
        Straightforward.

    \Case[\Rule{Ts-Abstr}]
        We have
        \[
            \STyping{\Gamma_2 |- \Abstr{x. e_1}: t_1 \to t_2}
            \qquad
            \STyping{\Gamma_2, \Set{x\colon t_1} |- e_1: t_2}
            \: .
        \]

        Since $ \SMoreGen{\Gamma_1 < \Gamma_2} $,
        we have $
            \SMoreGen{\Gamma_1, \Set{x\colon t_1} < \Gamma_2, \Set{x\colon t_1}}
        $; since $ \Mvar(\Gamma_1) \subseteq \Mvar(\Gamma_2) $,
        we have $
            \Mvar(\Gamma_1, \Set{x\colon t_1}) \subseteq
            \Mvar(\Gamma_2, \Set{x: t_1})
        $. We derive $ \STyping{\Gamma_1, \Set{x\colon t_1} |- e_1: t_2} $
        by the induction hypothesis and apply \Rule{Ts-Abstr} to conclude.

    \Cases[\Rule{Ts-Appl}, \Rule{Ts-Pair}, \Rule{Ts-Tag}, and \Rule{Ts-Subsum}]
        Straightforward application of the induction hypothesis.

    \Case[\Rule{Tk-Match}]
        We have
        \begin{gather*}
            \STyping{\Gamma_2 |- \Match{e_0 with (p_i \to e_i)_{i \in I}}: t} \\
            \STyping{\Gamma_2 |- e_0: t_0} \qquad
            t_0 \leq \textstyle\bigvee_{i \in I} \SAcc{p_i} \qquad
            t_i = (t_0 \setminus \bigvee_{j < i} \SAcc{p_j}) \land \SAcc{p_i} \\
            \forall i \in I . \enspace
            \STyping{\Gamma_2, \Gen_{\Gamma_2}(\SPatternEnv{t_i // p_i}) |-
                e_i: t_i'} \qquad
            t = \textstyle\bigvee_{i \in I} t_i'
            \enspace .
        \end{gather*}

        By the induction hypothesis,
        we derive $ \STyping{\Gamma_1 |- e_0: t_0} $.

        For any branch, note that $ \Mvar(\Gamma_1) \subseteq \Mvar(\Gamma_2) $
        implies $ \SMoreGen{\Gen_{\Gamma_1}(t) < \Gen_{\Gamma_2}(t)} $
        for any $ t $ by Lemma~\ref{lem:a-s-generalization}.
        Hence, we have $
            \SMoreGen{\Gamma_1, \Gen_{\Gamma_1}(\SPatternEnv{t_i // p_i}) <
                \Gamma_2, \Gen_{\Gamma_2}(\SPatternEnv{t_i // p_i})}
        $.
        Additionally, since $
            \Mvar(\Gen_{\Gamma_1}(\SPatternEnv{t_i // p_i})) \subseteq
            \Mvar(\Gamma_1) \subseteq \Mvar(\Gamma_2)
        $, we have $
            \Mvar(\Gamma_1, \Gen_{\Gamma_1}(\SPatternEnv{t_i // p_i})) \subseteq
            \Mvar(\Gamma_2, \Gen_{\Gamma_2}(\SPatternEnv{t_i // p_i}))
        $.

        Hence we may apply the induction hypothesis for all $ i $
        to derive $
            \STyping{\Gamma_1, \Gen_{\Gamma_1}(\SPatternEnv{t_i // p_i}) |-
                e_i: t_i'}
        $ and then apply \Rule{Ts-Match} to conclude.
    \qedhere
    \end{LCases}
\end{Proof}

\begin{Lemma}[Stability of typing under type substitutions]
\label{lem:a-s-typesubst}
    Let $ \theta $ be a type substitution.
    If $ \STyping{ \Gamma |- e: t } $,
    then $ \STyping{ \Gamma \theta |- e: t \theta } $.
\end{Lemma}

\begin{Proof}
    By induction on the derivation of $ \STyping{\Gamma |- e: t} $.
    We reason by cases on the last applied rule.

    \begin{LCases}
    \Case[\Rule{Ts-Var}]
        We have
        \begin{gather*}
            \STyping{\Gamma |- x: t} \qquad
            t \in \Inst(\Gamma(x)) \qquad
            \Gamma(x) = \Scheme{A. t_x} \qquad
            t = t_x \theta_x \qquad
            \Dom(\theta_x) \subseteq A
        \end{gather*}
        and must show
        $
            \STyping{\Gamma \theta |- x: t \theta}
        $.

        By \textalpha-renaming we assume $
            A \cap (\Dom(\theta) \cup \Var(\theta)) = \varnothing $.
        Under this assumption, $ (\Gamma \theta)(x) = \Scheme{A. t_x \theta} $.
        We must show that $ t \theta = t_x \theta \theta_x' $
        for a substitution $ \theta_x' $
        such that $ \Dom(\theta_x') \subseteq A $.

        Let $
            \theta_x' = \SubstC[\alpha \theta_x \theta / \alpha | \alpha \in A]
        $. We show that $
            t \theta \theta_x' = t_x \theta_x \theta = t \theta
        $, by showing that, for every $ \alpha $, $
            \alpha \theta \theta_x' = \alpha \theta_x \theta
        $. If $ \alpha \in A $, then $
            \alpha \theta \theta_x' = \alpha \theta_x' =
            \alpha \theta_x \theta
        $ ($ \theta $ is not defined on the variables in $ A $).
        If $ \alpha \notin A $, then $
            \alpha \theta \theta_x' = \alpha \theta
        $ ($ \theta $ never produces any variable in $ A $)
        and $ \alpha \theta_x \theta = \alpha \theta $
        as $ \alpha \notin \Dom(\theta_x) $.

    \Case[\Rule{Ts-Const}]
        Straightforward.

    \Case[\Rule{Ts-Abstr}]
        We have
        \[
            \STyping{\Gamma |- \Abstr{x. e_1}: t_1 \to t_2} \qquad
            \STyping{\Gamma, \Set{x\colon t_1} |- e_1: t_2}
            \: .
        \]

        By the induction hypothesis we have $
            \STyping{\Gamma \theta, \Set{x\colon t_1 \theta} |- e_1: t_2 \theta}
        $.
        Then by \Rule{Ts-Abstr} we derive $
            \STyping{\Gamma \theta |- \Abstr{x. e_1}:
                (t_1 \theta) \to (t_2 \theta)}
        $, which is $
            \STyping{\Gamma \theta |- \Abstr{x. e_1}: (t_1 \to t_2) \theta}
        $.

    \Cases[\Rule{Ts-Appl}, \Rule{Ts-Pair}, and \Rule{Ts-Tag}]
        Straightforward application of the induction hypothesis.

    \Case[\Rule{Ts-Match}]
        We have
        \begin{gather*}
            \STyping{\Gamma |- \Match{e_0 with (p_i \to e_i)_{i \in I}}: t} \\
            \STyping{\Gamma |- e_0: t_0} \qquad
            t_0 \leq \textstyle\bigvee_{i \in I} \SAcc{p_i} \qquad
            t_i = (t_0 \setminus \bigvee_{j < i} \SAcc{p_j}) \land \SAcc{p_i} \\
            \forall i \in I . \enspace
            \STyping{\Gamma, \Gen_\Gamma(\SPatternEnv{t_i // p_i}) |- e_i: t_i'}
            \qquad
            t = \textstyle\bigvee_{i \in I} t_i'
        \end{gather*}
        and must show
        $
            \STyping{\Gamma \theta |- \Match{e_0 with (p_i \to e_i)_{i \in I}}:
                t \theta}
        $.

        We prove it by establishing, for some types $ \hat{t_0} $
        and $ \hat{t_i} $, $ \hat{t_i'} $ for each $ i $, that
        \begin{gather*}
            \STyping{\Gamma \theta |- e_0: \hat{t_0}} \qquad
            \hat{t_0} \leq \textstyle\bigvee_{i \in I} \SAcc{p_i} \qquad
            \hat{t_i} = (\hat{t_0} \setminus
                              \bigvee_{j < i} \SAcc{p_j}) \land
                              \SAcc{p_i} \\
            \forall i \in I . \enspace
            \STyping{\Gamma \theta,
                \Gen_{\Gamma \theta}(\SPatternEnv{\hat{t_i} // p_i}) |-
                e_i: \hat{t_i'}} \qquad
            \textstyle\bigvee_{i \in I} \hat{t_i'} \leq t \theta
            \enspace .
        \end{gather*}

        Let $
            A = \Set{\alpha_1, \dots, \alpha_n} =
            \Var(t_0) \setminus \Mvar(\Gamma)
        $.
        Let $ B = \Set{\beta_1, \dots, \beta_n} $ be a set of type variables
        such that $
            B \cap (\Dom(\theta) \cup \Var(\theta) \cup \Var(\Gamma)) =
            \varnothing
        $.
        Let $ \theta_0 = \SubstDots[\beta_1/\alpha_1 ... \beta_n/\alpha_n] $ and $ \theta' = \theta \circ \theta_0 $.

        By the induction hypothesis, using $ \theta' $,
        we derive $ \STyping{\Gamma \theta' |- e_0: t_0 \theta'} $.
        From it, we derive $ \STyping{\Gamma \theta |- e_0: t_0 \theta'} $
        by weakening (Lemma~\ref{lem:a-s-weakening});
        we prove the required premises below.
        We take $ \hat{t_0} = t_0 \theta' $:
        note that the exhaustiveness condition is satisfied
        because substitutions preserve subtyping
        (and all accepted types of patterns are closed).
        We have $ \hat{t_i} = t_i \theta' $ for all $ i $.

        For all branches, we have $
            \STyping{\Gamma, \Gen_\Gamma(\SPatternEnv{t_i // p_i}) |- e_i: t_i'}
        $ and, by the induction hypothesis using $ \theta $, we can derive $
            \STyping{\Gamma \theta,
                (\Gen_\Gamma(\SPatternEnv{t_i // p_i})) \theta |- e_i:
                t_i' \theta}
        $.

        We apply Lemma~\ref{lem:a-s-weakening}
        to derive $
            \STyping{\Gamma \theta,
                \Gen_{\Gamma \theta}(\SPatternEnv{t_i \theta' // p_i}) |- e_i:
                t_i' \theta}
        $ (we prove the required premises below).
        We take $ \hat{t_i'} = t_i' \theta $.

    \begin{LCases}
        \Item{Proof of $ \STyping{\Gamma \theta |- e_0: t_0 \theta'} $
            from $ \STyping{\Gamma \theta' |- e_0: t_0 \theta'} $}
            We prove this by Lemma~\ref{lem:a-s-weakening},
            which requires us to show
            $ \SMoreGen{\Gamma \theta < \Gamma \theta'} $
            and $ \Mvar(\Gamma \theta) \subseteq \Mvar(\Gamma \theta') $.
            We show this by showing,
            for every $ (x \colon \Scheme{A_x. t_x}) \in \Gamma $---assume
            by \textalpha-renaming $
                A_x \cap (\Dom(\theta) \cup \Var(\theta) \cup A \cup B) =
                \varnothing
            $---,
            $ t_x \theta \simeq t_x \theta' $,
            which implies both $
                \SMoreGen{\Scheme{A_x. t_x \theta} <
                \Scheme{A_x. t_x \theta'}}
            $ and $
                \Mvar(\Scheme{A_x. t_x \theta}) \subseteq
                \Mvar(\Scheme{A_x. t_x \theta'})
            $ (by Lemma~\ref{lem:a-s-generalization}
            and Lemma~\ref{lem:a-s-mvar}).

            We have $ t_x \theta_0 \simeq t_x $ by Lemma~\ref{lem:a-s-mvar2}:
            $ \Dom(\theta_0) \cap \Mvar(t_x) = \varnothing $
            because every $ \alpha \in \Mvar(t_x) $
            is either in $ A_x $ or $ \Mvar(\Gamma) $,
            and in both cases this means it cannot be in $ \Dom(\theta_0) $.
            Hence---since substitutions preserve subtyping---we have also
            $ t_x \theta' = t_x \theta_0 \theta \simeq t_x \theta $.

        \Item{Proof of $
                \STyping{\Gamma \theta,
                \Gen_{\Gamma \theta}(\SPatternEnv{t_i \theta' // p_i}) |- e_i:
                t_i' \theta}
            $ from $
                \STyping{\Gamma \theta,
                (\Gen_\Gamma(\SPatternEnv{t_i // p_i})) \theta |- e_i:
                t_i' \theta}
            $}
            To apply Lemma~\ref{lem:a-s-weakening},
            we must show
            \begin{gather*}
                \SMoreGen{
                    \Gen_{\Gamma \theta}(\SPatternEnv{t_i \theta' // p_i}) <
                    (\Gen_\Gamma(\SPatternEnv{t_i // p_i})) \theta}
                \\
                \Mvar(
                    \Gamma \theta,
                    \Gen_{\Gamma \theta}(\SPatternEnv{t_i \theta' // p_i})
                ) \subseteq \Mvar(
                    \Gamma \theta,
                    (\Gen_\Gamma(\SPatternEnv{t_i // p_i})) \theta
                )
                \: .
            \end{gather*}

            The latter holds because every variable in $
                \Mvar(\Gamma \theta,
                    \Gen_{\Gamma \theta}(\SPatternEnv{t_i \theta' // p_i}))
            $ is in $ \Mvar(\Gamma \theta) $.

            For the former, we prove that, for every $ x \in \Capt(p_i) $,
            \[
                \SMoreGen{
                    \Gen_{\Gamma \theta}((\SPatternEnv{t_i \theta' // p_i})(x))
                    <
                    (\Gen_\Gamma((\SPatternEnv{t_i // p_i})(x))) \theta}
                \: .
            \]

            Let
            \[
                t_x' = (\SPatternEnv{t_i \theta' // p_i})(x) \qquad
                t_x = (\SPatternEnv{t_i // p_i})(x)
                \: ;
            \]
            the statement becomes
            \[
                \SMoreGen{
                    \Gen_{\Gamma \theta}(t_x') < (\Gen_\Gamma(t_x)) \theta}
                \: .
            \]

            We have $ \Gen_\Gamma(t_x) = \Scheme{A_x. t_x} $,
            where $ A_x = \Var(t_x) \setminus \Mvar(\Gamma) $.
            Since $ \Var(t_x) \subseteq \Var(t_i) = \Var(t_0) $,
            $ A_x \subseteq A $.
            Let $ J = \SetC{j | \alpha_j \in A_x} $;
            thus $ J \subseteq \Set{1, \dots, n} $
            and $ A_x = A|_J = \SetC{\alpha_j | j \in J} $.
            Let $ B|_J = \SetC{\beta_j | j \in J} $.
            We have $ \Gen_\Gamma(t_x) = \Scheme{B|_J. t_x \theta_0} $
            by \textalpha-renaming
            (we are substituting also the $ \alpha_i $ such that
            $ i \notin J $, but it makes no difference as they not in $ t_x $).
            Thus---since $
                B \cap (\Dom(\theta) \cup \Var(\theta)) = \varnothing
            $---we have
            \[
                (\Gen_\Gamma(t_x)) \theta =
                \Scheme{B|_J. t_x \theta_0 \theta} = \Scheme{B|_J. t_x \theta'}
                \: .
            \]

            The instances of this type scheme are all types
            $ t_x \theta' \theta_x $, with $ \Dom(\theta_x) \subseteq B|_J $.
            Given such a type, we must construct an instance of $
                \Gen_{\Gamma \theta}(t_x')
            $ that is a subtype of it.
            Let $ \theta_x' $ be the restriction of $ \theta_x $
            to variables in $ \Var(t_x') \setminus \Mvar(\Gamma \theta) $.
            Then $
                t_x' \theta_x'
            $ is a valid instance of $ \Gen_{\Gamma \theta}(t_x') $.
            We prove $
                t_x' \theta_x' \leq
                t_x \theta' \theta_x
            $.

            We have $
                t_x' \theta_x' =
                t_x' \theta_x
            $: the two substitutions differ only on variables in $
                B|_J \setminus \Var(t_x')
            $ (variables which do not appear in the type at all)
            and on variables in $
                B|_J \cap \Mvar(\Gamma \theta)
            $ (which is empty, because $ B $ was chosen fresh).
            By Lemma~\ref{lem:a-s-patterns-subtyping},
            we have $
                t_x' = (\SPatternEnv{t_i \theta' // p_i})(x) \leq
                (\SPatternEnv{t_i // p_i})(x) \theta'
            $: hence, $
                t_x' \theta_x =
                    (\SPatternEnv{t_i \theta' // p_i})(x) \theta_x \leq
                (\SPatternEnv{t_i // p_i})(x) \theta' \theta_x
                    = t_x \theta' \theta_x
            $.
    \end{LCases}

    \Case[\Rule{Ts-Subsum}]
        The conclusion follows from the induction hypothesis
        since substitutions preserve subtyping.
        \qedhere
    \end{LCases}
\end{Proof}

\begin{Corollary}
\label{cor:a-s-typesubst2}
    Let $ \Gamma $ be a type environment and $ \theta $ a type substitution
    such that $ \Dom(\theta) \cap \Mvar(\Gamma) = \varnothing $.
    If $ \STyping{ \Gamma |- e: t } $,
    then $ \STyping{ \Gamma |- e: t \theta } $.
\end{Corollary}

\begin{Proof}
    From $ \STyping{ \Gamma |- e: t } $ we derive
    $ \STyping{ \Gamma \theta |- e: t \theta } $
    by Lemma~\ref{lem:a-s-typesubst}.
    Then, we show $ \SMoreGen{ \Gamma < \Gamma \theta } $
    and $ \Mvar(\Gamma) \subseteq \Mvar(\Gamma \theta) $,
    which allow us to apply Lemma~\ref{lem:a-s-weakening} to derive
    $ \STyping{ \Gamma |- e: t \theta } $.

    To show the two conditions above, we show that,
    for every $ (x \colon \Scheme{A. t}) \in \Gamma $---assume
    by \textalpha-renaming $
        A \cap (\Dom(\theta) \cup \Var(\theta)) = \varnothing
    $---,
    $ \SMoreGen{\Scheme{A. t} < \Scheme{A. t \theta}} $
    and $ \Mvar(\Scheme{A. t}) \subseteq \Mvar(\Scheme{A. t \theta}) $.

    We show $ t \simeq t \theta $, which implies both
    (by Lemma~\ref{lem:a-s-generalization}
    and Lemma~\ref{lem:a-s-mvar}).
    The equivalence holds by Lemma~\ref{lem:a-s-mvar2}:
    $ \Dom(\theta) \cap \Mvar(t) = \varnothing $
    because every $ \alpha \in \Mvar(t) $
    is either in $ A $ or $ \Mvar(\Gamma) $,
    and in both cases this means it cannot be in $ \Dom(\theta) $.
\end{Proof}

\begin{Lemma}[Expression substitution]
\label{lem:a-s-exprsubst}
    Let $ x_1 $, \dots, $ x_n $ be distinct variables
    and $ v_1 $, \dots, $ v_n $ values.
    Let $ \Gamma' = \Set{x_1\colon s_1, \dots, x_n\colon s_n} $
    and $ \varsigma = \SubstDots[v_1/x_1 ... v_n/x_n] $.

    If $ \STyping{ \Gamma, \Gamma' |- e: t } $ and,
    for all $ k \in \Set{1, \dots, n} $ and for all $ t_k \in \Inst(s_k) $,
    $ \STyping{ \Gamma |- v_k: t_k } $,
    then $ \STyping{ \Gamma |- e \varsigma: t } $.
\end{Lemma}

\begin{Proof}
    By induction on the derivation of $ \STyping{\Gamma, \Gamma' |- e: t} $.
    We reason by cases on the last applied rule.

    \begin{LCases}
    \Case[\Rule{Ts-Var}]
        We have
        \[
            \STyping{\Gamma, \Gamma' |- x: t} \qquad
            t \in \Inst((\Gamma, \Gamma')(x))
            \: .
        \]

        Either $ x = x_k $ for some $ k $ or not.
        In the latter case, $ x \varsigma = x $, $ x \notin \Dom(\Gamma') $
        and hence $ (\Gamma, \Gamma')(x) = \Gamma(x) $.
        Then, since $ t \in \Inst((\Gamma, \Gamma')(x)) $,
        $ t \in \Inst(\Gamma(x)) $ and we can apply \Rule{Ts-Var}.

        If $ x = x_k $, then $ (\Gamma, \Gamma')(x) = \Gamma'(x) = s_k $.
        We must then prove $ \STyping{\Gamma |- v_k: t} $,
        which we know by hypothesis since $ t \in \Inst(s_k) $.

    \Case[\Rule{Ts-Const}]
        Straightforward.

    \Case[\Rule{Ts-Abstr}]
        We have
        \[
            \STyping{\Gamma, \Gamma' |- \Abstr{x. e_1}: t_1 \to t_2} \qquad
            \STyping{\Gamma, \Gamma', \Set{x\colon t_1} |- e_1: t_2}
            \: .
        \]

        By \textalpha-renaming we can assume $ x \notin \Dom(\Gamma, \Gamma') $;
        then $ (\Abstr{x. e_1}) \varsigma = \Abstr{x. (e_1 \varsigma)} $
        and $
            \Gamma, \Gamma', \Set{x\colon t_1} =
            \Gamma, \Set{x\colon t_1}, \Gamma'
        $. Therefore we have $
            \STyping{\Gamma, \Set{x\colon t_1}, \Gamma' |- e_1: t_2}
        $ and hence $
            \STyping{\Gamma, \Set{x\colon t_1} |- e_1 \varsigma: t_2}
        $ by the induction hypothesis. We apply \Rule{Ts-Abstr} to conclude.

    \Cases[\Rule{Ts-Appl}, \Rule{Ts-Pair}, \Rule{Ts-Tag}, and \Rule{Ts-Subsum}]
        Straightforward application of the induction hypothesis.

    \Case[\Rule{Ts-Match}]
        We have
        \begin{gather*}
            \STyping{\Gamma, \Gamma' |-
                 \Match{e_0 with (p_i \to e_i)_{i \in I}}: t} \\
            \STyping{\Gamma, \Gamma' |- e_0: t_0} \qquad
            t_0 \leq \textstyle\bigvee_{i \in I} \SAcc{p_i} \qquad
            t_i = (t_0 \setminus \bigvee_{j < i} \SAcc{p_j}) \land
                  \SAcc{p_i} \\
            \forall i \in I . \enspace
            \STyping{\Gamma, \Gamma',
                \Gen_{\Gamma,\Gamma'}(\SPatternEnv{t_i // p_i}) |- e_i: t_i'}
            \qquad
            t = \textstyle\bigvee_{i \in I} t_i'
            \enspace .
        \end{gather*}

        We assume by \textalpha-renaming
        that no capture variable of any pattern is in the domain
        of $ \Gamma $ or $ \Gamma' $.
        Then, $
            (\Match{e_0 with (p_i \to e_i)_{i \in I}}) \varsigma =
            \Match{e_0 \varsigma{} with (p_i \to e_i \varsigma)_{i \in I}}
        $ and $
            \Gamma, \Gamma', \Gen_{\Gamma,\Gamma'}(\SPatternEnv{t_i // p_i}) =
            \Gamma, \Gen_{\Gamma,\Gamma'}(\SPatternEnv{t_i // p_i}), \Gamma'
        $ for any $ i $.

        By the induction hypothesis,
        we derive $ \STyping{\Gamma |- e_0 \varsigma: t_0} $ and $
            \STyping{\Gamma, \Gen_{\Gamma,\Gamma'}(\SPatternEnv{t_i // p_i}) |-
                e_i \varsigma: t_i'}
        $ for all $ i $.
        From the latter, we prove $
            \STyping{\Gamma, \Gen_\Gamma(\SPatternEnv{t_i // p_i}) |-
                e_i \varsigma: t_i'}
        $ by weakening (Lemma~\ref{lem:a-s-weakening}):
        we have $
            \SMoreGen{\Gen_\Gamma(\SPatternEnv{t_i // p_i}) <
                \Gen_{\Gamma,\Gamma'}(\SPatternEnv{t_i // p_i})} $
        by Lemma~\ref{lem:a-s-generalization}---since $
            \Mvar(\Gamma) \subseteq \Mvar(\Gamma,\Gamma')
        $ – and clearly we have $
            \Mvar(\Gamma, \Gen_\Gamma(\SPatternEnv{t_i // p_i})) \subseteq
            \Mvar(\Gamma, \Gen_{\Gamma,\Gamma'}(\SPatternEnv{t_i // p_i}))
        $ since $
            \Mvar(\Gen_\Gamma(\SPatternEnv{t_i // p_i})) \subseteq \Mvar(\Gamma)
        $.
    \qedhere
    \end{LCases}
\end{Proof}

\begin{Theorem}[Progress]
\label{thm:a-s-progress}
    Let $ e $ be a well-typed, closed expression
    (i.e., $ \STyping{ \varnothing |- e: t } $ holds for some $ t $).
    Then, either $ e $ is a value
    or there exists an expression $ e' $ such that $ \Smallstep{ e ~> e' } $.
\end{Theorem}

\begin{Proof}
By hypothesis we have $ \STyping{\EmptyEnv |- e: t} $.
The proof is by induction on its derivation;
we reason by cases on the last applied rule.

    \begin{LCases}
    \Case[\Rule{Ts-Var}]
        This case does not occur because variables are not closed.

    \Case[\Rule{Ts-Const}]
        In this case $ e $ is a constant $ c $ and therefore a value.

    \Case[\Rule{Ts-Abstr}]
        In this case $ e $ is an abstraction $ \Abstr{x. e_1} $.
        Since it is also closed, it is a value.

    \Case[\Rule{Ts-Appl}]
        We have
        \[
            \STyping{ \EmptyEnv |- \Appl{e_1}{e_2}: t} \qquad
            \STyping{ \EmptyEnv |- e_1: t' \to t} \qquad
            \STyping{ \EmptyEnv |- e_2: t'}
            \: .
        \]

        By the induction hypothesis, each of $ e_1 $ and $ e_2 $
        either is a value or may reduce.
        If $ \Smallstep{e_1 ~> e_1'} $,
        then $ \Smallstep{\Appl{e_1}{e_2} ~> \Appl{e_1'}{e_2}} $.
        If $ e_1 $ is a value and $ \Smallstep{e_2 ~> e_2'} $,
        then $ \Smallstep{\Appl{e_1}{e_2} ~> \Appl{e_1}{e_2'} } $.

        If both are values then, by Lemma~\ref{lem:a-s-generation},
        $ e_1 $ has the form $ \Abstr{x. e_3} $ for some $ e_3 $.
        Then, we can apply \Rule{R-Appl}
        and $ \Smallstep{\Appl{e_1}{e_2} ~> e_3 \SubstSingle[e_2/x]} $.

    \Case[\Rule{Ts-Pair}]
        We have
        \[
            \STyping{\EmptyEnv |- (e_1, e_2): t_1 \times t_2} \qquad
            \STyping{\EmptyEnv |- e_1: t_1} \qquad
            \STyping{\EmptyEnv |- e_2: t_2}
            \: .
        \]

        By the induction hypothesis,
        each of $ e_1 $ and $ e_2 $ either is a value or may reduce.
        If $ \Smallstep{e_1 ~> e_1'} $,
        then $ \Smallstep{(e_1, e_2) ~> (e_1', e_2)} $.
        If $ e_1 $ is a value and $ \Smallstep{e_2 ~> e_2'} $,
        then $ \Smallstep{(e_1, e_2) ~> (e_1, e_2')} $.
        If both are values, then $ (e_1, e_2) $ is also a value.

    \Case[\Rule{Ts-Tag}]
        We have
        \[
            \STyping{\EmptyEnv |- \Tagg(e_1): \Tagg(t_1)} \qquad
            \STyping{\EmptyEnv |- e_1: t_1}
            \: .
        \]

        Analogously to the previous case, by the induction hypothesis we have
        that either $ e_1 $ is a value or $ \Smallstep{e_1 ~> e_1'} $.
        In the former case, $ \Tagg(e_1) $ is a value as well.
        In the latter, we have $ \Smallstep{\Tagg(e_1) ~> \Tagg(e_1')} $.

    \Case[\Rule{Ts-Match}]
        We have
        \[
            \STyping{\EmptyEnv |-
                \Match{e_0 with (p_i \to e_i)_{i \in I}}: t} \qquad
            \STyping{\EmptyEnv |- e_0: t_0} \qquad
            t_0 \leq \textstyle\bigvee_{i \in I} \SAcc{p_i}
            \: .
        \]

        By the inductive hypothesis,
        either $ e_0 $ is a value or it may reduce.
        In the latter case, if $ \Smallstep{e_0 ~> e_0'} $, then $
            \Smallstep{\Match{e_0 with (p_i \to e_i)_{i \in I}} ~>
                \Match{e_0' with (p_i \to e_i)_{i \in I}}} $.

        If $ e_0 $ is a value, on the other hand,
        the expression may reduce by application of \Rule{R-Match}.
        Since $ t_0 \leq \textstyle\bigvee_{i \in I} \SAcc{p_i} $,
        $ \STyping{\EmptyEnv |- e_0: \bigvee_{i \in I} \SAcc{p_i}} $
        holds by subsumption.
        Hence, since $ e_0 $ is a value,
        $ \STyping{\EmptyEnv |- e_0: \SAcc{p_i}} $ holds for at least one $ i $
        (by Lemma~\ref{lem:a-s-patterns-union});
        for each such $ i $ we have $ \Matching{e_0 / p_i} = \varsigma_i $
        (by Lemma~\ref{lem:a-s-patterns-accepted}).
        Let $ j $ be the least of these $ i $;
        then $
            \Smallstep{ \Match{e_0 with (p_i \to e_i)_{i \in I}} ~>
                e_j \varsigma_j}
        $.

        \Case[\Rule{Ts-Subsum}]
            Straightforward application of the induction hypothesis.
        \qedhere
    \end{LCases}
\end{Proof}

\begin{Theorem}[Subject reduction]
\label{thm:a-s-subject}
    Let $ e $ be an expression
    and $ t $ a type such that $ \STyping{ \Gamma |- e: t } $.
    If $ \Smallstep{ e ~> e' } $, then $ \STyping{ \Gamma |- e': t } $.
\end{Theorem}

\begin{Proof}
    By induction on the derivation of $ \STyping{ \Gamma |- e: t} $.
    We reason by cases on the last applied rule.

    \begin{LCases}
    \Cases[\Rule{Ts-Var}, \Rule{Ts-Const}, and \Rule{Ts-Abstr}]
        These cases do not occur:
        variables, constants, and abstractions never reduce.

    \Case[\Rule{Ts-Appl}]
        We have
        \[
            \STyping{\Gamma |- \Appl{e_1}{e_2}: t} \qquad
            \STyping{\Gamma |- e_1: t' \to t} \qquad
            \STyping{\Gamma |- e_2: t'}
            \: .
        \]

        $ \Smallstep{\Appl{e_1}{e_2} ~> e'} $ occurs in any of three ways:
        $(i)$ $ \Smallstep{e_1 ~> e_1'} $ and $ e' = \Appl{e_1'}{e_2} $;
        $(ii)$ $ e_1 $ is a value, $ \Smallstep{e_2 ~> e_2'} $
            and $ e' = \Appl{e_1}{e_2'} $;
        $(iii)$ both $ e_1 $ and $ e_2 $ are values,
            $ e_1 $ is of the form $ \Abstr{x. e_3} $,
            and $ e' = e_3 \SubstSingle[e_2/x] $.

        In the first case, we derive by the induction hypothesis
        that $ \STyping{\Gamma |- e_1': t' \to t} $
        and conclude by applying \Rule{Ts-Appl} again.
        The second case is analogous.

        In the third case,
        we know by Lemma~\ref{lem:a-s-generation}
        that $ \STyping{\Gamma, \Set{x\colon t'} |- e_3: t} $.
        We also know that $ e_2 $ is a value
        such that $ \STyping{\Gamma |- e_2: t'} $.
        Then, by Lemma~\ref{lem:a-s-exprsubst},
        $ \STyping{\Gamma |- e_3 \SubstSingle[e_2/x]: t} $.

    \Case[\Rule{Ts-Pair}]
        We have
        \[
            \STyping{\Gamma |- (e_1, e_2): t_1 \times t_2} \qquad
            \STyping{\Gamma |- e_1: t_1} \qquad
            \STyping{\Gamma |- e_2: t_2}
            \: .
        \]

        $ \Smallstep{(e_1, e_2) ~> e'} $ occurs
        either because $ \Smallstep{e_1 ~> e_1'} $ and $ e' = (e_1', e_2) $,
        or because $ e_1 $ is a value, $ \Smallstep{e_2 ~> e_2'} $,
        and $ e' = (e_1, e_2') $.
        In either case, the induction hypothesis allows us to derive
        that the type of the component that reduces is preserved;
        therefore, we can apply \Rule{Ts-Pair} again to conclude.

    \Case[\Rule{Ts-Tag}]
        Analogously to the previous case,
        a variant expression only reduces if its argument does,
        so we apply the induction hypothesis and \Rule{Ts-Tag} to conclude.

    \Case[\Rule{Ts-Match}]
        We have
        \begin{gather*}
            \STyping{\Gamma |- \Match{e_0 with (p_i \to e_i)_{i \in I}}: t} \\
            \STyping{\Gamma |- e_0: t_0} \qquad
            t_0 \leq \textstyle\bigvee_{i \in I} \SAcc{p_i} \qquad
            t_i = (t_0 \setminus \bigvee_{j < i} \SAcc{p_j}) \land
                \SAcc{p_i} \\
            \forall i \in I . \enspace
            \STyping{\Gamma, \Gen_\Gamma(\SPatternEnv{t_i // p_i}) |- e_i: t_i'}
                \qquad
            t = \textstyle\bigvee_{i \in I} t_i'
            \enspace .
        \end{gather*}

        The reduction $
            \Smallstep{\Match{e_0 with (p_i \to e_i)_{i \in I}} ~> e'}
        $ occurs
        either because $ \Smallstep{e_0 ~> e_0'} $
        and $ e' = \Match{e_0' with (p_i \to e_i)_{i \in I}} $
        or because $ e_0 $ is a value and $ e' = e_j \varsigma $,
        where $ \Matching{e_0 / p_j} = \varsigma $
        and, for all $ i < j $,
        $ \Matching{e_0 / p_i} = \MatchFail $.
        In the former case, we apply the induction hypothesis
        and conclude by \Rule{Ts-Match}.

        In the latter case, $ \varsigma $ is a substitution
        from the capture variables of $ p_j $ to values. We can derive
        \[
            \STyping{\Gamma |- e_0: \SAcc{p_j}} \qquad
            \forall i < j . \enspace
            \STyping{\Gamma |- e_0: \lnot \SAcc{p_i}}
        \]
        by Lemma~\ref{lem:a-s-patterns-accepted}
        and thence $ \STyping{\Gamma |- e_0: t_j} $
        by Lemma~\ref{lem:a-s-patterns-intersection}.
        Therefore, by Lemma~\ref{lem:a-s-patterns-correct}, we have that,
        for all $ x \in \Capt(p_j) $,
        $ \STyping{\Gamma |- x \varsigma: (\SPatternEnv{t_j // p_j})(x)} $.
        Let $ \Gamma' = \SPatternEnv{t_j // p_j} $.

        We show that, additionally,
        $ \STyping{\Gamma |- x \varsigma: t_x} $ holds
        for every $ t_x \in \Inst(\Gen_\Gamma(\Gamma'(x))) $.
        Every such $ t_x $ is equal to $ \Gamma'(x) \theta $
        for a $ \theta $ such that $
            \Dom(\theta) \subseteq \Var(\Gamma'(x)) \setminus \Mvar(\Gamma)
        $.
        Then, $ \STyping{\Gamma |- x \varsigma: \Gamma'(x) \theta} $ holds
        by Corollary~\ref{cor:a-s-typesubst2}, since $
            \Dom(\theta) \cap \Mvar(\Gamma) = \varnothing $
        (the substitution does not change any meaningful
        variable of $ \Gamma $).

        From $ \STyping{\Gamma, \Generalize_\Gamma(\Gamma') |- e_j: t_j'} $
        and from the fact that we have $ \STyping{\Gamma |- x \varsigma: t_x} $
        for all $ x \in \Capt(p_j) $
        and all $ t_x \in \Inst(\Gen_\Gamma(\Gamma'(x))) $,
        we derive $ \STyping{\Gamma |- e_j \varsigma: t_j'} $
        by Lemma~\ref{lem:a-s-exprsubst} and then conclude by subsumption.

    \Case[\Rule{Ts-Subsum}]
        Straightforward application of the induction hypothesis.
    \qedhere
    \end{LCases}
\end{Proof}

\begin{Corollary}[Type soundness]
\label{cor:a-s-soundness}
    Let $ e $ be a well-typed, closed expression,
    that is, such that $ \STyping{ \varnothing |- e: t } $ holds for some $ t $.
    Then, either $ e $ diverges or it reduces to a value $ v $
    such that $ \STyping{ \varnothing |- v: t } $.
\end{Corollary}

\begin{Proof}
    Consequence of Theorem~\ref{thm:a-s-progress} and
    Theorem~\ref{thm:a-s-subject}.
\end{Proof}

\subsubsection{Completeness of \VariantsS{} with respect to \VariantsK}

\begin{figure}

    \begin{mathpar}
        \MyInfer[R-Fix]
            { }
            { \Smallstep{ \Fix{} (\Abstr{x. e}) ~>
                e \SubstSingle[{ \Fix{} (\Abstr{x. e}) }/x] } }
            {}
        \\
        \MyInfer[Tk-Fix]
            { \KTyping{ K; \Gamma |- e: \tau \to \tau } }
            { \KTyping{ K; \Gamma |- \Fix e: \tau } }
            {}
        \and
        \MyInfer[Ts-Fix]
            { \STyping{ \Gamma |- e: t \to t } }
            { \STyping{ \Gamma |- \Fix e: t } }
            {}
    \end{mathpar}
    
    \caption{Rules for the fixed-point combinator.}
    \label{fig:proofs-fix}
\end{figure}

In the proof of completeness, we consider a calculus and type systems
extended with the addition of a fixed-point combinator $ \Fix $:
this simplifies the proof
(as it allows us to assume that all arrow types are inhabited)
and it would be desirable anyway in order to use the system in pratice.
We add a new production $ \Fix e $ to the grammar defining expressions,
a new production $ \Fix E $ to the grammar of evaluation contexts,
and the new reduction rule \RuleRef{R-Fix} in Figure~\ref{fig:proofs-fix}.
We extend \VariantsK{} and \VariantsS{} with the addition, respectively,
of the rules \RuleRef{Tk-Fix} and \RuleRef{Ts-Fix}
in Figure~\ref{fig:proofs-fix}.

As mentioned in Section~\ref{sec:variants-s-comparisons},
we prove completeness of \VariantsS{} with respect to \VariantsK{}
using inductive techniques which do not account for the presence
of recursion in kinds: we therefore have to restrict ourselves
to only consider kinding environments which do not feature recursion,
(the \emph{non-recursive} environments defined below).
We conjecture that coinductive techniques could be used instead to prove
the result for general kinding environments.

\begin{Definition}[Non-recursive kinding environments]
\label{def:a-sk-nonrecursive}
    We say that a kinding environment $ K $ is \emph{non-recursive} if,
    for all $ \alpha $ such that $ K(\alpha) = (L, U, T) $,
    we have $ \alpha \notin \bigcup_{\Tagg\colon \tau \in T} \Var_K(\tau) $.
\end{Definition}

\begin{Definition}
\label{def:a-sk-measure}
    We define a function $ w $ which,
    given a \Kin-type $ \tau $ in a non-recursive kinding environment $ K $,
    yields the \emph{measure} $ w(\tau, K) $ of $ \tau $ in $ K $.
    It is defined by the following equations.
    \begin{align*}
        w(\alpha, K) & =
            \begin{cases}
                1 + \sum_{\Tagg\colon \tau \in T} w(\tau, K)
                    & \text{if } K(\alpha) = (L, U, T) \\
                1 & \text{otherwise}
            \end{cases} \\
        w(b, K)
            & = 1 \\
        w(\tau_1 \to \tau_2, K)
            & = w(\tau_1, K) + w(\tau_2, K) + 1 \\
        w(\tau_1 \times \tau_2, K)
            & = w(\tau_1, K) + w(\tau_2, K) + 1
    \end{align*}
\end{Definition}

\begin{Definition}[Translation of types]
    Given a \Kin-type $ \tau $ in a non-recursive kinding environment $ K $,
    its \emph{translation} is the \Sem-type $ \STransl_K[\tau] $
    defined inductively
    by the rules in Figure~\ref{fig:variants-s-translation-app}.

    We define the translation of type schemes as
    $ \STransl_K[\KScheme{A. K' > \tau}] = \Scheme{A. \STransl_K,K'[\tau]} $
    and of type environments by translating each type scheme pointwise.
\end{Definition}

\begin{figure*}

\begin{align*}
  \STransl_K[\alpha]
            & =
            \begin{cases}
                \alpha
                    & \text{if } K(\alpha) = \KUnkind \\
                (\SLow_K(L, T) \lor \alpha) \land \SUpp_K(U, T)
                    & \text{if } K(\alpha) = (L, U, T)
            \end{cases}
        \\
        \STransl_K[b]
            & =
            b \\
        \STransl_K[\tau_1 \to \tau_2]
            & =
            \STransl_K[\tau_1] \to \STransl_K[\tau_2] \\
        \STransl_K[\tau_1 \times \tau_2]
            & =
            \STransl_K[\tau_1] \times \STransl_K[\tau_2]
    \end{align*}
    \begin{align*}
    \text{where:} \\[2mm]
    \SLow_K(L, T)
            & =
            \textstyle\bigvee_{\Tagg \in L}
                \TTagg(\textstyle\bigwedge_{\Tagg\colon \tau \in T}
                \STransl_K[\tau])
        \\
        \SUpp_K(U, T)
            & =
            \begin{cases}
                \textstyle\bigvee_{\Tagg \in U}
                    \TTagg(\textstyle\bigwedge_{\Tagg\colon \tau \in T}
                    \STransl_K[\tau])
                    & \text{if } U \neq \Tags
                \\[5pt]
                \textstyle\bigvee_{\Tagg \in \Domain(T)}
                    \TTagg(\textstyle\bigwedge_{\Tagg\colon \tau \in T}
                    \STransl_K[\tau])
                    \: \lor \:
                    (\AnyVariantType \setminus
                    \textstyle\bigvee_{\Tagg \in \Domain(T)}
                    \TTagg(\AnyType))
                    & \text{if } U = \Tags
            \end{cases}
    \end{align*}

    \caption{Translation of \Kin-types to \Sem-types.}
    \label{fig:variants-s-translation-app}
\end{figure*}

\begin{Lemma}
\label{lem:a-sk-variables}
    For any \Kin-type $ \tau $ in a non-recursive kinding environment $ K $,
    we have $ \Var(\STransl_K[\tau]) \subseteq \Var_K(\tau) $.
    Likewise, for any \Kin-scheme $ \sigma $
    and \Kin-type environment $ \Gamma $,
    we have $ \Var(\STransl_K[\sigma]) \subseteq \Var_K(\sigma) $
    and $ \Var(\STransl_K[\Gamma]) \subseteq \Var_K(\Gamma) $.
\end{Lemma}

\begin{Proof}
    The translation does not introduce new variables, therefore we can show $
        \Var(\STransl_K[\tau]) \subseteq \Var_K(\tau)
    $ by induction on $ w(\tau, K) $.
    We extend this straightforwardly to type schemes and environments.
\end{Proof}

\begin{Lemma}
\label{lem:a-sk-patterns}
    Let $ p $ be a pattern and $ t \leq \SAcc{p} $ an \Sem-type.
    If $ \KPatternTyping{K |- p: \tau => \Gamma} $
    and $ t \leq \STransl_K[\tau] $,
    then, for all $ x \in \Capt(p) $,
    $ (\SPatternEnv{t // p})(x) \leq \STransl_K[\Gamma(x)] $.
\end{Lemma}

\begin{Proof}
    By structural induction on $ p $.

    \begin{LCases}
    \Cases[$ p = \Wildcard $ and $ p = c $]
        There is nothing to prove since $ \Capt(p) = \varnothing $.

    \Case[$ p = x $]
        We have
        \[
            \KPatternTyping{K |- p: \tau => \Set{x\colon \tau}} \qquad
            \SPatternEnv{t // x} = \Set{x\colon t}
        \]
        and must prove $
            \Set{x\colon t}(x) \leq \STransl_K[\Set{x\colon \tau}(x)]
        $, that is, $ t \leq \STransl_K[\tau] $, which is true by hypothesis.

    \Case[$ p = (p_1, p_2) $]
        We have
        \begin{gather*}
            \KPatternTyping{K |- p: \tau_1 \times \tau_2 => \Gamma} \qquad
            \Gamma = \Gamma_1 \cup \Gamma_2 \qquad
            \forall i . \enspace
            \KPatternTyping{K |- p_i: \tau_i => \Gamma_i} \\
            \SPatternEnv{t // p} =
                \SPatternEnv{\SPiFst(t) // p_1} \cup
                \SPatternEnv{\SPiSnd(t) // p_2}
            \: .
        \end{gather*}

        Since $ t \leq \STransl_K[\tau_1] \times \STransl_K[\tau_2] $,
        by Property~\ref{pro:a-s-productproj}
        we have $ \SPiIth(t) \leq \STransl_K[\tau_i] $.
        Likewise, $ \SPiIth(t) \leq \SAcc{p_i} $.
        We apply the induction hypothesis to conclude.

    \Case[$ p = \Tagg(p_1) $]
        We have
        \begin{gather*}
            \KPatternTyping{K |- p: \alpha => \Gamma} \qquad
            \KPatternTyping{K |- p_1: \tau_1 => \Gamma} \\
            K(\alpha) = (L, U, T) \qquad
            (\Tagg \in U \text{ implies } \Tagg\colon \tau_1 \in T) \qquad
            \SPatternEnv{t // p} = \SPatternEnv{\SPiTagg(t) // p_1}
            \: .
        \end{gather*}

        Since $ t \leq \SAcc{\Tagg(p_1)} = \TTagg(\SAcc{p_1}) $,
        by Property~\ref{pro:a-s-variantproj}
        we have $ \SPiTagg(t) \leq \SAcc{p_1} $.
        We next prove $ \SPiTagg(t) \leq \STransl_K[\tau_1] $,
        which allows us to apply the induction hypothesis and conclude.

        The translation of $ \alpha $ is $
            \STransl_K[\alpha] = (\SLow_K(L, T) \lor \alpha) \land \SUpp_K(U, T)
        $. We have $ t \leq \STransl_K[\alpha] $
        and hence $ t \leq \SUpp_K(U, T) $.
        Since $ t \leq \TTagg(\Any) $, $
            t \leq \SUpp_K(U, T) \land \TTagg(\Any)
        $. We distribute the intersection over the summands
        of $ \SUpp_K(U, T) $, which is a union.

        If $ \Tagg \notin U $ (in which case $ U \neq \Tags $),
        then all summands have the form $ \Tagg_1(\tau') $
        and for each $ \Tagg_1 $ we have $ \Tagg_1 \neq \Tagg $:
        hence, the intersection is empty and thus we have $
            t \leq \Empty \simeq \TTagg(\Empty)
        $. Then $ \SPiTagg(t) \leq \Empty \leq \STransl_K[\tau_1] $.

        If $ \Tagg \in U $, then necessarily $ \Tagg \in \Dom(T) $
        holds as well. In that case the intersection $
            \SUpp_K(U, T) \land \TTagg(\Any)
        $ is equivalent to $
            \Tagg(\bigwedge_{\Tagg\colon \tau' \in T} \STransl_K[\tau'])
        $. Hence $
            t \leq \Tagg(\bigwedge_{\Tagg\colon \tau' \in T} \STransl_K[\tau'])
        $ and $
            \SPiTagg(t) \leq
                \bigwedge_{\Tagg\colon \tau' \in T} \STransl_K[\tau']
        $. Since $ \Tagg\colon \tau_1 \in T $, $
            \bigwedge_{\Tagg\colon \tau' \in T} \STransl_K[\tau'] \leq
            \STransl_K[\tau_1]
        $, from which follows $ \SPiTagg(t) \leq \STransl_K[\tau_1] $.

    \Case[$ p = \PAnd{p_1 \& p_2} $]
        We directly apply the induction hypothesis to both sub-patterns
        and conclude.

    \Case[$ p = \POr{p_1 | p_2} $]
        We have
        \begin{gather*}
            \KPatternTyping{K |- p: \tau => \Gamma} \qquad
            \forall i . \enspace
            \KPatternTyping{K |- p_i: \tau => \Gamma} \\
            \SPatternEnv{t // p} =
                \SPatternEnv{(t \land \SAcc{p_1}) // p_1}
                \SPatternEnvOr
                \SPatternEnv{(t \setminus \SAcc{p_1}) // p_2}
            \: .
        \end{gather*}

        Since $ t \land \SAcc{p_1} $ and $ t \setminus \SAcc{p_1} $
        are subtypes of $ t $, they are also subtypes of $ \STransl_K[\tau] $.
        We can apply the induction hypothesis and, for each $ x $,
        derive both that $
            (\SPatternEnv{t \land \SAcc{p_1} // p_1})(x) \leq
            \STransl_K[\Gamma(x)]
        $ and that $
            (\SPatternEnv{t \setminus \SAcc{p_1} // p_2})(x) \leq
            \STransl_K[\Gamma(x)]
        $. Hence, $ (\SPatternEnv{t // p})(x) \leq \STransl_K[\Gamma(x)] $.
        \qedhere
    \end{LCases}
\end{Proof}

\begin{Lemma}[Translation of type substitutions]
\label{lem:a-sk-typesubst}
    Let $ K $, $ K' $ be two non-recursive kinding environments such that
    $ \Dom(K') \cap (\Dom(K) \cup \Var_\varnothing(K)) = \varnothing $.
    Let $ \theta $ be a \Kin-type substitution
    such that $ \Dom(\theta) \subseteq \Dom(K') $
    and $ \KAdmissible{K, K' |- \theta: K} $.

    Let $ \theta' $ be the \Sem-type substitution
    defined as $
        \SubstC[\STransl_K[\alpha \theta]/\alpha | \alpha \in \Dom(K')]
    $.
    For every \Kin-type $ \tau $,
    we have $ \STransl_K, K'[\tau] \theta' \simeq \STransl_K[\tau \theta] $.
\end{Lemma}

\begin{Proof}
    By complete induction on $ w(\tau, (K, K')) $.
    We proceed by cases on $ \tau $ and assume that the lemma holds
    for all $ \tau' $ such that $ w(\tau', (K, K')) < w(\tau, (K, K')) $.

    \begin{LCases}
    \Case[$ \tau = \alpha $, with $ (K, K')(\alpha) = \KUnkind $]
        We have $ \STransl_K, K'[\alpha] = \alpha $,
        hence $ \STransl_K, K'[\alpha] \theta' = \alpha \theta' $.
        Either $ \alpha \in \Dom(K) $ or $ \alpha \in \Dom(K') $
        (the domains are disjoint). In the former case,
        $ \alpha \theta = \alpha $ and $ \alpha \theta' = \alpha $.
        Thus we have $
            \STransl_K, K'[\alpha] \theta' = \alpha = \STransl_K[\alpha \theta]
        $. In the latter, $ \alpha \theta' = \STransl_K[\alpha \theta] $ holds
        by definition of $ \theta' $.

    \Case[$ \tau = \alpha $, with $ K(\alpha) = (L, U, T) $ and
          $ \alpha \notin \Dom(K') $]
        We have $ \STransl_K, K'[\alpha] = \STransl_K[\alpha] $
        because no variable in the kind of $ \alpha $ is in $ \Dom(K') $.
        For the same reason, since the translation does not add variables,
        $ \STransl_K, K'[\alpha] \theta' = \STransl_K[\alpha] $.
        Additionally, $ \alpha \theta = \alpha $, so also
        $ \STransl_K[\alpha \theta] = \STransl_K[\alpha] $.

    \Case[$ \tau = \alpha $, with $ K'(\alpha) = (L', U', T') $]
        Because $ \KAdmissible{K, K' |- \theta: K} $,
        we know that $ \alpha \theta $ is some variable $ \beta $
        such that $ K(\beta) = (L, U, T) $
        and $ \KKindEntail{(L, U, T) |= (L', U', T' \theta)} $.

        We have
        \[
            \STransl_K[\alpha \theta] =
            \STransl_K[\beta] =
                (\SLow_K(L, T) \lor \beta) \land \SUpp_K(U, T)
        \]
        and
        \begin{align*}
            \STransl_K, K'[\alpha] \theta' & =
            \big(
                (\SLow_{K, K'}(L', T') \lor \alpha) \land \SUpp_{K, K'}(U', T')
            \big) \theta' \\
            & = (\SLow_{K, K'}(L', T') \theta' \lor \alpha \theta') \land
                \SUpp_{K, K'}(U', T') \theta' \\
            & = \Big(\SLow_{K, K'}(L', T') \theta' \lor
                \big((\SLow_K(L, T) \lor \beta) \land \SUpp_K(U, T)\big)
                \Big) \land
                \SUpp_{K, K'}(U', T') \theta'
            \: .
        \end{align*}

        Let us define
        \begin{align*}
            l & = \SLow_K(L, T) &
            u & = \SUpp_K(U, T) \\
            l' & = \SLow_{K, K'}(L', T') \theta' &
            u' & = \SUpp_{K, K'}(U', T') \theta'
        \end{align*}
        and assume that the following hold (we prove them below):
        \begin{gather*}
            l \leq u \qquad
            l' \leq u' \qquad
            l' \leq l \qquad
            u \leq u'
            \: .
        \end{gather*}
        Then we have also $ l' \leq u $ by transitivity.
        Whenever $ t \leq t' $, we have $ t \land t' \simeq t $
        and $ t \lor t' \simeq t' $.

        Thus we have the following equivalences:
        \begin{align*}
            \STransl_K, K'[\alpha] \theta'
            & = (l' \lor ((l \lor \beta) \land u)) \land u'
                \\
            & \simeq (l' \land u') \lor ((l \lor \beta) \land u \land u')
                & & \text{distributivity} \\
            & \simeq l' \lor ((l \lor \beta) \land u)
                & & \text{$ l' \leq u' $ and $ u \leq u' $} \\
            & \simeq (l' \lor l \lor \beta) \land (l' \lor u)
                & & \text{distributivity} \\
            & \simeq (l \lor \beta) \land u
                & & \text{$ l' \leq l $ and $ l' \leq u $}
        \end{align*}
        by which we conclude.

        We now prove our four assumptions. The first, $ l \leq u $,
        holds because $ L \subseteq U $ and $ L \subseteq \Dom(T) $:
        hence each branch of $ l $ appears in $ u $ as well.
        The second is analogous.

        For the other assumptions, note that $
            \STransl_K, K'[\tau'] \theta' \simeq \STransl_K[\tau' \theta]
        $ holds for all $ \tau' $ in the range of $ T' $.
        To prove $ l' \leq l $, note that $ L' \subseteq L $
        and $ T' \theta \subseteq T $. In $ l' $, we distribute the application
        of $ \theta' $ over all the summands of the union and inside all
        variant type constructors. Then, we show $
            \Tagg(\bigwedge_{\Tagg\colon \tau' \in T'}
                \STransl_K, K'[\tau'] \theta') \leq l
        $ for each $ \Tagg \in L' $. We have $
            \Tagg(\bigwedge_{\Tagg\colon \tau' \in T'}
                \STransl_K, K'[\tau'] \theta') \simeq
            \Tagg(\bigwedge_{\Tagg\colon \tau' \in T'}
                \STransl_K[\tau' \theta]) =
            \Tagg(\bigwedge_{\Tagg\colon \tau' \theta \in T' \theta}
                \STransl_K[\tau' \theta])
        $. Since $ L' \subseteq L $, there is a summand of $ l $ with the
        same tag. Since $ \Tagg $ is in the lower bound, it has a single type
        in both $ T $ and $ T' $ and, since $ T' \theta \subseteq T $,
        the type it has in $ T $ must be $ \tau' \theta $.

        To prove $ u \leq u' $, note that $ U \subseteq U' $.
        If $ U = \Tags $, then $ U' = \Tags $.
        Then both $ u $ and $ u' $ are unions of two types:
        the union of tags mentioned respectively in $ T $ and $ T' $
        and the rest. For each $ \Tagg $, if $ \Tagg \notin \Dom(T) $,
        then $ \Tagg \notin \Dom(T') $, in which case both $ u $ and $ u' $
        admit it with any argument type. If $ \Tagg \in \Dom(T) $,
        either $ \Tagg \in \Dom(T') $ or not. In the former case,
        $ u $ admits a smaller argument type than $ u' $ because
        $ T' \theta \subseteq T $. The same occurs in the latter case,
        since $ u' $ admits $ \Tagg $ with any argument type.

        If $ U \neq \Tags $, then $ U' $ could be $ \Tags $ or not.
        In either case we can prove, for each $ \Tagg \in U $, that $ u' $
        admits $ \Tagg $ with a larger argument type than $ u $ does.

    \Case[$ \tau = b $]
        Straightforward, since a basic type is translated into itself
        and is never affected by substitutions.

    \Case[$ \tau = \tau_1 \to \tau_2 $]
        By the induction hypothesis we have $
            \STransl_K, K'[\tau_i] \theta' \simeq \STransl_K[\tau_i \theta]
        $ for both $ i $. Then
        \begin{align*}
            \STransl_K, K'[\tau_1 \to \tau_2] \theta' & =
                (\STransl_K, K'[\tau_1] \theta') \to
                (\STransl_K, K'[\tau_2] \theta') \simeq
            \STransl_K[\tau_1 \theta] \to
                  \STransl_K[\tau_2 \theta] \\ & =
            \STransl_K[(\tau_1 \theta) \to (\tau_2 \theta)] =
            \STransl_K[(\tau_1 \to \tau_2) \theta]
            \: .
        \end{align*}

    \Case[$ \tau = \tau_1 \times \tau_2 $]
        Analogous to the previous case.
    \qedhere
    \end{LCases}
\end{Proof}

\begin{Lemma}
\label{lem:a-sk-backwards}
    If $ \STyping{ \varnothing |- v: \STransl_K[\tau] } $,
    then there exists a value $ v' $
    such that $ \KTyping{ K; \varnothing |- v': \tau } $
    and, for every pattern $ p $,
    $ \Matching{v / p} = \MatchFail \iff \Matching{v' / p} = \MatchFail $.
\end{Lemma}

\begin{Proof}
    By structural induction on $ v $.

    Note that values are always typed by an application of the typing rule
    corresponding to their form
    (\Rule{Ts-Const}, \Rule{Ts-Abstr}, \Rule{Ts-Pair}, or \Rule{Ts-Tag})
    to appropriate premises, possibly followed by applications
    of \Rule{Ts-Subsum}. Hence, if $ \STyping{\EmptyEnv |- v: t} $,
    there is a type $ t' \leq t $ such that $ \STyping{\EmptyEnv |- v: t'} $
    and that the last typing rule used to derive $
        \STyping{\EmptyEnv |- v: t'}
    $ is one of the four above, given by the form of $ v $.

    \begin{LCases}
    \Case[$ v = c $]
        We have $ \STransl_K[\tau] \geq c $. Hence $ \tau = b_c $,
        as the translation of any other $ \tau $ is disjoint from $ c $.
        Then we can take $ v' = v $.

    \Case[$ v = (v_1, v_2) $]
        We have $ \STransl_K[\tau] \geq t_1 \times t_2 $ for some $ t_1 $
        and $ t_2 $. Hence $ \tau = \tau_1 \times \tau_2 $: any other $ \tau $
        would translate to a type disjoint from all products. Therefore $
            \STyping{\EmptyEnv |- v:
                \STransl_K[\tau_1] \times \STransl_K[\tau_2]}
        $. By Lemma~\ref{lem:a-s-generation} we have $
            \STyping{\EmptyEnv |- v_i: \STransl_K[\tau_i]}
        $ for both $ i $; then by the induction hypothesis we find $ v_i' $
        for both $ i $ and let $ v' = (v_1', v_2') $.

    \Case[$ v = \Tagg(v_1) $]
        We have $ \STransl_K[\tau] \geq \TTagg(t_1) $ and $
            \STyping{\EmptyEnv |- v: \TTagg(t_1)}
        $ for some $ t_1 \nleq \Empty $
        (since $ t_1 $ types the value $ v_1 $).
        Therefore, by the same reasoning as above, $ \tau = \alpha $
        with $ K(\alpha) = (L, U, T) $. Since $
            \STransl_K[\tau] \geq \TTagg(t_1)
        $, we have $ \Tagg \in L $ and therefore $ \Tagg\colon \tau_1 \in T $
        for some $ \tau_1 $ such that $ t_1 \leq \STransl_K[\tau_1] $.
        Then we have $ \STyping{\EmptyEnv |- v_1: \STransl_K[\tau_1]} $;
        we may apply the induction hypothesis to find a value $ v_1' $
        and let $ v' = \Tagg(v_1') $.

    \Case[$ v = \Abstr{x. e} $]
        Note that an abstraction is only accepted by patterns which accept any value, so any two abstractions fail to match exactly the same patterns.

        We have $ \STyping{\EmptyEnv |- v: t_1 \to t_2} $
        for some $ t_1 \to t_2 \leq \STransl_K[\tau] $.
        Hence we know $ \tau $ is of the form $ \tau_1 \to \tau_2 $;
        thus we have $
            \STyping{\EmptyEnv |- v: \STransl_K[\tau_1] \to \STransl_K[\tau_2]}
        $. We take $ v' $ to be the function $
            \Abstr{x. \Appl
                {\Appl{ \Fix }{ (\Abstr{f. \Abstr{x. \Appl{f}{x}}}) }}
                {x}}
        $, which never terminates and can be assigned any arrow type.
    \qedhere
    \end{LCases}
\end{Proof}

\begin{Lemma}
\label{lem:a-sk-exhaustiveness}
    Let $ K $ be a kinding environment, $ \tau $ a \Kin-type,
    and $ P $ a set of patterns.
    If $ \KExhaustive{K |- \tau < P} $,
    then $ \STransl_K[\tau] \leq \bigvee_{p \in P} \SAcc{p} $.
\end{Lemma}

\begin{Proof}
    By contradiction, assume that $ \KExhaustive{K |- \tau < P} $ holds
    but $ \STransl_K[\tau] \nleq \bigvee_{p \in P} \SAcc{p} $.
    The latter condition implies that there exists a value $ v $
    in the interpretation of $ \STransl_K[\tau] $ which is not
    in the interpretation of $ \bigvee_{p \in P} \SAcc{p} $.
    Because the definition of accepted type is exact with respect
    to the semantics of pattern matching, we have $
        \Matching{v / p} = \MatchFail
    $ for all $ p \in P $. We also have $
        \STyping{\EmptyEnv |- v: \STransl_K[\tau]}
    $ since $ v $ is in the interpretation of that type
    (typing is complete with respect to the interpretation if we restrict
    ourselves to translations of \Kin-types).

    By Lemma~\ref{lem:a-sk-backwards}, from $ v $ we can build a value $ v' $
    such that $ \KTyping{K; \EmptyEnv |- v': \tau} $ and,
    for every pattern $ p $, $
        \Matching{v / p} = \MatchFail \iff \Matching{v' / p} = \MatchFail
    $. We reach a contradiction, since $ \KExhaustive{K |- \tau < P} $ and
    $ \KTyping{K; \EmptyEnv |- v': \tau} $ imply that there exists a $ p \in P $
    such that $
        \Matching{v' / p} \neq \MatchFail
    $, whereas we have $ \Matching{v / p} = \MatchFail $
    for all $ p \in P $.
\end{Proof}

\begin{Theorem}[Preservation of typing]
\label{thm:a-sk-preservation}
    Let $ e $ be an expression,
    $ K $ a non-recursive kinding environment,
    $ \Gamma $ a \Kin-type environment,
    and $ \tau $ a \Kin-type.
    If $ \KTyping{ K; \Gamma |- e: \tau } $,
    then $ \STyping{ \STransl_K[\Gamma] |- e: \STransl_K[\tau] } $.
\end{Theorem}

\begin{Proof}
    By induction on the derivation of $ \KTyping{K; \Gamma |- e: \tau} $.
    We reason by cases on the last applied rule.

    \begin{LCases}
    \Case[\Rule{Tk-Var}]
        We have
        \begin{gather*}
            \KTyping{K; \Gamma |- x: \tau} \qquad
            \tau \in \Inst_K(\Gamma(x)) \qquad \text{hence} \\
            \Gamma(x) = \KScheme{A. K_x > \tau_x} \qquad
            \tau = \tau_x \theta \qquad
            \Dom(\theta) \subseteq A \qquad
            \KAdmissible{K, K_x |- \theta: K}
        \end{gather*}
        and must show $ \STyping{\STransl_K[\Gamma] |- x: \STransl_K[\tau]} $.
        Since $ \STransl_K[\Gamma](x) = \Scheme{A. \STransl_K, K_x[\tau_x]} $,
        by \Rule{Ts-Var} we can derive $ \STransl_K, K_x[\tau_x] \theta' $
        for any \Sem-type substitution $ \theta' $
        with $ \Dom(\theta') \subseteq A $.

        Consider the \Sem-type substitution $
            \theta' = \SubstC[\STransl_K[\alpha \theta]/\alpha | \alpha \in A]
        $. We have $
            \STransl_K, K_x[\tau_x] \theta' \simeq \STransl_K[\tau_x \theta]
        $ by Lemma~\ref{lem:a-sk-typesubst}
        (we can assume the conditions on the domain of $ K_x $ to hold
        by renaming the variables in $ A $).
        Hence, we derive $ \STransl_K, K_x[\tau_x] \theta' $ by \Rule{Ts-Var}
        and then $ \STransl_K[\tau_x \theta] $ by subsumption.

    \Case[\Rule{Tk-Const}]
        We have
        \[
            \KTyping{K; \Gamma |- c: b_c} \qquad
            \STransl_K[b_c] = b_c
        \]
        and may derive $ \STyping{\STransl_K[\Gamma] |- c: c} $
        by \Rule{Ts-Const} and $ \STyping{\STransl_K[\Gamma] |- c: b_c} $
        by subsumption.

    \Case[\Rule{Tk-Abstr}]
        We have
        \[
            \KTyping{K; \Gamma |- \Abstr{x. e_1}: \tau_1 \to \tau_2} \qquad
            \KTyping{K; \Gamma, \Set{x\colon \tau_1} |- e_1: \tau_2} \qquad
            \STransl_K[\tau_1 \to \tau_2] =
                \STransl_K[\tau_1] \to \STransl_K[\tau_2]
            \: .
        \]
        By the induction hypothesis we derive $
            \STyping{\STransl_K[\Gamma], \Set{x\colon \STransl_K[\tau_1]} |-
                e_1: \STransl_K[\tau_2]}
        $, then we apply \Rule{Ts-Abstr}.

    \Cases[\Rule{Tk-Appl}, \Rule{Tk-Pair}, and \Rule{Tk-Fix}]
        Straightforward application of the induction hypothesis.

    \Case[\Rule{Tk-Tag}]
        We have
        \begin{gather*}
            \KTyping{K; \Gamma |- \Tagg(e_1): \alpha} \qquad
            \KTyping{K; \Gamma |- e_1: \tau_1} \qquad
            K(\alpha) = (L, U, T) \qquad
            \Tagg \in L \qquad
            \Tagg\colon \tau_1 \in T \\
            \STransl_K[\alpha] =
                (\SLow_K(L, T) \lor \alpha) \land \SUpp_K(U, T)
            \: .
        \end{gather*}

        We derive $ \STyping{\STransl_K[\Gamma] |- e_1: \STransl_K[\tau_1]} $
        by the induction hypothesis, then $
            \STyping{ \STransl_K[\Gamma] |- \Tagg(e_1):
                \Tagg(\STransl_K[\tau_1])}
        $ by \Rule{Ts-Tag}. We show that $
            \Tagg(\STransl_K[\tau_1]) \leq \STransl_K[\alpha]
        $ holds: hence, we may derive the supertype by subsumption.

        Since $ \Tagg \in L $ and hence $ \Tagg \in \Dom(T) $,
        both $ \SLow_K(L, T) $ and $ \SUpp_K(U, T) $ contain a summand $
            \Tagg(\bigwedge_{\Tagg\colon \tau' \in T} \STransl_K[\tau'])
        $. Since $ \Tagg\colon \tau_1 \in T $
        and no other type may be associated to $ \Tagg $,
        the intersection has a single factor $ \STransl_K[\tau_1] $.
        Thus we have both $ \Tagg(\STransl_K[\tau_1]) \leq \SLow_K(L, T) $
        and $ \Tagg(\STransl_K[\tau_1]) \leq \SUpp_K(U, T) $;
        hence, $ \Tagg(\STransl_K[\tau_1]) \leq \STransl_K[\alpha] $.

    \Case[\Rule{Tk-Match}]
        We have
        \begin{gather*}
            \KTyping{K; \Gamma |- \Match{e_0 with (p_i \to e_i)_{i \in I}}:
                \tau} \\
            \KTyping{K; \Gamma |- e_0: \tau_0} \qquad
            \KExhaustive{K |- \tau_0 < \SetC{p_i | i \in I}} \\
            \forall i \in I . \enspace
            \KPatternTyping{K |- p_i: \tau_0 => \Gamma_i} \qquad
            \KTyping{K; \Gamma, \Gen_{K;\Gamma}(\Gamma_i) |- e_i: \tau}
        \end{gather*}
        and must show
        \[
            \STyping{\STransl_K[\Gamma] |-
                \Match{e_0 with (p_i \to e_i)_{i \in I}}: \STransl_K[\tau]}
        \]
        which we prove by establishing, for some types $ t_0 $
        and $ t_i $, $ t_i' $ for each $ i $, that
        \begin{gather*}
            \STyping{\STransl_K[\Gamma] |- e_0: t_0} \qquad
            t_0 \leq \textstyle\bigvee_{i \in I} \SAcc{p_i} \qquad
            t_i = (t_0 \setminus \textstyle\bigvee_{j < i} \SAcc{p_j})
                  \land \SAcc{p_i} \\
            \forall i \in I . \enspace
            \STyping{\STransl_K[\Gamma],
                \Gen_{\STransl_K[\Gamma]}(\SPatternEnv{t_i // p_i}) |- e_i:
                t_i'}
            \qquad
            \textstyle\bigvee_{i \in I} t_i' \leq \STransl_K[\tau]
            \: .
        \end{gather*}
        and then applying \Rule{Ts-Match}, followed by \Rule{Ts-Subsum}
        if necessary.

        By the induction hypothesis we derive $
            \STyping{\STransl_K[\Gamma] |- e_0: \STransl_K[\tau_0]}
        $ and hence have $ t_0 = \STransl_K[\tau_0] $.
        By Lemma~\ref{lem:a-sk-exhaustiveness},
        we have $ t_0 \leq \textstyle\bigvee_{i \in I} \SAcc{p_i} $.
        For every branch, $ t_i \leq t_0 $ and $ t_i \leq \SAcc{p_i} $:
        therefore, we can apply Lemma~\ref{lem:a-sk-patterns} and derive that
        $ (\SPatternEnv{t_i // p_i})(x) \leq \STransl_K[\Gamma_i(x)] $ holds
        for every $ x \in \Capt(p_i) $.

        For each branch, we derive $
            \STyping{\STransl_K[\Gamma],
                \STransl_K[\Gen_{K;\Gamma}(\Gamma_i)] |- e_i: \STransl_K[\tau]}
        $ by the induction hypothesis. We derive $
            \STyping{\STransl_K[\Gamma],
                \Gen_{\STransl_K[\Gamma]}(\SPatternEnv{t_i // p_i}) |- e_i:
                \STransl_K[\tau]}
        $ by Lemma~\ref{lem:a-s-weakening} by proving $
            \SMoreGen{
                \STransl_K[\Gamma],
                    \Gen_{\STransl_K[\Gamma]}(\SPatternEnv{t_i // p_i}) <
                \STransl_K[\Gamma], \STransl_K[\Gen_{K;\Gamma}(\Gamma_i)]}
        $ and $
            \Mvar(\STransl_K[\Gamma],
                \Gen_{\STransl_K[\Gamma]}(\SPatternEnv{t_i // p_i}))
            \subseteq
            \Mvar(\STransl_K[\Gamma],
                \STransl_K[\Gen_{K;\Gamma}(\Gamma_i)])
        $. The latter is straightforward. For the former, for each
        $ x \in \Capt(p_i) $---say $ \Gamma_i(x) = \tau_x $
        and $ (\SPatternEnv{t_i // p_i})(x) = t_x $---we must show $
            \SMoreGen{\Gen_{\STransl_K[\Gamma]}(t_x) <
                \STransl_K[\Gen_{K;\Gamma}(\tau_x)]}
        $. This holds because $ t_x \leq \STransl_K[\tau_x] $ and because,
        by Lemma~\ref{lem:a-sk-variables},
        $ \Var(\STransl_K[\Gamma]) \subseteq \Var_K(\Gamma) $.

        We can thus choose $ t_i' = \STransl_K[\tau] $ for all branches,
        satisfying $
            \textstyle\bigvee_{i \in I} t_i' \leq \STransl_K[\tau]
        $.
        \qedhere
    \end{LCases}
\end{Proof}

\subsection{Type reconstruction}

\subsubsection{Definition of type reconstruction for \VariantsS{}}

\begin{Definition}[Constraints]
\label{def:a-sr-constraints}
    A \emph{constraint} $ c $ is a term
    inductively generated by the following grammar:
    \[
      \Bnf{ c
        \= \SRSub{t < t}
        \| \SRSub{x < t}
        \| \SRDef{ \Gamma | C }
        \| \SRLet{ C | (\Gamma_i \In C_i)_{i \in I} } }
    \]
    where $ C $ ranges over \emph{constraint sets},
    that is, finite sets of constraints, and
    where the range of every type environment $ \Gamma $ in constraints
    of the form \Keyw{def} or \Keyw{let}
    only contains types (i.e., trivial type schemes).
\end{Definition}

We give different definitions of constraint generation and rewriting here
than those in Section~\ref{recon},
because we keep track explicitly of the new variables introduced
during the derivation, rather than informally requiring them to be fresh.
For instance, in $ \SRConstrGenA{e: t => C | A} $,
$ A $ is the set of variables which appear in $ C $ but not in $ t $.
We will omit it for soundness proofs, where it is not relevant.

We use the symbol $ \SRDisj $ to denote the union of two disjoint sets.
Therefore, when we write $ A_1 \SRDisj A_2 $,
we require $ A_1 $ and $ A_2 $ to be disjoint.
When we require this for sets of type variables,
the condition is always satisfiable by an appropriate choice of variables,
since there is an infinite supply to choose from.

\begin{figure*}

    \begin{mathpar}
        \MyInfer
            { }
            { \SRPatternEnvA{ t /// \Wildcard => \varnothing, \varnothing |
                \varnothing } }
            {}
        \and
        \MyInfer
            { }
            { \SRPatternEnvA{ t /// x => \Set{x\colon t}, \varnothing |
                \varnothing } }
            {}
        \and
        \MyInfer
            { }
            { \SRPatternEnvA{ t /// c => \varnothing, \varnothing |
                \varnothing } }
            {}
        \\
        \MyInfer
            { \SRPatternEnvA{ \alpha_1 /// p_1 => \Gamma_1, C_1 | A_1 } \\
              \SRPatternEnvA{ \alpha_2 /// p_2 => \Gamma_2, C_2 | A_2 } }
            { \SRPatternEnvA{ t /// (p_1,p_2) =>
                \Gamma_1 \cup \Gamma_2,
                C_1 \cup C_2 \cup \Set{\SRSub{t < \alpha_1 \times \alpha_2}} |
                A_1 \SRDisj A_2 \SRDisj \Set{\alpha_1, \alpha_2} }
            }
            {A_1, A_2, \alpha_1, \alpha_2 \Fresh t}
        \and
        \MyInfer
            { \SRPatternEnvA{ \alpha /// p => \Gamma, C | A } }
            { \SRPatternEnvA{ t /// \Tagg(p) =>
                \Gamma, C \cup \Set{\SRSub{ t < \Tagg(\alpha)}} |
                A \SRDisj \Set{\alpha} } }
            {A, \alpha \Fresh t}
        \and
        \MyInfer
            { \SRPatternEnvA{ t /// p_1 => \Gamma_1, C_1 | A_1 } \\
              \SRPatternEnvA{ t /// p_2 => \Gamma_2, C_2 | A_2 } }
            { \SRPatternEnvA{ t /// \PAnd{p_1 \& p_2} =>
                \Gamma_1 \cup \Gamma_2, C_1 \cup C_2 |
                A_1 \SRDisj A_2 } }
            {}
        \and
        \MyInfer
            { (\SRPatternEnvA{ t \land \SAcc{p_1}) /// p_1 =>
                \Gamma_1, C_1 | A_1 } \\
              (\SRPatternEnvA{ t \setminus \SAcc{p_1}) /// p_2 =>
                \Gamma_2, C_2 | A_2 } }
            { \SRPatternEnvA{ t /// \POr{p_1 | p_2} =>
                \SetC{x\colon \Gamma_1(x) \lor \Gamma_2(x) | x \in \Capt(p_1)},
                C_1 \cup C_2 | A_1 \SRDisj A_2 } }
            {}
    \end{mathpar}

    \caption{Constraint generation for pattern environments.}
    \label{fig:reconstruction-constraint-generation-matching-a}
\end{figure*}

\begin{Definition}[Freshness]
\label{def:a-sr-fresh}
    We say that a type variable $ \alpha $ is \emph{fresh}
    with respect to a set of type variables $ A $,
    and write $ \alpha \Fresh A $, if $ \alpha \notin A $.
    We write $ A \Fresh A' $ if $ \forall \alpha \in A . \: \alpha \Fresh A' $.

    We extend this to define freshness with respect to
    types, type environments, and type substitutions:
    we write $ \alpha \Fresh t $ if $ \alpha \Fresh \Var(t) $,
    $ \alpha \Fresh \Gamma $ if $ \alpha \Fresh \Var(\Gamma) $,
    and $ \alpha \Fresh \theta $
    if $ \alpha \Fresh (\Dom(\theta) \cup \Var(\theta)) $.
\end{Definition}

\begin{Definition}[Environment generation for pattern matching]
\label{def:a-sr-patterenv}
    The environment generation relation for pattern matching
    $ \SRPatternEnvA{t /// p => \Gamma, C | A} $
    is defined by the rules in
    Figure~\ref{fig:reconstruction-constraint-generation-matching-a}.
\end{Definition}

\begin{figure*}

    \begin{mathpar}
        \MyInfer[TRs-Var]
            {  }
            { \SRConstrGenA{ x: t => \Set{ \SRSub{x < t} } | \varnothing } }
            {}
        \and
        \MyInfer[TRs-Const]
            {  }
            { \SRConstrGenA{ c: t => \Set{ \SRSub{c < t} } | \varnothing } }
            {}
        \and
        \MyInfer[TRs-Abstr]
            { \SRConstrGenA{ e: \beta => C | A } }
            { \SRConstrGenA{ \Abstr{x. e}: t => \Set{
                \SRDef{ \Set{x\colon \alpha} | C }, \:
                \SRSub{\alpha \to \beta < t} } |
                A \SRDisj \Set{\alpha, \beta} } }
            {A, \alpha, \beta \Fresh t}
        \and
        \MyInfer[TRs-Appl]
            { \SRConstrGenA{ e_1: \alpha \to \beta => C_1 | A_1 } \\
              \SRConstrGenA{ e_2: \alpha           => C_2 | A_2 } }
            { \SRConstrGenA{ \Appl{e_1}{e_2}: t    =>
                C_1 \cup C_2 \cup \Set{\SRSub{\beta < t}} |
                A_1 \SRDisj A_2 \SRDisj \Set{\alpha, \beta} } }
            {A_1, A_2, \alpha, \beta \Fresh t}
        \and
        \MyInfer[TRs-Pair]
            { \SRConstrGenA{ e_1: \alpha_1 => C_1 | A_1 } \\
              \SRConstrGenA{ e_2: \alpha_2 => C_2 | A_2 } }
            { \SRConstrGenA{ (e_1, e_2): t =>
                C_1 \cup C_2 \cup
                \Set{ \SRSub{ \alpha_1 \times \alpha_2 < t} } |
                A_1 \SRDisj A_2 \SRDisj \Set{\alpha_1, \alpha_2}
              } }
            {A_1, A_2, \alpha_1, \alpha_2 \Fresh t}
        \and
        \MyInfer[TRs-Tag]
            { \SRConstrGenA{ e: \alpha => C | A } }
            { \SRConstrGenA{ \Tagg(e): t =>
                C \cup \Set{ \SRSub{\TTagg(\alpha) < t} } |
                A \SRDisj \Set{\alpha} } }
            {A, \alpha \Fresh t}
        \and
        \MyInfer[TRs-Match]
            { \SRConstrGenA{ e_0: \alpha => C_0 | A_0 } \\
              t_i = (\alpha \setminus \textstyle\bigvee_{j < i} \SAcc{p_j})
                \land \SAcc{p_i} \\\\
              \forall i \in I \\
              \SRPatternEnvA{ t_i /// p_i => \Gamma_i, C_i | A_i } \\
              \SRConstrGenA{ e_i: \beta => C_i' | A_i' } \\\\
              C_0' = C_0 \cup (\textstyle\bigcup_{i \in I} C_i) \cup
                \Set{ \SRSub{\alpha < \textstyle\bigvee_{i \in I} \SAcc{p_i}} }
              \\\\
              A = A_0 \SRDisj (\textstyle\SRBigDisj_{i \in I} A_i)
                \SRDisj (\textstyle\SRBigDisj_{i \in I} A_i')
                \SRDisj \Set{\alpha, \beta} 
            }
            { \SRConstrGenA{
                \Match{e_0 with (p_i \to e_i)_{i \in I}}: t => \Set{
                \SRLet{ C_0' | (\Gamma_i \In C_i')_{i \in I}}, \:
                \SRSub{\beta < t} } | A }
            }
            {A \Fresh t}
    \end{mathpar}

    \caption{Constraint generation rules with explicit variable introduction.}
    \label{fig:reconstruction-constraint-generation-a}
\end{figure*}

\begin{Definition}[Constraint generation]
\label{def:a-sr-constraintgen}
    The constraint generation relation $ \SRConstrGenA{e: t => C | A} $
    is defined by the rules in
    Figure~\ref{fig:reconstruction-constraint-generation-a}.
\end{Definition}

\begin{figure*}

    \begin{mathpar}
        \MyInfer
            { \forall i \in I \\
              \SRConstrRewrA{ \Gamma |- c_i => D_i | A_i } }
            { \SRConstrRewrA{ \Gamma |- \SetC{c_i | i \in I} =>
                \textstyle\bigcup_{i \in I} D_i |
                \SRBigDisj_{i \in I} A_i } }
            {}
        \and
        \MyInfer
            { }
            { \SRConstrRewrA{ \Gamma |- \SRSub{t < t'} => \Set{\SRSub{t < t'}} |
                \varnothing }
            }
            {} 
        \and
        \MyInfer
            { \Gamma(x) = \Scheme{ \Set{\alpha_1, \dots, \alpha_n}. t_x } }
            { \SRConstrRewrA{ \Gamma |- \SRSub{x < t} => \Set{
                \SRSub{t_x \SubstDots[\beta_1/\alpha_1 ... \beta_n/\alpha_n] <
                    t} } | \Set{\beta_1, \dots, \beta_n} }
            }
            {}
        \and
        \MyInfer
            { \SRConstrRewrA{ \Gamma, \Gamma' |- C => D | A } }
            { \SRConstrRewrA{ \Gamma |- \SRDef{ \Gamma' | C } => D | A } }
            {}
        \and
        \MyInfer
            { \SRConstrRewrA{ \Gamma |- C_0 => D_0 | A_0 } \\
              \theta_0 \in \Tally(D_0) \\\\
              \forall i \in I \\
              \SRConstrRewrA{
                \Gamma, \Gen_{\Gamma \theta_0}(\Gamma_i \theta_0) |-
                C_i => D_i | A_i } \\\\
              A = A_0 \SRDisj (\textstyle\SRBigDisj_{i \in I} A_i)
                \SRDisj \Var(\theta_0)
            }
            { \SRConstrRewrA{ \Gamma |-
                \SRLet{ C_0 | (\Gamma_i \In C_i)_{i \in I} } =>
                    \SREq(\theta_0) \cup \textstyle\bigcup_{i \in I} D_i | A }
            }
            {}
    \end{mathpar}

    \caption{Constraint rewriting rules with explicit variable introduction.}
    \label{fig:reconstruction-constraint-rewriting-poly-a}
\end{figure*}

Note that all rules include a constraint of the form $ \SRSub{(\cdot) < t} $.
We add this constraint everywhere to streamline the proofs;
in practice, it can be dropped from \RuleRef{TRs-Appl} and \RuleRef{TRs-Match}
by using directly $ t $ instead of $ \beta $ to generate constraints
for the sub-expressions.

\begin{Definition}[Type-constraint set]
\label{def:a-sr-typeconstraints}
    A \emph{type-constraint set} $ D $ is a set
    of constraints of the form $ \SRSub{t < t'} $,
    where $ t $ and $ t' $ are types.

    We say that a type substitution $ \theta $ satisfies a type-constraint set
    $ D $, written $ \SRSat{ \theta |- D } $, if
    $ t \theta \leq t' \theta $ holds for every $ \SRSub{t < t'} $ in $ D $.
\end{Definition}

\begin{Definition}[Equivalent type-constraint set]
\label{def:a-sr-equiv}
    The \emph{equivalent type-constraint set} $ \SREq(\theta) $
    of a type substitution $ \theta $
    is defined as
    \[
        \SREq(\theta)
            =
            \textstyle\bigcup_{\alpha \in \Dom(\theta)}
        \Set{\SRSub{\alpha < \alpha \theta}, \,
          \SRSub{\alpha \theta < \alpha}}
        \: .
    \]
\end{Definition}

\begin{Definition}[Constraint rewriting]
\label{def:a-sr-constraintrewr}
    The constraint rewriting relation $ \SRConstrRewrA{\Gamma |- c => D | A} $
    between type environments, constraints or constraint sets,
    and type-constraint sets is defined by the rules in
    Figure~\ref{fig:reconstruction-constraint-rewriting-poly-a}.
\end{Definition}

\begin{Definition}[Tallying problem]
\label{def:a-sr-tallying}
    Let $ D $ be a type-constraint set.
    A type substitution $ \theta $ is a solution
    to the \emph{tallying problem} of $ D $
    if it satisfies $ D $, that is, if $ \SRSat{\theta |- D} $.
\end{Definition}

\begin{Property}[Tallying algorithm]
\label{pro:a-sr-tally}
    There exists a terminating algorithm $ \Tally $ such that,
    for any type-constraint set $ D $,
    $ \Tally(D) $ is a finite, possibly empty, set of type substitutions.
\end{Property}

\begin{Theorem}[Soundness and completeness of $ \Tally $]
\label{thm:a-sr-tally-soundcomplete}
    Let $ D $ be a type-constraint set. For any type substitution $ \theta $:
    \begin{itemize}
        \item if $ \theta \in \Tally(D) $, then $ \SRSat{\theta |- D} $;
        \item if $ \SRSat{ \theta |- D } $, then $
            \Exists \theta' \in \Tally(D), \theta'' .
            \Forall \alpha \notin \Var(\theta') .
            \alpha \theta \simeq \alpha \theta' \theta''
          $.
    \end{itemize}

    Furthermore, if $ \theta \in \Tally(D) $,
    then $ \Dom(\theta) $ is the set of variables appearing in $ D $
    and $ \Var(\theta) $ is a set of fresh variables of the same cardinality.
    In the completeness property above, for $ \theta'' $ we can take
    $ \theta \cup \theta''' $ where $ \Dom(\theta''') = \Var(\theta')$.
\end{Theorem}

\subsubsection{Properties of type reconstruction for \VariantsS{}}

\begin{Lemma}
\label{lem:a-sr-variables}
    Given a constraint set $ C $,
    we write $ \Var(C) $ for the set of variables appearing in it.
    The following properties hold:
    \begin{itemize}
        \item whenever $ \SRPatternEnvA{t /// p => \Gamma, C | A} $,
            we have $ \Var(C) \subseteq \Var(t) \cup A $,
            $ \Var(\Gamma) \subseteq \Var(t) \cup A $, and $ A \Fresh t $;
        \item whenever $ \SRConstrGenA{e: t => C | A} $,
            we have $ \Var(C) \subseteq \Var(t) \cup A $ and $ A \Fresh t $;
        \item whenever $ \SRConstrRewrA{\Gamma |- C => D | A} $,
            we have $ \Var(D) \subseteq \Var(C) \cup \Var(\Gamma) \cup A $.
    \end{itemize}
\end{Lemma}

\begin{Proof}
    Straightforward proofs by induction on the derivations.
\end{Proof}

\begin{Lemma}[Correctness of environment reconstruction]
\label{lem:a-sr-patterns-correct}
    Let $ p $ be a pattern and $ t $, $ t' $ two types,
    with $ t' \leq \SAcc{p} $.
    Let $ \SRPatternEnv{t /// p => \Gamma, C} $.
    If $ \theta $ is a type substitution
    such that $ \SRSat{\theta |- C} $ and $ t' \leq t \theta $,
    then, for all $ x \in \Capt(p) $,
    $ (\SPatternEnv{t' // p})(x) \leq \Gamma(x) \theta $.
\end{Lemma}

\begin{Proof}
    By structural induction on $ p $.

    \begin{LCases}
    \Cases[$ p = \Wildcard $ and $ p = c $]
        There is nothing to prove since $ \Capt(p) = \varnothing $.

    \Case[$ p = x $]
        We have
        \[
            \SRPatternEnv{t /// x => \Set{x\colon t}, \varnothing} \qquad
            (\SPatternEnv{t' // x})(x) = t'
        \]
        and must show $ t' \leq t \theta $, which we know by hypothesis.

    \Case[$ p = (p_1, p_2) $]
        We have
        \[
            \SRPatternEnv{t /// p => \Gamma_1 \cup \Gamma_2,
                C_1 \cup C_2 \cup \Set{\SRSub{t < \alpha_1 \times \alpha_2}}}
            \qquad
            \forall i . \enspace
                \SRPatternEnv{\alpha_i /// p_i => \Gamma_i, C_i}
            \: .
        \]

        Each $ x \in \Capt(p) $ is either in $ \Capt(p_1) $ or in
        $ \Capt(p_2) $. Let $ x \in \Capt(p_i) $; then, we must show $
            (\SPatternEnv{\SPiIth(t') // p_i})(x) \leq \Gamma_i(x) \theta
        $. This follows from the induction hypothesis, since $
            t' \leq t \theta \leq \alpha_1 \theta \times \alpha_2 \theta
        $ implies $ \SPiIth(t') \leq \alpha_i \theta $
        by Property~\ref{pro:a-s-productproj}.

    \Case[$ p = \Tagg(p_1) $]
        We have
        \[
            \SRPatternEnv{t /// \Tagg(p) => \Gamma,
                C \cup \Set{\SRSub{t < \Tagg(\alpha)}}} \qquad
            \SRPatternEnv{\alpha /// p => \Gamma, C}
            \: .
        \]

        Analogous to the previous case. We can apply the induction hypothesis, because $ t' \leq t \theta \leq \Tagg(\alpha) \theta $
        implies $ \SPiTagg(t') \leq \alpha \theta $
        by Property~\ref{pro:a-s-variantproj}.

    \Case[$ p = \PAnd{p_1 \& p_2} $]
        Every $ x \in \Capt(p) $ is either in $ \Capt(p_1) $ or in
        $ \Capt(p_2) $. Let $ x \in \Capt(p_i) $;
        then, we apply the induction hypothesis to $ p_i $ to conclude.

    \Case[$ p = \POr{p_1 | p_2} $]
        We have
        \begin{gather*}
            \SRPatternEnv{t /// \POr{p_1 | p_2} =>
                \SetC{x\colon \Gamma_1(x) \lor \Gamma_2(x) | x \in \Capt(p_1)},
                C_1 \cup C_2}
            \\
            (\SRPatternEnv{t \land \SAcc{p_1}) /// p_1 => \Gamma_1, C_1} \qquad
            (\SRPatternEnv{t \setminus \SAcc{p_1}) /// p_2 => \Gamma_2, C_2}
            \: .
        \end{gather*}

        By the induction hypothesis applied to both $ p_1 $ and $ p_2 $ we derive, for all $ x $,
        \[
            (\SPatternEnv{t' \land \SAcc{p_1} // p_1})(x) \leq
                \Gamma_1(x) \theta \qquad
            (\SPatternEnv{t' \setminus \SAcc{p_1} // p_2})(x) \leq
                \Gamma_2(x) \theta
        \]
        from which we can conclude
        \[
            (\SPatternEnv{t' // p})(x) =
            (\SPatternEnv{t' \land \SAcc{p_1} // p_1})(x) \lor
                (\SPatternEnv{t' \setminus \SAcc{p_1} // p_2})(x) \leq
            \Gamma_1(x) \theta \lor \Gamma_2(x) \theta
            \: .
            \qedhere
        \]
    \end{LCases}
\end{Proof}

\begin{Lemma}[Precise solution to environment reconstruction constraints]
\label{lem:a-sr-patterns-precise}
    Let $ p $ be a pattern, $ t $ a type, and $ \theta $ a type substitution
    such that $ t \theta \leq \SAcc{p} $.
    Let $ \SRPatternEnvA{t /// p => \Gamma, C | A} $,
    with $ A \Fresh \Dom(\theta) $.

    There exists a type substitution $ \theta' $
    such that $ \Domain(\theta') = A $,
    that $ \SRSat{(\theta \cup \theta') |- C} $,
    and that, for all $ x \in \Capt(p) $,
    $ \Gamma(x) (\theta \cup \theta') \leq (\SPatternEnv{t \theta // p})(x) $.
\end{Lemma}

\begin{Proof}
    By structural induction on $ p $.

    \begin{LCases}
    \Cases[$ p = \Wildcard $ and $ p = c $]
        In both cases we take $ \theta' = \EmptySubst $.

    \Case[$ p = x $]
        We have
        \begin{gather*}
            \SRPatternEnvA{t /// x =>
                \Set{x\colon t}, \varnothing | \varnothing}
            \: .
        \end{gather*}

        We take $ \theta' = \EmptySubst $
        and have $ t (\theta \cup \theta') \leq t \theta $.

    \Case[$ p = (p_1, p_2) $]
        We have
        \begin{gather*}
            \SRPatternEnvA{t /// (p_1,p_2) => \Gamma_1 \cup \Gamma_2,
                C_1 \cup C_2 \cup
                \Set{\SRSub{t < \alpha_1 \times \alpha_2}} |
                A_1 \SRDisj A_2 \SRDisj \Set{\alpha_1, \alpha_2}}
            \\
            \SRPatternEnvA{\alpha_1 /// p_1 => \Gamma_1, C_1 | A_1} \qquad
            \SRPatternEnvA{\alpha_2 /// p_2 => \Gamma_2, C_2 | A_2} \qquad
            A_1, A_2, \alpha_1, \alpha_2 \Fresh t
            \: .
        \end{gather*}

        Let $
            \theta^\star = \theta \cup \Subst[\SubstF{\SPiFst(t \theta)/\alpha_1}, \SubstF{\SPiSnd(t \theta)/\alpha_2}]
        $. We have $ t \theta' = t \theta $ and $
            t \theta' \leq \SAcc{(p_1,p_2)} = \SAcc{p_1} \times \SAcc{p_2}
        $; thus, by Property~\ref{pro:a-s-productproj}, $
            \SPiIth(t \theta') \leq \SAcc{p_i}
        $. We also have $ A_i \Fresh \Dom(\theta^\star), \alpha_i $ for both $ i $,
        since $ \Set{\alpha_1,\alpha_2} $ is disjoint from each $ A_i $.

        We can therefore apply the induction hypothesis to $ p_i $,
        $ \alpha_i $, and $ \theta^\star $, for both $ i $.
        We derive from each that there is a substitution $ \theta_i' $
        with domain $ A_i $, such that $
            \SRSat{(\theta^\star \cup \theta_i') |- C_i}
        $ and, for all $ x \in \Capt(p_i) $, $
            \Gamma_i(x)(\theta^\star \cup \theta_i') \leq
            (\SPatternEnv{\alpha_i \theta^\star // p_i})(x)
        $.

        We take $
            \theta' = \Subst[\SubstF{\SPiFst(t \theta)/\alpha_1},
                \SubstF{\SPiSnd(t \theta)/\alpha_2}]
                \cup \theta_1' \cup \theta_2'
        $. We have $
            \SRSat{(\theta \cup \theta') |-
                C_1 \cup C_2 \cup \Set{\SRSub{t < \alpha_1 \times \alpha_2}}}
        $ since it satisfies $ C_1 $ and $ C_2 $ and since $
            t\theta \leq (\alpha_1 \times \alpha_2) \theta' =
            \SPiFst(t \theta) \times \SPiSnd(t \theta)
        $.

    \Case[$ p = \Tagg(p_1) $]
        We have
        \begin{gather*}
            \SRPatternEnvA{t /// \Tagg(p_1) => \Gamma_1,
                C_1 \cup \Set{\SRSub{t < \Tagg(\alpha)}} |
                A_1 \SRDisj \Set{\alpha}} \qquad
            \SRPatternEnvA{\alpha /// p_1 => \Gamma_1, C_1 | A_1} \qquad
            A_1, \alpha \Fresh t
            \: .
        \end{gather*}

        Analogously to the previous case, we construct $
            \theta^\star = \theta \cup \SubstSingle[\SPiTagg(t \theta)/\alpha]
        $ and apply the induction hypothesis to $ p_1 $, $ \alpha $,
        and $ \theta^\star $. We derive $ \theta_1' $ and take $
            \theta' = \SubstSingle[\SPiTagg(t \theta)/\alpha] \cup \theta_1'
        $.

    \Case[$ p = \PAnd{p_1 \& p_2} $]
        We have
        \begin{gather*}
            \SRPatternEnvA{t /// \PAnd{p_1 \& p_2} =>
                \Gamma_1 \cup \Gamma_2,
                C_1 \cup C_2 |
                A_1 \SRDisj A_2} \qquad
            \SRPatternEnvA{t /// p_1 => \Gamma_1, C_1 | A_1} \qquad
            \SRPatternEnvA{t /// p_2 => \Gamma_2, C_2 | A_2}
            \: .
        \end{gather*}

        For both $ i $, we apply the induction hypothesis to $ p_i $, $ t $,
        and $ \theta $ to derive $ \theta_i' $.
        We take $ \theta' = \theta_1' \cup \theta_2' $.

    \Case[$ p = \POr{p_1 | p_2} $]
        We have
        \begin{gather*}
            \SRPatternEnvA{t /// \POr{p_1 | p_2} =>
                \SetC{x\colon \Gamma_1(x) \lor \Gamma_2(x) | x \in \Capt(x)},
                C_1 \cup C_2 | A_1 \SRDisj A_2} \\
            \SRPatternEnvA{(t \land \SAcc{p_1}) /// p_1 =>
                \Gamma_1, C_1 | A_1} \qquad
            \SRPatternEnvA{(t \setminus \SAcc{p_1}) /// p_2 =>
                \Gamma_2, C_2 | A_2}
            \: .
        \end{gather*}

        We apply the induction hypothesis to $ p_1 $, $ t \land \SAcc{p_1} $,
        and $ \theta $ to derive $ \theta_1' $. We apply it to $ p_2 $,
        $ t \setminus \SAcc{p_1} $, and $ \theta $ to derive $ \theta_2' $;
        here, note that $ t \theta \leq \SAcc{p_1} \lor \SAcc{p_2} $
        implies $ t \theta \setminus \SAcc{p_1} \leq \SAcc{p_2} $.

        We take $ \theta' = \theta_1' \cup \theta_2' $. We have
        $ \SRSat{(\theta \cup \theta') |- C} $ since it satisfies $ C_1 $
        and $ C_2 $. Furthermore, for all $ x $, we have $
            \Gamma_1 (x) (\theta \cup \theta_1') \leq
            (\SPatternEnv{t \theta \land \SAcc{p_1} // p_1})(x)
        $ and $
            \Gamma_2 (x) (\theta \cup \theta_2') \leq
            (\SPatternEnv{t \theta \setminus \SAcc{p_1} // p_2})(x)
        $. Then, $
            \Gamma(x)(\theta \cup \theta') =
            \Gamma_1 (x) (\theta \cup \theta') \lor
                \Gamma_2 (x) (\theta \cup \theta') =
            \Gamma_1 (x) (\theta \cup \theta_1') \lor
                \Gamma_2 (x) (\theta \cup \theta_2')
        $, since $ A_1 $ and $ A_2 $ are disjoint and both are disjoint from
        $ \Var(t) $. Finally, $
            \Gamma_1 (x) (\theta \cup \theta_1') \lor
                \Gamma_2 (x) (\theta \cup \theta_2') \leq
            (\SPatternEnv{t \theta // p})(x)
        $.
    \qedhere
    \end{LCases}
\end{Proof}

\begin{Theorem}[Soundness of constraint generation and rewriting]
\label{thm:a-sr-soundness}
    Let $ e $ be an expression, $ t $ a type,
    and $ \Gamma $ a type environment.
    If $ \SRConstrGen{ e: t => C } $, $ \SRConstrRewr{ \Gamma |- C => D } $,
    and $ \SRSat{ \theta |- D } $,
    then $ \STyping{ \Gamma \theta |- e: t \theta } $.
\end{Theorem}

\begin{Proof}
    By structural induction on $ e $.

    \begin{LCases}
    \Case[$ e = x $]
        We have
        \begin{gather*}
            \SRConstrGen{x: t => \Set{\SRSub{x < t}}} \\
            \SRConstrRewr{\Gamma |- \Set{\SRSub{x < t}} =>
                \Set{\SRSub{
                    t_x \SubstDots[\beta_1/\alpha_1 ... \beta_n/\alpha_n] < t}}}
            \qquad
            \Gamma(x) = \Scheme{\Set{\alpha_1, \dots, \alpha_n}. t_x}
            \: .
        \end{gather*}

        Let $ A = \Set{\alpha_1, \dots, \alpha_n} $.
        By \textalpha-renaming we assume $ A \Fresh \theta $; then we have
        $ (\Gamma \theta)(x) = (\Scheme{A. t_x})\theta =
          \Scheme{A. (t_x \theta)} $.
        Consider the substitution
        $ \theta_x = \SubstDots[\beta_1 \theta/\alpha_1 ...
                                \beta_n \theta/\alpha_n]
        $. It has domain $ A $, so we can derive
        $ \STyping{\Gamma \theta |- x: t_x \theta \theta_x} $.

        We show $
            t_x \theta \theta_x =
            t_x \SubstDots[\beta_1/\alpha_1 ... \beta_n/\alpha_n] \theta
        $ by showing $
            \alpha \theta \theta_x =
            \alpha \SubstDots[\beta_1/\alpha_1 ... \beta_n/\alpha_n] \theta
        $ holds for all $ \alpha \in \Var(t_x) $. Either $ \alpha \in A $ or not. In the first case, $ \alpha = \alpha_i $ for some $ i $; then
        $
            \alpha \theta \theta_x = \alpha \theta_x = \beta_i \theta
        $ and $
            \alpha \SubstDots[\beta_1/\alpha_1 ... \beta_n/\alpha_n] \theta
            = \beta_i \theta
        $.
        In the latter, $ \alpha \neq \alpha_i $ for all $ i $; then $
            \alpha \theta \theta_x = \alpha \theta
        $, since $
            \Var(\alpha \theta) \cap \Dom(\theta_x) = \varnothing
        $ and $
            \alpha \SubstDots[\beta_1/\alpha_1 ... \beta_n/\alpha_n] \theta
            = \alpha \theta
        $.

        Therefore we derive $
            \STyping{\Gamma \theta |- x:
                t_x \SubstDots[\beta_1/\alpha_1 ... \beta_n/\alpha_n] \theta}
        $ by \Rule{Ts-Var}. Finally, since $
            \SRSat{\theta |- \SRSub{
                t_x \SubstDots[\beta_1/\alpha_1 ... \beta_n/\alpha_n] < t}}
        $, we derive $
            \STyping{\Gamma \theta |- x: t \theta}
        $ by subsumption.

    \Case[$ e = c $]
        We have
        \begin{gather*}
            \SRConstrGen{c: t => \Set{\SRSub{c < t}}} \qquad
            \SRConstrRewr{\Gamma |- \Set{\SRSub{c < t}} =>
                \Set{\SRSub{c < t}}}
            \: .
        \end{gather*}
        Analogously to the previous case, we first apply \Rule{Ts-Const}
        and then conclude by subsumption.

    \Case[$ e = \Abstr{x. e_1} $]
        We have
        \begin{gather*}
            \SRConstrGen{\Abstr{x. e_1}: t => \Set{
                \SRDef{ \Set{x\colon \alpha} | C_1 },
                \SRSub{\alpha \to \beta < t} }} \qquad
            \SRConstrGen{e_1: \beta => C_1} \\
            \SRConstrRewr{\Gamma |-
                \Set{ \SRDef{ \Set{x\colon \alpha} | C_1 },
                    \SRSub{\alpha \to \beta < t} } =>
                D_1 \cup \Set{ \SRSub{\alpha \to \beta < t} }} \qquad
            \SRConstrRewr{\Gamma, \Set{x\colon \alpha} |- C_1 => D_1}
            \: .
        \end{gather*}

        By the induction hypothesis we derive $
            \STyping{\Gamma \theta, \Set{x\colon \alpha \theta} |- e_1:
                \beta \theta}
        $. We apply \Rule{Ts-Abstr} to derive $
            \STyping{\Gamma \theta |- \Abstr{x. e_1}: (\alpha \to \beta) \theta}
        $. Since $ \SRSat{\theta |- D} $, we have $
            (\alpha \to \beta) \theta \leq t \theta
        $. Hence, we derive by subsumption $
            \STyping{\Gamma \theta |- \Abstr{x. e_1}: t \theta}
        $.

    \Case[$ e = \Appl{e_1}{e_2} $]
        We have
        \begin{gather*}
            \SRConstrGen{\Appl{e_1}{e_2}: t =>
                C_1 \cup C_2 \cup \Set{\SRSub{\beta < t}}} \qquad
            \SRConstrGen{e_1: \alpha \to \beta => C_1} \qquad
            \SRConstrGen{e_2: \alpha => C_2} \\
            \SRConstrRewr{\Gamma |-
                C_1 \cup C_2 \cup \Set{\SRSub{\beta < t}} =>
                D_1 \cup D_2 \cup \Set{\SRSub{\beta < t}}} \qquad
            \SRConstrRewr{\Gamma |- C_1 => D_1} \qquad
            \SRConstrRewr{\Gamma |- C_2 => D_2}
            \: .
        \end{gather*}

        We derive $
            \STyping{\Gamma \theta |- e_1: (\alpha \theta) \to (\beta \theta)}
        $ and $
            \STyping{\Gamma \theta |- e_2: \alpha \theta}
        $ by the induction hypothesis. Then by \Rule{Ts-Appl} we derive $
            \STyping{\Gamma \theta |- \Appl{e_1}{e_2}: \beta \theta}
        $, and finally---since $ \beta \theta \leq t \theta $---we conclude
        by subsumption.

    \Case[$ e = (e_1, e_2) $]
        We have
        \begin{gather*}
            \SRConstrGen{(e_1, e_2): t =>
                C_1 \cup C_2 \cup
                    \Set{\SRSub{ \alpha_1 \times \alpha_2 < t }}} \qquad
            \SRConstrGen{e_1: \alpha_1 => C_1} \qquad
            \SRConstrGen{e_2: \alpha_2 => C_2} \\
            \SRConstrRewr{\Gamma |-
                C_1 \cup C_2 \cup
                    \Set{\SRSub{ \alpha_1 \times \alpha_2 < t }} =>
                 D_1 \cup D_2 \cup
                \Set{\SRSub{ \alpha_1 \times \alpha_2 < t }}} \\ \qquad
            \SRConstrRewr{\Gamma |- C_1 => D_1} \qquad
            \SRConstrRewr{\Gamma |- C_2 => D_2}
            \: .
        \end{gather*}

        We have $ \STyping{\Gamma \theta |- e_i: \alpha_i \theta} $
        for both $ i $ by the induction hypothesis. Then, we derive $
            \STyping{\Gamma \theta |- (e_1, e_2):
                (\alpha_1 \times \alpha_2) \theta}
        $ by \Rule{Ts-Pair}, and finally conclude by subsumption.

    \Case[$ e = \Tagg(e_1) $]
        We have
        \begin{gather*}
            \SRConstrGen{\Tagg(e_1): t => C} \qquad
            C = C_1 \cup \Set{\SRSub{ \Tagg(\alpha) < t }} \qquad
            \SRConstrGen{e_1: \alpha => C_1} \\
            \SRConstrRewr{\Gamma |- C => D} \qquad
            D = D_1 \cup \Set{\SRSub{ \Tagg(\alpha) < t }} \qquad
            \SRConstrRewr{\Gamma |- C_1 => D_1}
            \: .
        \end{gather*}

        Analogous to the previous case. We apply the induction hypothesis,
        then \Rule{Ts-Tag}, then subsumption.

    \Case[$ e = \Match{e_0 with (p_i \to e_i)_{i \in I}} $]
        We have
        \begin{align*}
            & \SRConstrGen{\Match{e_0 with (p_i \to e_i)_{i \in I}}: t => C}
            \\ & \qquad
            C = \Set{
                \SRLet{ C_0' | (\Gamma_i \In C_i')_{i \in I} },
                \SRSub{ \beta < t }} \\ & \qquad
            C_0' = C_0 \cup (\textstyle\bigcup_{i \in I} C_i) \cup
                \Set{ \SRSub{ \alpha < \textstyle\bigvee_{i \in I} \SAcc{p_i}} }
            \qquad
            \SRConstrGen{e_0: \alpha => C_0} \\ & \qquad
            \forall i \in I \qquad
            t_i = (\alpha \setminus \textstyle\bigvee_{j < i} \SAcc{p_j})
                \land \SAcc{p_i}
            \qquad
            \SRPatternEnv{t_i /// p_i => \Gamma_i, C_i} \qquad
            \SRConstrGen{e_i: \beta => C_i'} \\ &
            \SRConstrRewr{\Gamma |- C => D} \\ & \qquad
            \SRConstrRewr{\Gamma |- C_0' => D_0'} \qquad
            \SRConstrRewr{\Gamma |- C_0 => D_0} \qquad
            D_0' = D_0 \cup (\textstyle\bigcup_{i \in I} C_i) \cup
                \Set{\SRSub{ \alpha < \textstyle\bigvee_{i \in I} \SAcc{p_i}}}
            \\ & \qquad
            \theta_0 \in \Tally(D_0') \qquad
            \forall i \in I . \enspace
            \SRConstrRewr{\Gamma,
                          \Gen_{\Gamma \theta_0}(\Gamma_i \theta_0) |-
                          C_i' => D_i'} \\ & \qquad
            D = \SRDe(\theta_0) \cup (\textstyle\bigcup_{i \in I} D_i')
                \cup \Set{\SRSub{\beta < t}}
        \end{align*}
        and we must show
        $ \STyping{\Gamma \theta |- \Match{e_0 with (p_i \to e_i)_{i \in I}}:
            t \theta} $.

        We prove it by establishing, for some types $ \hat{t_0} $
        and $ \hat{t_i} $, $ \hat{t_i'} $ for each $ i $, that
        \begin{gather*}
            \STyping{\Gamma \theta |- e_0: \hat{t_0}} \qquad
            \hat{t_0} \leq \textstyle\bigvee_{i \in I} \SAcc{p_i} \qquad
            \hat{t_i} = (\hat{t_0} \setminus \bigvee_{j < i} \SAcc{p_j}) \land
                \SAcc{p_i} \\
            \forall i \in I . \qquad
            \STyping{\Gamma \theta,
                \Gen_{\Gamma \theta}(\SPatternEnv{\hat{t_i} // p_i}) |- e_i:
                \hat{t_i'}} \qquad
            \textstyle\bigvee_{i \in I} \hat{t_i'} \leq t \theta
            \enspace .
        \end{gather*}

        Since $ \theta_0 \in \Tally(D_0') $, $ \SRSat{\theta_0 |- D_0'} $
        and thus $ \SRSat{\theta_0 |- D_0} $. Then, from
        \[
            \SRConstrGen{e_0: \alpha => C_0} \qquad
            \SRConstrRewr{\Gamma |- C_0 => D_0} \qquad
            \SRSat{\theta_0 |- D_0}
        \]
        we derive $ \STyping{\Gamma \theta_0 |- e_0: \alpha \theta_0} $
        by the induction hypothesis.

        Let $
            A = \Var(\alpha \theta_0) \setminus \Mvar(\Gamma \theta_0)
              = \Set{\alpha_1, \dots, \alpha_n}
        $. Let $
            B = \Set{\beta_1, \dots, \beta_n}
        $ be a set of type variables such that
        $ B \Fresh \Gamma, \theta, \theta_0 $
        and let $
            \theta^\star = \SubstDots[\beta_1/\alpha_1 ... \beta_n/\alpha_n]
        $.
        We derive
        $ \STyping{\Gamma \theta_0 |- e_0: \alpha \theta_0 \theta^\star} $
        by Corollary~\ref{cor:a-s-typesubst2},
        since $ \theta^\star $ does not act on meaningful variables
        of $ \Gamma \theta_0 $.
        By Lemma~\ref{lem:a-s-typesubst}, we derive $
            \STyping{\Gamma \theta_0 \theta |- e_0:
                \alpha \theta_0 \theta^\star \theta}
        $; by Lemma~\ref{lem:a-s-weakening}, $
            \STyping{\Gamma \theta |- e_0: \alpha \theta_0 \theta^\star \theta}
        $ (we prove the required premises below).

        We take $ \hat{t_0} = \alpha \theta_0 \theta^\star \theta $. We have $
            \alpha \theta_0 \theta^\star \theta \leq
            \textstyle\bigvee_{i \in I} \SAcc{p_i}
        $ because $ \SRSat{ \theta_0 |- D_0' } $ implies $
            \alpha \theta_0 \leq
            \textstyle\bigvee_{i \in I} \SAcc{p_i}
        $ and because subtyping is preserved by substitutions
        (recall that the accepted types of patterns are closed). We also have $
            \hat{t_i} = t_i \theta_0 \theta^\star \theta
        $ for all $ i $.

        For each branch $ i $, from
        \[
            \SRConstrGen{e_i: \beta => C_i'} \qquad
            \SRConstrRewr{\Gamma, \Gen_{\Gamma \theta_0}(\Gamma_i \theta_0) |-
                C_i' => D_i'} \qquad
            \SRSat{\theta |- D_i'}
        \]
        we derive $
                \STyping{\Gamma \theta,
                    (\Gen_{\Gamma \theta_0}(\Gamma_i \theta_0)) \theta |-
                    e_i: \beta \theta}
        $ by the induction hypothesis.
        We derive by Lemma~\ref{lem:a-s-weakening} $
            \STyping{\Gamma \theta,
                \Gen_{\Gamma \theta}(\SPatternEnv{\hat{t_i} // p_i}) |-
                e_i: \beta \theta}
        $ (we prove the premises below).
        Thus we have $ \hat{t_i'} = \beta \theta $ for every branch;
        we apply \Rule{Ts-Match} to derive $
            \STyping{\Gamma \theta |- \Match{e_0 with (p_i \to e_i)_{i \in I}}:
                \beta \theta}
        $, then subsumption to derive $
            \STyping{\Gamma \theta |- \Match{e_0 with (p_i \to e_i)_{i \in I}}:
                t \theta}
        $.

    \Item{Proof of $
            \STyping{\Gamma \theta |- e_0: \alpha \theta_0 \theta^\star \theta}
        $ from $
            \STyping{\Gamma \theta_0 \theta |- e_0:
                \alpha \theta_0 \theta^\star \theta}
        $}
        To apply Lemma~\ref{lem:a-s-weakening}, we must show
        \[
            \SMoreGen{ \Gamma \theta < \Gamma \theta_0 \theta }
            \qquad
            \Mvar(\Gamma \theta) \subseteq
            \Mvar(\Gamma \theta_0 \theta)
            \: .
        \]

        To prove $
            \SMoreGen{ \Gamma \theta < \Gamma \theta_0 \theta }
        $, consider an arbitrary $ (x\colon \Scheme{A_x. t_x}) \in \Gamma $.
        By \textalpha-renaming, we assume
        $ A_x \Fresh \theta, \theta_0 $; then, we must prove $
            \SMoreGen{ \Scheme{A_x. t_x \theta} <
                \Scheme{A_x. t_x \theta_0 \theta} }
        $. For every $ \gamma $, $
            \gamma \theta \simeq \gamma \theta_0 \theta
        $ since $ \SRSat{\theta |- \SRDe(\theta_0)} $.
        Hence, $ t_x \theta \simeq t_x \theta_0 \theta $.

        Since $ t_x \theta \simeq t_x \theta_0 \theta $
        implies $ \Mvar(t_x \theta) = \Mvar(t_x \theta_0 \theta) $
        by Lemma~\ref{lem:a-s-mvar},
        this also shows $
            \Mvar(\Gamma \theta) \subseteq \Mvar(\Gamma \theta_0 \theta)
        $.

    \Item{Proof of $
            \STyping{\Gamma \theta,
                \Gen_{\Gamma \theta}(\SPatternEnv{\hat{t_i} // p_i}) |-
                e_i: \beta \theta}
        $ from $
            \STyping{\Gamma \theta,
                (\Gen_{\Gamma \theta_0}(\Gamma_i \theta_0)) \theta |-
                e_i: \beta \theta}
        $}
        By Lemma~\ref{lem:a-s-weakening}, we can prove the result by showing
        \begin{gather*}
            \SMoreGen{\Gamma \theta,
                \Gen_{\Gamma \theta}(\SPatternEnv{\hat{t_i} // p_i}) <
                \Gamma \theta,
                (\Gen_{\Gamma \theta_0}(\Gamma_i \theta_0)) \theta}
            \\
            \Mvar(\Gamma \theta,
                \Gen_{\Gamma \theta}(\SPatternEnv{\hat{t_i} // p_i})) \subseteq
            \Mvar(\Gamma \theta,
                (\Gen_{\Gamma \theta_0}(\Gamma_i \theta_0)) \theta)
            \: .
        \end{gather*}

         The second condition is straightforward.
         For the first, we prove, for every $ x \in \Capt(p_i) $, $
            \SMoreGen{\Gen_{\Gamma \theta}(
                (\SPatternEnv{\hat{t_i} // p_i})(x)) <
                (\Gen_{\Gamma \theta_0}(\Gamma_i \theta_0(x))) \theta}
        $. Let $
            \Gamma_i(x) = t_x
        $. Then, $
            \Gen_{\Gamma \theta_0}(\Gamma_i \theta_0(x)) =
            \Scheme{A. t_x \theta_0}
        $, where $ A $ is $
            \Var(\alpha \theta_0) \setminus \Mvar(\Gamma \theta_0)
        $ as defined above (not all variables in $ A $ appear in $
            t_x \theta_0
        $, but schemes are defined disregarding useless quantification).
        By \textalpha-renaming, we have $
            \Gen_{\Gamma \theta_0}(\Gamma_i \theta_0(x)) =
            \Scheme{B. t_x \theta_0 \theta^\star}
        $ and, since $ B \Fresh \theta $, $
            (\Gen_{\Gamma \theta_0}(\Gamma_i \theta_0(x))) \theta =
            \Scheme{B. t_x \theta_0 \theta^\star \theta}
        $.

        Since $
            \hat{t_i} \leq \hat{t_0} =
            \alpha \theta_0 \theta^\star \theta
        $ and since $
            \SRSat{\theta \circ \theta^\star \circ \theta_0 |- C_i}
        $ (because $ \SRSat{\theta_0 |- C_i} $),
        by Lemma~\ref{lem:a-sr-patterns-correct} we have $
            (\SPatternEnv{\hat{t_i} // p_i})(x) \leq
            t_x \theta_0 \theta^\star \theta
         $. Then, $
            \SMoreGen{
                \Gen_{\Gamma \theta}((\SPatternEnv{\hat{t_i} // p_i})(x)) <
                \Scheme{B. t_x \theta_0 \theta^\star \theta}}
        $ holds because all variables in $ B $ may be quantified
        when generalizing $
            (\SPatternEnv{\hat{t_i} // p_i})(x)
        $, since no $ \beta_i $ appears in $ \Gamma \theta $.
    \qedhere
    \end{LCases}
\end{Proof}

\begin{Theorem}[Completeness of constraint generation and rewriting]
\label{thm:a-sr-completeness}
    Let $ e $ be an expression, $ t $ a type,
    and $ \Gamma $ a type environment.
    Let $ \theta $ be a type substitution
    such that $ \STyping{ \Gamma \theta |- e: t \theta } $.

    Let $ \SRConstrGenA{ e: t => C | A } $,
    with $ A \Fresh \Gamma, \Dom(\theta) $.
    There exist a type-constraint set $ D $,
    a set of fresh type variables $ A' $, and a type substitution $ \theta' $,
    with $ \Dom(\theta') = A \cup A' $,
    such that $ \SRConstrRewrA{ \Gamma |- C => D | A' } $
    and $ \SRSat{ (\theta \cup \theta') |- D } $.
\end{Theorem}

\begin{Proof}
    By structural induction on $ e $.

    \begin{LCases}
    \Case[$ e = x $]
        We have
        \begin{gather*}
            \SRConstrGenA{x: t => \Set{\SRSub{x < t}} | \varnothing} \\
            \STyping{\Gamma \theta |- x: t \theta} \qquad
            (\Gamma \theta) (x) = \Scheme{A_x. t_x} \qquad
            \Dom(\theta_x) \subseteq A_x \qquad
            t_x \theta_x \theta \leq t \theta \: .
        \end{gather*}

        Given $ A_x = \Set{\alpha_1, \dots, \alpha_n} $,
        we pick a set $ A' = \Set{\beta_1, \dots, \beta_n} $
        of fresh variables.
        Let $ \hat{\theta} = \SubstC[\beta_i/\alpha_i | \alpha_i \in A_x] $.
        We have $
            \SRConstrRewrA{\Gamma |- \Set{\SRSub{x < t}} =>
                \Set{\SRSub{t_x \hat{\theta} < t}} | A'}
        $.

        We pick $
            \theta' =
            \SubstC[\alpha_i \theta_x \theta/\beta_i | \beta_i \in A']
        $. It remains to prove that $
            \SRSat{ (\theta \cup \theta') |-
                \Set{\SRSub{t_x \hat{\theta} < t}} }
        $, that is, that
        \[
            t_x \hat{\theta} (\theta \cup \theta') \leq t (\theta \cup \theta')
            = t \theta
        \]
        (the equality above holds because the variables in $ A' $ are fresh).

        We prove $
            t_x \hat{\theta} (\theta \cup \theta') = t_x \theta_x \theta
        $ (from which we can conclude because
        $ t_x \theta_x \theta \leq t \theta $).
        We prove it by showing $
            \gamma \hat{\theta} (\theta \cup \theta') = \gamma \theta_x \theta
        $ for every $ \gamma \in \Var(t_x) $.
        If $ \gamma \in A_x $, then $ \gamma = \alpha_i $ for some $ i $.
        Then, $ \gamma \hat{\theta} = \beta_i $ and $
            \gamma \hat{\theta} (\theta \cup \theta') = \alpha_i \theta_x \theta
        $.
        If $ \gamma \notin A_x $, then $
            \gamma \hat{\theta} (\theta \cup \theta') = \gamma \theta
        $ (the variables $ \theta' $ is defined on do not appear in $ t_x $);
        likewise, $ \gamma \theta_x \theta = \gamma \theta $
        since $ \theta_x $ is only defined on variables in $ A_x $.

    \Case[$ e = c $]
        We have
        \[
            \SRConstrGenA{c: t => \Set{\SRSub{c < t}} | \varnothing} \qquad
            \STyping{\Gamma \theta |- c: t \theta} \qquad
            c \leq t \theta
            \: .
        \]

        We have $
            \SRConstrRewrA{\Gamma |- \Set{\SRSub{c < t}} =>
                \Set{\SRSub{c < t}} | \varnothing}
        $. Let $ \theta' = \EmptySubst $.
        We have $ \SRSat{ (\theta \cup \theta') |- \Set{\SRSub{c < t}} } $
        because $ c \theta = c \leq t \theta $.

    \Case[$ e = \Abstr{x. e_1} $]
        We have
        \begin{gather*}
            \SRConstrGenA{\Abstr{x. e_1}: t =>
                \Set{ \SRDef{ \Set{x\colon \alpha} | C_1 },
                    \SRSub{\alpha \to \beta < t} } |
                A_1 \SRDisj \Set{\alpha, \beta}} \qquad
            \SRConstrGenA{e_1: \beta => C_1 | A_1} \qquad
            A_1, \alpha, \beta \Fresh t \\
            \STyping{\Gamma \theta |- \Abstr{x. e_1}: t \theta} \qquad
            \STyping{\Gamma \theta, \Set{x\colon t_1} |- e_1: t_2} \qquad
            t_1 \to t_2 \leq t \theta
            \: .
        \end{gather*}

        Let $
            \theta^\star =
            \theta \cup \Subst[\SubstF{t_1/\alpha}, \SubstF{t_2/\beta}]
        $. Note that $ \Gamma \theta^\star = \Gamma \theta $
        and $ t \theta^\star = t \theta $,
        because $ \Set{\alpha_1, \alpha_2} \Fresh \Gamma, t $.

        We have $
            \STyping{(\Gamma, \Set{x\colon \alpha}) \theta^\star |- e_1:
                \beta \theta^\star}
        $, $ \SRConstrGenA{e_1: \beta => C_1 | A_1} $,
        and $ A_1 \Fresh \Dom(\theta^\star) $.
        By the induction hypothesis, therefore, $
            \SRConstrRewrA{\Gamma, \Set{x\colon \alpha} |- C_1 => D_1 | A_1'}
        $ and $
            \SRSat{ (\theta^\star \cup \theta_1') |- D_1 }
        $, for some $ D_1 $, $ A_1' $, $ \theta_1' $
        such that $ \Dom(\theta_1') = A_1 \cup A_1' $
        and that the variables in $ A_1' $ are fresh.

        $ \SRConstrRewrA{ \Gamma, \Set{x\colon \alpha} |- C_1 => D_1 | A_1' } $
        implies $
            \SRConstrRewrA{ \Gamma |- \SRDef{\Set{x\colon \alpha} | C_1} =>
                D_1 | A_1' }
        $. Hence, we have $
            \SRConstrRewrA{ \Gamma |- C =>
                D = D_1 \cup \Set{\SRSub{\alpha \to \beta < t}} | A_1' }
        $. Let $
            \theta' =
            \Subst[\SubstF{t_1/\alpha}, \SubstF{t_2/\beta}] \cup \theta_1'
        $. It is defined on the correct domain and it solves the constraints,
        since it solves $ D_1 $ and since $
            (\alpha \to \beta) \theta' = t_1 \to t_2 \leq t \theta
        $.

    \Case[$ e = \Appl{e_1}{e_2} $]
        We have
        \begin{gather*}
            \SRConstrGenA{\Appl{e_1}{e_2}: t =>
                C_1 \cup C_2 \cup \Set{\SRSub{\beta < t}} |
                A_1 \SRDisj A_2 \SRDisj \Set{\alpha, \beta}} \\ \qquad
            \SRConstrGenA{e_1: \alpha \to \beta => C_1 | A_1} \qquad
            \SRConstrGenA{e_2: \alpha => C_2 | A_2} \qquad
            A_1, A_2, \alpha, \beta \Fresh t \\
            \STyping{\Gamma \theta |- \Appl{e_1}{e_2}: t \theta} \qquad
            \STyping{\Gamma \theta |- e_1: t_1 \to t_2} \qquad
            \STyping{\Gamma \theta |- e_2: t_1} \qquad
            t_2 \leq t \theta
            \: .
        \end{gather*}

        Let $
            \theta^\star =
            \theta \cup \Subst[\SubstF{t_1/\alpha}, \SubstF{t_2/\beta}]
        $. Note that $ \Gamma \theta^\star = \Gamma \theta $
        and $ t \theta^\star = t \theta $,
        since $ \alpha, \beta \Fresh \Gamma, t $.

        We have $
            \STyping{\Gamma \theta |- e_1: (\alpha \to \beta) \theta^\star}
        $, $ \SRConstrGenA{e_1: \alpha \to \beta => C_1 | A_1} $,
        and $ A_1 \Fresh \Dom(\theta^\star) $.
        By the induction hypothesis, therefore, $
            \SRConstrRewrA{ \Gamma |- C_1 => D_1 | A_1' }
        $ and $ \SRSat{ (\theta^\star \cup \theta_1') |- D_1 } $,
        for some $ D_1 $ and $ \theta_1' $
        with $ \Dom(\theta_1') = A_1 \cup A_1' $.

        Likewise, by applying the induction hypothesis to the derivation
        for $ e_2 $, we derive $ \SRConstrRewrA{\Gamma |- C_2 => D_2 | A_2'} $
        and $ \SRSat{(\theta^\star \cup \theta_2') |- D_2} $,
        for some $ D_2 $ and $ \theta_2' $
        with $ \Dom(\theta_2') = A_2 \cup A_2' $.

        We can thus conclude that $
            \SRConstrRewrA{\Gamma |- C =>
                D = D_1 \cup D_2 \cup \Set{\SRSub{\beta < t}} | A_1' \cup A_2'}
        $. Let $
            \theta' = \Subst[\SubstF{t_1/\alpha}, \SubstF{t_2/\beta}] \cup
                \theta_1' \cup \theta_2'
        $. It is defined on the correct domain and $ \theta \cup \theta' $
        solves the constraints: it solves both $ D_1 $ and $ D_2 $,
        and $
            \beta (\theta \cup \theta') = \beta \theta' = t_2 \leq t \theta =
            t (\theta \cup \theta')
        $.

    \Case[$ e = (e_1, e_2) $]
        We have
        \begin{gather*}
            \SRConstrGenA{(e_1, e_2): t =>
                C_1 \cup C_2 \cup
                \Set{\SRSub{\alpha_1 \times \alpha_2 < t}} |
                A_1 \SRDisj A_2 \SRDisj \Set{\alpha_1, \alpha_2}} \\ \qquad
            \SRConstrGenA{e_1: \alpha_1 => C_1 | A_1} \qquad
            \SRConstrGenA{e_2: \alpha_2 => C_2 | A_2} \qquad
            A_1, A_2, \alpha_1, \alpha_2 \Fresh t \\
            \STyping{\Gamma \theta |- (e_1, e_2): t \theta} \qquad
            \STyping{\Gamma \theta |- e_1: t_1} \qquad
            \STyping{\Gamma \theta |- e_2: t_2} \qquad
            t_1 \times t_2 \leq t \theta
            \: .
        \end{gather*}

        Analogous to the previous case. We define $
            \theta^\star =
            \theta \cup  \Subst[\SubstF{t_1/\alpha_1}, \SubstF{t_2/\alpha_2}]
        $ and proceed as above.

    \Case[$ e = \Tagg(e_1) $]
        We have
        \begin{gather*}
            \SRConstrGenA{\Tagg(e_1): t =>
                C_1 \cup \Set{\SRSub{\Tagg(\alpha) < t}} |
                A_1 \SRDisj \Set{\alpha}} \qquad
            \SRConstrGenA{e_1: \alpha => C_1 | A_1} \qquad
            A_1, \alpha \Fresh t \\
            \STyping{\Gamma \theta |- \Tagg(e): t \theta} \qquad
            \STyping{\Gamma \theta |- e_1: t_1} \qquad
            \Tagg(t_1) \leq t \theta
            \: .
        \end{gather*}

        Analogous to the two previous cases.
        Here we define $ \theta^\star = \theta \cup \SubstSingle[t_1/\alpha] $.

    \Case[$ e = \Match{e_0 with (p_i \to e_i)_{i \in I}} $]
        We have
        \begin{align*}
            & \SRConstrGenA{\Match{e_0 with (p_i \to e_i)_{i \in I}}: t =>
                \Set{
                    \SRLet{ C_0' | (\Gamma_i \In C_i') },
                    \SRSub{\beta < t}
                } | A} \\ & \qquad
            \SRConstrGenA{e_0: \alpha => C_0 | A_0} \qquad
            t_i = (\alpha \setminus \textstyle\bigvee_{j < i} \SAcc{p_j}) \land
                \SAcc{p_i} \\ & \qquad
            \forall i \in I \qquad
            \SRPatternEnvA{t_i /// p_i => \Gamma_i, C_i | A_i} \qquad
            \SRConstrGenA{e_i: \beta => C_i' | A_i'} \\ & \qquad
            C_0' = C_0 \cup (\textstyle\bigcup_{i \in I} C_i) \cup
                \Set{\SRSub{\alpha <  \textstyle\bigvee_{i \in I} \SAcc{p_i}}}
            \\ & \qquad
            A = A_0 \SRDisj (\textstyle\SRBigDisj_{i \in I} A_i) \SRDisj
                (\textstyle\SRBigDisj_{i \in I} A_i') \SRDisj
                \Set{\alpha, \beta}
            \\ &
            \STyping{\Gamma \theta |- \Match{e_0 with (p_i \to e_i)_{i \in I}}:
                t \theta} \\ & \qquad
            \STyping{\Gamma \theta |- e_0: t_0} \qquad
            t_0 \leq \textstyle\bigvee_{i \in I} \SAcc{p_i} \qquad
            t_i^\star = (t_0 \setminus \textstyle\bigvee_{j < i} \SAcc{p_j})
                \land \SAcc{p_i} \\ & \qquad
            \forall i \in I \qquad
            \STyping{\Gamma \theta,
                \Gen_{\Gamma \theta}(\SPatternEnv{t_i^\star // p_i}) |- e_i:
                t_i'} \qquad
            t' = \textstyle\bigvee_{i \in I} t_i' \leq t \theta
            \: .
        \end{align*}

        Let $
            \theta^\star =
                \theta \cup \SubstSingle[t_0/\alpha]
        $. Then we have
        \[
            \SRConstrGenA{e_0: \alpha => C_0 | A_0} \qquad
            \STyping{\Gamma \theta^\star |- e_0: \alpha \theta^\star} \qquad
            A_0 \Fresh \Gamma, \Dom(\theta^\star)
        \]
        and, by the induction hypothesis, we find
        $ D_0 $, $ A_0' $ (containing fresh variables), and $ \theta_0' $
        such that
        \[
            \SRConstrRewrA{\Gamma |- C_0 => D_0 | A_0'} \qquad
            \SRSat{\theta^\star \cup \theta_0' |- D_0} \qquad
            \Dom(\theta_0') = A_0 \cup A_0'
            \: .
        \]
        From $ \SRConstrRewrA{\Gamma |- C_0 => D_0 | A_0'} $
        we can derive
        \[
            \SRConstrRewrA{\Gamma |- C_0' =>
                D_0' =
                D_0 \cup (\textstyle\bigcup_{i \in I} C_i) \cup
                \Set{\SRSub{\alpha <  \textstyle\bigvee_{i \in I} \SAcc{p_i}}} | A_0'}
        \]
        because subtyping constraints are always rewritten to themselves.

        For each branch $ i $, note that $ t_i \theta^\star = t_i^\star $.
        By Lemma~\ref{lem:a-sr-patterns-precise},
        we can find $ \theta_i^\star $
        such that
        \[
            \Dom(\theta_i^\star) = A_i \qquad
            \SRSat{\theta^\star \cup \theta_i^\star |- C_i} \qquad
            \forall x \in \Capt(p_i) . \enspace
                \Gamma (x) (\theta^\star \cup \theta_i^\star) \leq
                (\SPatternEnv{t_i^\star // p_i}) (x)
            \: .
        \]
        Note also that $
            \SRSat{\theta^\star |-
                \SRSub{\alpha <  \textstyle\bigvee_{i \in I} \SAcc{p_i}}}
        $.
        We therefore have $
            \SRSat{\theta^\star \cup \theta_0'
                \cup (\bigcup_{i \in I} \theta_i^\star) |- D_0'}
        $. Let $
            \theta^{\star\star} = \theta^\star \cup
                \theta_0' \cup (\bigcup_{i \in I} \theta_i^\star)
        $.

        By the properties of tallying,
        if $ \Var(D_0') = \Set{\alpha_1, \dots, \alpha_n} $
        and given a set $ B = \Set{\alpha_1', \dots, \alpha_n'} $
        of fresh variables,
        there exist two substitutions $ \theta_0 \in \Tally(D_0') $
        and $ \theta_0'' $ such that
        \begin{gather*}
            \Dom(\theta_0) = \Var(D_0') \qquad
            \Var(\theta_0) = B \qquad
            \Dom(\theta_0'') = B \\
            \forall \gamma \notin \Var(\theta_0) . \enspace
            \gamma \theta_0 (\theta^{\star\star} \cup \theta_0'') \simeq
                \gamma \theta^{\star\star}
            \: .
        \end{gather*}

        Let $
            \theta^\top =
                \theta^{\star\star} \cup
                \SubstSingle[t'/\beta] \cup \theta_0''
        $. To apply the induction hypothesis for a branch $ i $, we need
        \[
            \SRConstrGenA{e_i: \beta => C_i' | A_i'} \qquad
            \STyping{
                (\Gamma, \Gen_{\Gamma \theta_0}(\Gamma_i \theta_0))
                \theta^\top |- e_i: \beta \theta^\top} \qquad
            A_i' \Fresh \Gamma, \Gen_{\Gamma \theta_0}(\Gamma_i \theta_0),
                \Dom(\theta^\top)
            \: .
        \]

        We derive the typing judgment above by subsumption and by weakening
        (we prove the premises below).
        As for the freshness condition, note that the variables in
        $ \Gamma \theta_0 $ are all either in $ \Gamma $
        or in $ \Var(\theta_0) $;
        in the latter case, they are fresh by our choice of $ B $.

        By applying the induction hypothesis to each branch $ i $,
        we therefore find $ D_i $, $ A_i'' $ (of fresh variables),
        and $ \theta_i' $ such that
        \[
            \SRConstrRewrA{
                \Gamma, \Gen_{\Gamma \theta_0}(\Gamma_i \theta_0) |-
                C_i' => D_i | A_i'' } \qquad
            \SRSat{\theta^\top \cup \theta_i' |- D_i} \qquad
            \Dom(\theta_i') = A_i' \cup A_i''
            \: .
        \]

        Hence, we have
        \[
            \SRConstrRewrA{ \Gamma \theta |-
                \Set{
                    \SRLet{ C_0' | (\Gamma_i \In C_i') },
                    \SRSub{\beta < t}
                } =>
                \SREq(\theta_0) \cup (\textstyle\bigcup_{i \in I} D_i) \cup
                    \Set{\SRSub{\beta < t}} | A' }
            \: ,
        \]
        where $ A' = A_0' \cup (\bigcup_{i \in I} A_i'') \cup \Var(\theta_0) $.

        We take $
            \theta' = [\SubstF{t_0/\alpha}, \SubstF{t'/\beta}] \cup
                \theta_0' \cup (\bigcup_{i \in I} \theta_i^\star) \cup
                \theta_0'' \cup (\bigcup_{i \in I} \theta_i')
        $. It has the correct domain; we must only show
        \[
            \SRSat{\theta \cup \theta' |-
                \SREq(\theta_0) \cup (\textstyle\bigcup_{i \in I} D_i) \cup
                    \Set{\SRSub{\beta < t}}}
            \: .
        \]

        The last constraint is satisfied since
        $ \beta (\theta \cup \theta') = t' \leq t \theta $.
        Constraints in $ \SREq(\theta_0) $ are of the form
        $ \alpha \leq \alpha \theta_0 $ or $ \alpha \theta_0 \leq \alpha $,
        for $ \alpha \in \Dom(\theta_0) $.
        Since these $ \alpha $ are not in $ \Var(\theta_0) $, we have $
            \alpha \theta_0 (\theta^{\star\star} \cup \theta_0'') \simeq
            \alpha \theta^{\star\star}
        $ and hence $
            \alpha \theta_0 (\theta \cup \theta') \simeq
            \alpha (\theta \cup \theta')
        $.
        For each $ i $, since $ \SRSat{\theta^\top \cup \theta_i' |- D_i} $,
        we have also $ \SRSat{\theta \cup \theta' |- D_i} $ (the other
        substitutions we add are not defined on the variables in $ D_i $).

    \Item{Proof of $
            \STyping{
                (\Gamma, \Gen_{\Gamma \theta_0}(\Gamma_i \theta_0))
                \theta^\top |- e_i: \beta \theta^\top}
        $ from $
            \STyping{\Gamma \theta,
                \Gen_{\Gamma \theta}(\SPatternEnv{t_i^\star // p_i}) |- e_i:
                t_i'}
        $}
        From $
            \STyping{\Gamma \theta,
                \Gen_{\Gamma \theta}(\SPatternEnv{t_i^\star // p_i}) |- e_i:
                t_i'}
        $, we derive $
            \STyping{\Gamma \theta,
                \Gen_{\Gamma \theta}(\SPatternEnv{t_i^\star // p_i}) |- e_i:
                \beta \theta^\top}
        $ by subsumption, since $ t_i' \leq t' = \beta \theta^\top $.
        We then apply Lemma~\ref{lem:a-s-weakening},
        which requires us to show the two premises
        \begin{gather*}
            \SMoreGen{
                (\Gamma, \Gen_{\Gamma \theta_0}(\Gamma_i \theta_0))
                \theta^\top
                <
                \Gamma \theta,
                \Gen_{\Gamma \theta}(\SPatternEnv{t_i^\star // p_i}) } \\
            \Mvar(
                (\Gamma, \Gen_{\Gamma \theta_0}(\Gamma_i \theta_0))
                \theta^\top
            ) \subseteq \Mvar(
                \Gamma \theta,
                \Gen_{\Gamma \theta}(\SPatternEnv{t_i^\star // p_i})
            )
            \: .
        \end{gather*}

        Note that $ \Gamma \theta = \Gamma \theta^\top $
        since the two substitutions only differ on variables
        introduced by constraint generation or tallying.
        Simplifying, we need to show
        \[
            \SMoreGen{
                \Gen_{\Gamma \theta_0}(\Gamma_i \theta_0) \theta^\top <
                \Gen_{\Gamma \theta}(\SPatternEnv{t_i^\star // p_i}) }
            \qquad
            \Mvar( (\Gen_{\Gamma \theta_0}(\Gamma_i \theta_0)) \theta^\top )
                \subseteq \Mvar(\Gamma \theta)
            \: .
        \]

        To prove the former, consider $ x \in \Capt(p_i) $
        and let $ \Gamma_i(x) = t_x $.
        We must show $
            \SMoreGen{
                \Gen_{\Gamma \theta_0}(t_x \theta_0) \theta^\top <
                \Gen_{\Gamma \theta}((\SPatternEnv{t_i^\star // p_i})(x)) }
        $.
        We have
        \[
            \Gen_{\Gamma \theta_0}(t_x \theta_0) = \Scheme{B_x. t_x \theta_0}
            \qquad
            B_x = \Var(t_x \theta_0) \setminus \Mvar(\Gamma \theta_0)
            \: .
        \]
        Note that all variables in $ \Var(t_x \theta_0) $ are
        in $ \Var(\theta_0) $: this is because all variables in $ \Var(t_x) $
        occur in $ D_0' $ ($ \alpha $ occurs in the exhaustiveness constraint,
        variables introduced by pattern environment generation occur in $ C_i $)
        and hence are in the domain of $ \theta_0 $.
        Then, $ B_x \subseteq B $: its elements are some of the $ \alpha_i' $
        in $ B $. Consider a set $
            B_x'  = \SetC{\alpha_i'' | \alpha_i' \in B_x}
        $ of fresh variables and the renaming $
            \tilde{\theta} = \SubstC[\alpha_i''/\alpha_i' | \alpha_i' \in B_x]
        $: we have
        \[
            \Gen_{\Gamma \theta_0}(t_x \theta_0) =
                \Scheme{B_x'. t_x \theta_0 \tilde{\theta}}
            \qquad
            B_x = \Var(t_x \theta_0) \setminus \Mvar(\Gamma \theta_0)
        \]
        and, since the variables in $ B_x' $ are fresh,
        \[
            (\Gen_{\Gamma \theta_0}(t_x \theta_0)) \theta^\top =
                \Scheme{B_x'. t_x \theta_0 \tilde{\theta} \theta^\top}
            \: .
        \]

        Consider an arbitrary instance $
            (\SPatternEnv{t_i^\star // p_i})(x) \hat{\theta}
        $ of $
            \Gen_{\Gamma \theta}((\SPatternEnv{t_i^\star // p_i})(x))
        $; we have $
            \Dom(\hat{\theta}) \subseteq
                \Var((\SPatternEnv{t_i^\star // p_i})(x)) \setminus
                \Mvar(\Gamma \theta)
        $.
        We must show that there exists an instance of $
            (\Gen_{\Gamma \theta_0}(t_x \theta_0)) \theta^\top
        $ which is a subtype of it.
        We take the instance $
            t_x \theta_0 \tilde{\theta} \theta^\top \check{\theta}
        $, with $
            \check{\theta} =
                \SubstC[\beta_i \theta^\top \hat{\theta}/\beta_i' |
                    \beta_i' \in B_x']
        $.
        We have $
            t_x \theta_0 \tilde{\theta} \theta^\top \check{\theta} =
            t_x \theta_0 \theta^\top \hat{\theta}
        $: for each $ \alpha_i' \in \Var(t_x \theta_0) $,
        \[
            \alpha_i' \tilde{\theta} \theta^\top \check{\theta} =
            \alpha_i'' \theta^\top \check{\theta} =
            \alpha_i'' \check{\theta}
        \]
        (since all $ \alpha_i'' $ are fresh),
        and $ \alpha_i'' \check{\theta} = \alpha_i' \theta^\top \hat{\theta} $.
        We have $
            t_x \theta_0 \theta^\top \hat{\theta} \simeq
            t_x \theta^\top \hat{\theta}
        $ and $
            t_x \theta^\top \hat{\theta} \leq
            (\SPatternEnv{t_i^\star // p_i})(x) \hat{\theta}
        $ since $
            t_x \theta^\top \leq (\SPatternEnv{t_i^\star // p_i})(x)
        $.

        As for the condition on variables, we have $
            \Mvar( (\Gen_{\Gamma \theta_0}(\Gamma_i \theta_0)) \theta^\top )
            \subseteq
            \Var( (\Gen_{\Gamma \theta_0}(\Gamma_i \theta_0)) \theta^\top )
        $. Since $
            \Var( \Gen_{\Gamma \theta_0}(\Gamma_i \theta_0) ) \subseteq
            \Mvar(\Gamma \theta_0)
        $, $
            \Var( (\Gen_{\Gamma \theta_0}(\Gamma_i \theta_0)) \theta^\top )
            \subseteq \Mvar(\Gamma \theta_0 \theta^\top) = \Mvar(\Gamma \theta)
        $.
    \qedhere
    \end{LCases}
\end{Proof}

\subsection{Extensions}

We give full definitions for the three variants of the \VariantsS{} system
that we have sketched in Section~\ref{extensions}.

\subsubsection{Overloaded functions}
\label{sssec:a-extensions-overloaded}

To remove the restriction on the use of intersection types for functions,
we change the typing rule \Rule{Ts-Abstr}: we allow the derivation of an
intersection of arrow types for a \textlambda-abstraction if each of these
types is derivable. The modified rule is the following.
\begin{mathpar}
  \MyInfer[Ts-Abstr]
    { \forall j \in J .
        \enspace \Typing{ \Gamma, \Set{x\colon t_j'} |- e: t_j } }
    { \Typing{ \Gamma |- \Abstr{x. e}:
        \textstyle\bigwedge_{j \in J} t_j' \to t_j } }
    {}
\end{mathpar}

Furthermore, we change the typing rule for pattern matching
so that redundant branches are excluded from typing.
This is necessary to use intersections effectively for pattern matching:
in practice, to be able to assign to a function defined by pattern matching
one arrow type for each branch.
\begin{mathpar}
    \MyInfer[Ts-Match]
        { \STyping{ \Gamma |- e_0: t_0 } \\
          t_0 \leq \textstyle\bigvee_{i \in I} \SAcc{p_i} \\
          t_i = (t_0 \setminus \textstyle\bigvee_{j < i} \SAcc{p_j})
            \land \SAcc{p_i} \\\\
          \forall i \in I \quad
          {\begin{cases}
            t_i' = \Empty & \text{if $ t_i \leq \Empty $} \\
            \STyping{
                \Gamma, \Gen_\Gamma(\SPatternEnv{ t_i // p_i }) |- e_i: t_i'}
                & \text{otherwise}
          \end{cases}}
        }
        { \STyping{ \Gamma |- \Match{ e_0 with (p_i \to e_i)_{i \in I} }:
            \textstyle\bigvee_{i \in I} t_i' } }
        {}
\end{mathpar}

Finally, we also change the rule \Rule{Ts-Var} for variables:
we allow a variable to be typed with any intersection of instantiations,
rather than just with a single instantiation.
\begin{mathpar}
    \MyInfer[Ts-Var]
        { \forall i \in I . \: t_i \in \Inst(\Gamma(x)) }
        { \STyping{ \Gamma |- x: \textstyle\bigwedge_{i \in I} t_i } }
        {}
\end{mathpar}
This allows us to instantiate type schemes which express parametric polymorphism
(for instance, $ \Scheme{\alpha. \alpha \to \alpha} $)
into types which express ad hoc polymorphism
(e.g., $ (\Keyw{bool} \to \Keyw{bool}) \land (\Keyw{int} \to \Keyw{int}) $).

\subsubsection{Refining the type of expressions in pattern matching}

The extension we present here improves the typing of pattern matching
by introducing more precise types for some variables in the matched expression
when typing the branches.
These refined types take into account which patterns have been selected
and which have not;
they are introduced for variables that appear in the matched expression,
possibly below pairs or variant constructors,
but not inside applications or \Keyw{match} constructs.

We reuse pattern environment generation to describe the derivation of these
refined types.
However, we need to introduce a new production for patterns
to use when translating expressions to patterns:
\[
    \Bnf{p \= \cdots \| \SMPatternPair<p, p>}
    \: .
\]
Patterns of the form $ \SMPatternPair<p, p> $ should not occur in programs;
they are only for internal use in the type system.
Unlike normal pair patterns, these patterns may include repeated variables.

We need not define the dynamic semantics of these patterns, as it won't be used.
We define their accepted type as $
    \SAcc{\SMPatternPair<p_1, p_2>} = \SAcc{p_1} \times \SAcc{p_2}
$ and environment generation as
\[
    \SPatternEnv{t // \SMPatternPair<p_1, p_2>} =
        \SPatternEnv{\SPiFst(t) // p_1} \SPatternEnvAnd
        \SPatternEnv{\SPiSnd(t) // p_2}
    \: ,
\]
where $ \SPatternEnvAnd $, defined as
\[
    (\Gamma \SPatternEnvAnd \Gamma')(x) =
        \begin{cases}
            \Gamma(x)
                & \text{if $ x \in \Dom(\Gamma) \setminus \Dom(\Gamma') $} \\
            \Gamma'(x)
                & \text{if $ x \in \Dom(\Gamma') \setminus \Dom(\Gamma) $} \\
            \Gamma(x) \land \Gamma'(x)
                & \text{if $ x \in \Dom(\Gamma) \cap \Dom(\Gamma') $}
        \end{cases}
\]
is the pointwise intersection of type environments.

We define a translation $ \SExprPattern{\cdot} $ of expressions to patterns.
It preserves variables and variants, converts pairs to the new form,
and turns everything else into a wildcard.
\[
    \SExprPattern{e} =
        \begin{cases}
            x & \text{if $ e = x $} \\
            \SMPatternPair<\SExprPattern{e_1}, \SExprPattern{e_2}>
                & \text{if $ e = (e_1, e_2) $} \\
            \Tagg(\SExprPattern{e_1})
                & \text{if $ e = \Tagg(e_1) $} \\
            \Wildcard & \text{otherwise}
        \end{cases}
\]

We change the typing rule for pattern matching as follows.
\begin{mathpar}
    \MyInfer[Ts-Match]
    { \STyping{ \Gamma |- e_0: t_0 } \\
      t_0 \leq \textstyle\bigvee_{i \in I} \SAcc{p_i}
        \land \SAcc{\SExprPattern{e_0}} \\
      t_i = (t_0 \setminus \textstyle\bigvee_{j < i} \SAcc{p_j})
        \land \SAcc{p_i} \\\\
      \forall i \in I \\
      \STyping{ \Gamma, \Gen_\Gamma(\SPatternEnv{ t_i // \SExprPattern{e_0} }),
        \Gen_\Gamma(\SPatternEnv{ t_i // p_i }) |-
        e_i: t_i' } }
    { \STyping{ \Gamma |- \Match{ e_0 with (p_i \to e_i)_{i \in I} }:
        \textstyle\bigvee_{i \in I} t_i' } }
    {}
\end{mathpar}

The main difference is the addition of the type environment $
    \Gen_\Gamma(\SPatternEnv{ t_i // \SExprPattern{e_0} })
$ which provides the refined types for the variables in $ \SExprPattern{e_0} $.
This environment is added before the usual one for the pattern $ p_i $:
hence, the capture variables of $ p_i $ still take precedence.

We also add the requirement $ t_0 \leq \SAcc{\SExprPattern{e_0}} $
to ensure $ \SPatternEnv{t_i // \SAcc{e_0}} $ is well defined.
This is not restrictive because any well-typed $ e $ can be typed with
a subtype of $ \SAcc{\SExprPattern{e}} $.


\subsubsection{Applicability to OCaml}\label{ocaml}

We change the semantics of pattern matching to include undefined results.
These occur when matching constants of different basic types
or when matching different constructors (for instance, a constant and a pair).
We use the following definition.

\begin{Definition}[Semantics of pattern matching]
\label{def:a-ext-matching}
    We write $ \Matching{v/p} $ for the result of matching
    a value $ v $ against a pattern $ p $.
    We have either $ \Matching{v/p} = \varsigma $,
    where $ \varsigma $ is a substitution defined
    on the variables in $ \Capt(p) $,
    $ \Matching{v/p} = \MatchFail $, or $ \Matching{v/p} = \MatchUndef $.
    In the first case, we say that $ v $ matches $ p $
    (or that $ p $ accepts $ v $);
    in the second, we say that matching fails;
    in the third, we say that it is undefined.

    The definition of $ \Matching{v/p} $ is given inductively
    in Figure~\ref{fig:variants-semantics-matching-undef}.
\end{Definition}

\begin{figure}

    \begin{align*}
        \Matching{v / \Wildcard}
            & = \EmptySubst
            \\
        \Matching{v / x}
            & = \SubstSingle[v / x]
            \\
        \Matching{v / c}
            & =
            \begin{cases}
                \EmptySubst & \text{if } v = c \\
                \MatchFail &
                    \text{if $ v \in \Constants $, $ b_v = b_c $,
                        and $ v \neq c $} \\
                \MatchUndef  & \text{otherwise}
            \end{cases}
            \\
        \Matching{v / (p_1, p_2)}
            & =
            \begin{cases}
                \varsigma_1 \cup \varsigma_2 &
                    \text{if } v = (v_1, v_2)
                    \text{ and } \forall i . \:
                    \Matching{v_i / p_i} = \varsigma_i \\
                \MatchFail &
                    \text{if } v = (v_1, v_2)
                    \text{, } \exists i . \:
                    \Matching{v_i / p_i} = \MatchFail
                    \text{, and } \forall i . \:
                    \Matching{v_i / p_i} \neq \MatchUndef \\
                \MatchUndef & \text{otherwise}
            \end{cases}
            \\
        \Matching{v / \Tagg(p_1)}
            & =
            \begin{cases}
                \varsigma_1 &
                    \text{if } v = \Tagg(v_1)
                    \text{ and } \Matching{v_1 / p_1} = \varsigma_1 \\
                \MatchFail &
                    \text{if } v = \Tagg(v_1)
                    \text{ and } \Matching{v_1 / p_1} = \MatchFail
                    \text{ or if } v = \Tagg_1(v_1)
                    \text{ and } \Tagg_1 \neq \Tagg \\
                \MatchUndef & \text{otherwise}
            \end{cases}
            \\
        \Matching{v / \PAnd{p_1 \& p_2}}
            & =
            \begin{cases}
                \varsigma_1 \cup \varsigma_2 &
                    \text{if } \forall i . \: \Matching{v / p_i} = \varsigma_i
                    \\
                \MatchFail &
                    \text{if } \exists i . \: \Matching{v / p_i} = \MatchFail
                    \text{ and }
                    \forall i . \: \Matching{v / p_i} \neq \MatchUndef \\
                \MatchUndef & \text{otherwise}
            \end{cases}
            \\
        \Matching{v / \POr{p_1 | p_2}}
            & =
            \begin{cases}
                \Matching{v / p_1} &
                    \text{if } \Matching{v / p_1} \neq \MatchFail \\
                \Matching{v / p_2} & \text{otherwise}
            \end{cases}
    \end{align*}

    \caption{Semantics of pattern matching including undefined results.}
    \label{fig:variants-semantics-matching-undef}
\end{figure}

Recall that the function $ b_{(\cdot)} $ (used here for $ \Matching{v/c} $)
assigns a basic type $ b_c $ to each constant $ c $.

The notions of reduction are unchanged, but the rule \Rule{R-Match}
is made more restrictive by the changed definition of $ \Matching{v/p} $:
a \Keyw{match} expression reduces only if matching succeeds for a branch
and fails---but is never undefined---for all previous branches.
The type system should therefore ensure that, in a well-typed expression $
    \Match{v with (p_i \to e_i)_{i \in I}}
$, $ \Matching{v / p_i} = \MatchUndef $ never happens.
While this is true for \VariantsK, \VariantsS{} has to be restricted
to ensure this.

We first define the \emph{compatible type} $ \SCompat{p} $ of a pattern $ p $
inductively as follows:
    \begin{align*}
        \SCompat{\Wildcard} = \SCompat{x} & = \AnyType &
        \SCompat{c} & = b_c \\
        \SCompat{(p_1, p_2)} & = \SCompat{p_1} \times \SCompat{p_2} &
        \SCompat{\Tagg(p)} & =
            \Tagg(\SCompat{p}) \lor (\AnyVariantType \setminus \Tagg(\Any)) \\
        \SCompat{\PAnd{p_1 \& p_2}} & = \SCompat{p_1} \land \SCompat{p_2} &
        \SCompat{\POr{p_1 | p_2}} & = \SCompat{p_1} \lor \SCompat{p_2}
        \: ,
    \end{align*}
where $ \AnyVariantType $ is the top type for variants,
defined in Section~\ref{sec:variants-s-comparisons},
Footnote~\ref{fn:anyvariant}.
For all well-typed values $ v $, $ \STyping{\Gamma |- v: \SCompat{p}} $
holds if and only if $ \Matching{v / p} \neq \MatchUndef $.

We change the rule for pattern matching by requiring
the type $ t_0 $ we assign to the matched expression
to be a subtype of all compatible types $ \SCompat{p_i} $.
\begin{mathpar}
    \MyInfer[Ts-Match]
    { \STyping{ \Gamma |- e_0: t_0 } \\
      t_0 \leq \textstyle\bigvee_{i \in I} \SAcc{p_i}
        \land \textstyle\bigwedge_{i \in I} \SCompat{p_i} \\
      t_i = (t_0 \setminus \textstyle\bigvee_{j < i} \SAcc{p_j})
        \land \SAcc{p_i} \\\\
      \forall i \in I \\
      \STyping{ \Gamma, \Gen_\Gamma(\SPatternEnv{ t_i // p_i }) |-
        e_i: t_i' } }
    { \STyping{ \Gamma |- \Match{ e_0 with (p_i \to e_i)_{i \in I} }:
        \textstyle\bigvee_{i \in I} t_i' } }
    {}
\end{mathpar}

Note that this condition is somewhat more restrictive than necessary:
patterns which follow a catch-all (wildcard or variable) pattern---or
in general that are useless because previous patterns already cover all
cases---can be left out of the intersection.
The precise condition would be
\[
    t_0 \leq
        \textstyle\bigvee_{i \in I} \Big(\SAcc{p_i} \land
        \textstyle\bigwedge_{j < i} \SCompat{p_j} \Big)
    \: ,
\]
but we choose the simpler condition
since they only differ in case there is redundancy.

\end{document}